\documentclass[11pt]{article}
\usepackage{amssymb,amsmath,amsthm,graphicx,psfrag,ulem}

\pdfoutput=1

\usepackage{graphicx,subfigure}
\usepackage{epsfig}
\usepackage{amsmath}
\usepackage{amsfonts}
\usepackage{amssymb}
\usepackage[usenames]{color}
\usepackage[a4paper,left=2.cm,right=2.cm,top=2.5cm,bottom=2.5cm]{geometry}

\newcommand{\beq}{\begin{equation}}
\newcommand{\eeq}{\end{equation}}
\newcommand{\be}{\begin{equation}}
\newcommand{\ee}{\end{equation}}
\newcommand{\beqa}{\begin{eqnarray}}
\newcommand{\eeqa}{\end{eqnarray}}
\newcommand{\beqar}{\begin{eqnarray*}}
\newcommand{\eeqar}{\end{eqnarray*}}
\newcommand{\bea}{\begin{eqnarray}}
\newcommand{\eea}{\end{eqnarray}}

\newcommand{\lp}{\left(}
\newcommand{\rp}{\right)}
\def\im{{\rm i}}

\def\CD{{ \cal D }}

\def\ii{\textrm{i}}
\def\Y{\mathbb{Y}}

\def\dr{\frac{\partial}{\partial r}}

\begin{document}

\allowdisplaybreaks

\normalem

\title{An instability of higher-dimensional rotating black holes}

\vskip1cm
\author{\'Oscar J. C. Dias${}^{\dagger a}$, Pau Figueras${}^{\S b}$, Ricardo Monteiro${}^{\dagger c}$, Harvey S. Reall${}^{\dagger d}$, Jorge E. Santos${}^{\dagger e}$\\ \\ ${}^{\dagger}$ DAMTP, Centre for Mathematical Sciences, University of Cambridge, \\ Wilberforce Road, Cambridge CB3 0WA, UK \\ ${}^\S$ Centre for Particle Theory \& Department of Mathematical Sciences, Science Laboratories, \\ University of Durham, South Road, Durham DH1 3LE, UK \\ \\ \small{ ${}^{a}$ O.Dias@damtp.cam.ac.uk, ${}^{b}$ Pau.Figueras@durham.ac.uk, ${}^{c}$ R.J.F.Monteiro@damtp.cam.ac.uk,} \\ \small{ ${}^{d}$ H.S.Reall@damtp.cam.ac.uk, ${}^{e}$ J.E.Santos@damtp.cam.ac.uk}}
\date{25 January 2010}

\maketitle

\begin{abstract}
\noindent
We present the first example of a linearized gravitational instability of an asymptotically flat vacuum black hole. We study perturbations of a Myers-Perry black hole with equal angular momenta in an odd number of dimensions. We find no evidence of any instability in five or seven dimensions, but in nine dimensions, for sufficiently rapid rotation, we find perturbations that grow exponentially in time. The onset of instability is associated with the appearance of time-independent perturbations which generically break all but one of the rotational symmetries. This is interpreted as evidence for the existence of a new 70-parameter family of black hole solutions with only a single rotational symmetry. We also present results for the Gregory-Laflamme instability of rotating black strings, demonstrating that rotation makes black strings more unstable.
\end{abstract}


\tableofcontents

\section{Introduction}

In four spacetime dimensions, the Kerr solution is believed to be stable against small perturbations of the metric: if it is perturbed then it will eventually settle down to another Kerr solution with slightly different mass and angular momentum. Although this has not been proved rigorously, evidence in favour of this picture arises from studies of linearized gravitational perturbations of the Kerr solution \cite{Teukolsky:1972my,Whiting:1988vc}, and from numerical studies of black hole formation \cite{Stergioulas:2009zz}.

In higher dimensions, the stability of the Schwarzschild solution against linearized gravitational perturbations has been established \cite{Gibbons:2002pq,Kodama:2003jz,Ishibashi:2003ap}. However, it has been argued that the $D$-dimensional generalization of the Kerr solution, discovered by Myers and Perry \cite{Myers:1986un}, should exhibit an instability for sufficiently large angular momentum. The MP solution is parameterized by its mass $M$ and $\lfloor (D-1)/2 \rfloor$ angular momenta (where $\lfloor \rfloor$ stands for the smallest integer part). In Ref.~\cite{Emparan:2003sy}, it was shown that if only one angular momentum is non-vanishing then, in the limit in which this angular momentum is very large (using a scale set by $M$), the black hole locally resembles a black {\it brane}. However, black branes suffer from the classical Gregory-Laflamme instability \cite{Gregory:1993vy}. This suggests that ``ultraspinning" MP black holes will also be unstable. The argument can be generalized to cases in which several angular momenta become large while others remain small \cite{Emparan:2003sy}. 

To confirm the existence of the ultraspinning instability, and to identify the critical angular momentum beyond which an instability exists, requires a linear stability analysis of the Myers-Perry solution. For the Kerr solution, the equations governing linearized gravitational perturbations can be decoupled and reduced to a single scalar wave equation, which renders a stability analysis tractable \cite{Teukolsky:1972my}. However, no analogous decoupling has been discovered for the MP solution so the equations to be studied consist of many coupled linear partial differential equations, making the problem seem very difficult.

Recently there was significant progress with this problem. Consider a MP solution with a single non-vanishing angular momentum $J$. For given $M$, the ultraspinning instability is expected to occur only when $J$ exceeds some critical value $J_{\rm crit}$ determined by $M$. A solution with $J=J_{\rm crit}$ is expected to admit a stationary zero-mode, i.e. a time-independent perturbation that is regular on the future horizon and vanishes at infinity. Ref.~\cite{Dias:2009iu} determined $J_{\rm crit}$ and the stationary zero-mode for $D=7,8,9$. This was done by writing down a certain Ansatz for the form of the metric perturbation and solving the resulting PDEs numerically using a novel (in this context) method that we shall explain below. 

We regard the work of Ref.~\cite{Dias:2009iu} as very strong evidence for the existence of an ultraspinning instability beyond $J=J_{\rm crit}$. However, no actual instability, i.e. a perturbation growing in time, was demonstrated. In this paper, we shall demonstrate that some MP black holes do admit perturbations that grow exponentially in time, thereby providing the first example of a linearized gravitational instability of an asymptotically flat vacuum black hole solution.

We shall exploit the idea introduced in Ref.~\cite{Kunduri:2006qa} of considering MP solutions with enhanced symmetry. The generic MP solution has isometry group $R \times U(1)^n$ where $R$ corresponds to time translations and $n=\lfloor (D-1)/2 \rfloor$. However, this is enhanced when some of the angular momenta coincide. In particular, for odd $D$, the MP solution with all angular momenta equal has a much larger $R \times U(N+1)$ isometry group, where $D=2N+3$. Furthermore, the solution is cohomogeneity-1, i.e. it depends only on a single coordinate. The metric involves a fibration over complex projective space $CP^N$. Gravitational perturbations of this solution can be decomposed into scalar, vector and tensor types according to how they transform under isometries of $CP^N$. The tensors, which exist only for $N \ge 2$ ($D \ge 7$), were studied in Ref.~\cite{Kunduri:2006qa} and no evidence of any instability was found. The special case of $D=5$, for which only scalar perturbations exist, was studied in Ref.~\cite{Murata:2008yx}. Again, no evidence of any instability was found.

In this paper, we shall study scalar-type perturbations of these cohomogeneity-1 MP black holes. The symmetries enable the problem to be reduced to coupled linear ordinary differential equations (ODEs) which we solve numerically. We find no evidence of any instability for $D=5$ (consistent with Ref.~\cite{Murata:2008yx}) or $D=7$. However, for $D=9$, when $J$ exceeds a certain critical value $J_{\rm crit}$, there is a perturbation that grows exponentially in time, i.e. an instability. We believe that such an instability will exist for all (odd) $D \ge 9$ although we have demonstrated this only for $D=9$.

As expected, the onset of instability is associated with the appearance of a stationary zero-mode of the MP solution with $J=J_{\rm crit}$. This zero-mode is interesting for another reason.  It has been proved that a stationary, rotating black hole must admit a rotational isometry (i.e. a $U(1)$ isometry) \cite{Hollands:2006rj,Moncrief:2008mr}. However, all known higher-dimensional black holes have {\it multiple} rotational isometries (e.g. MP black holes have $\lfloor (D-1)/2 \rfloor$ commuting $U(1)$ isometries), i.e. more symmetry than one expects on the basis of general arguments. Therefore, it has been conjectured that there exist solutions with less symmetry than any known solution, specifically solutions with a single rotational symmetry \cite{Reall:2002bh}.

It was proposed in Ref.~\cite{Reall:2002bh} that one could seek evidence for the existence of such solutions in the same way that the first evidence was obtained for the existence of non-uniform black string solutions. For black strings, the static zero-mode associated with the onset of the Gregory-Laflamme instability was conjectured to describe the ``branching off" of a new family of non-uniform black string solutions from the already known branch of uniform solutions \cite{Horowitz:2001cz}. Perturbative \cite{Gubser:2001ac} and numerical \cite{Wiseman:2002zc} work subsequently confirmed that this was correct. For rotating black holes, the idea proposed in Ref.~\cite{Reall:2002bh} is to look for a stationary zero-mode of a MP solution. By analogy with the black string example, this could be interpreted as the branching off of a new family of solutions. If the zero-mode preserves only a single rotational symmetry then this would be evidence for the existence of new black holes with just one rotational symmetry. 

In our case, the stationary zero-mode generically preserves only a single rotational symmetry. Therefore we conjecture that there exists a family of stationary black hole solutions with just a single rotational symmetry, that bifurcates from the cohomogeneity-1 MP black hole solution at $J=J_{\rm crit}$. In fact, we find not just one stationary zero-mode, but a large family, corresponding to all scalar harmonics on $CP^N$ with a certain eigenvalue. For $D=9$, we shall argue that the new family of solutions will involve 70 independent parameters, considerably more than the 5 parameters required to specify the MP solution! If correct, this implies that any hope of specifying higher-dimensional black holes uniquely using just a few parameters is bound to fail. Of course, the new black hole solutions may turn out to be unstable themselves.

Recently, Ref.~\cite{Emparan:2009vd} has provided other evidence for the existence of higher-dimensional black holes with a single rotational symmetry. Using the ``blackfold" approach of Refs  \cite{Emparan:2009cs,Emparan:2009at}, approximate solutions were constructed for $D \ge 5$ that describe ``helical" black rings. The blackfold approximation is based on the fact that higher-dimensional black holes can have widely separated horizon scales. The results of the present paper concern black holes that lie outside the regime of validity of this approximation. Our results complement those of Ref.~\cite{Emparan:2009vd}  because we are presenting evidence for topologically spherical black holes with a single rotational symmetry whereas Ref.~\cite{Emparan:2009vd}  considered black rings.

A difference between our results and the expectations of Ref.~\cite{Emparan:2003sy} concerns the nature of the stationary zero-mode. In our case, this perturbation breaks all but one rotational symmetry. However, in Ref.~\cite{Emparan:2003sy}, it is argued that for single-spinning black holes the stationary zero-mode should preserve the symmetries of the background. The results of Ref.~\cite{Dias:2009iu} confirmed this expectation. Singly spinning black holes exhibit symmetry enhancement for $D \ge 6$. They are cohomogeneity-2 with isometry group $R \times U(1) \times SO(D-3)$ where $SO(D-3)$ has $S^{D-4}$ orbits.\footnote{
One can decompose metric perturbations of these solutions into scalar, vector and tensor types using this $SO(D-3)$ symmetry. The tensors, which exist only for $D \ge 7$ have been studied previously in Ref.~\cite{Kodama:2009bf} and show no evidence of any instability. However, tensor perturbations arise from deformations of the $S^{D-4}$ part of the metric, whereas the expected ultraspinning instability should arise from perturbations of the metric transverse to $S^{D-4}$. The results of Ref.~\cite{Dias:2009iu} indicate that the instability should be a scalar-type perturbation.}
The zero-mode constructed in Ref.~\cite{Dias:2009iu} preserves all of the symmetries. Our results raise the question of whether singly spinning MP black holes might admit further stationary zero-modes that break some of their symmetry, and provide further evidence for new black hole solutions with reduced symmetry.

A corollary of our approach is the first data for the Gregory-Laflamme instability \cite{Gregory:1993vy} for rotating black {\it strings}. We consider black strings obtained as the product of a cohomogeneity-1 MP black hole with a flat direction. We argue that such solutions are always classically unstable. Our numerical results demonstrate that the string becomes more unstable as the angular momentum increases: the instability becomes stronger (i.e. it occurs on a shorter time scale) and the critical wavelength of unstable modes decreases, as the angular momentum increases.

An important feature of cohomogeneity-1 MP black holes is that they exhibit an upper bound on their angular momentum. Solutions saturating this bound are extreme black holes. Because of this upper bound, it is not obvious that such black holes should exhibit the ultraspinning behaviour discussed in Ref.~\cite{Emparan:2003sy}. Therefore we should explain why we undertook this project.

Ref.~\cite{Emparan:2003sy} observed that the transition to ``black brane-like" behaviour of singly spinning MP black holes can be seen in the thermodynamics: a certain thermodynamic quantity changes from ``Kerr black hole-like" to ``black brane-like" at a critical value of the spin. This was generalized in Ref.~\cite{Dias:2009iu}, where it is argued that the quantity of interest is the Hessian matrix $H_{ij} = (\partial^2 (-S)/\partial J_i \partial J_j)_M$ where $S$ is the black hole entropy. This is positive definite for small angular momenta. However, it can become indefinite as the angular momenta increase. Hence, one can {\it define} an ultraspinning black hole to be one for which this Hessian fails to be positive definite.

For $D=5$ MP black holes, $H_{ij}$ is always positive definite \cite{Dias:2009iu} so such black holes are never ultraspinning. However, for $D \ge 6$, singly spinning MP black holes with large enough angular momentum are, of course, ultraspinning and the numerical results of Ref.~\cite{Dias:2009iu} supply strong evidence that the instability appears only when $H_{ij}$ fails to be positive definite.

The present paper was motivated by the observation that, for $D \ge 7$, cohomogeneity-1 MP black holes do satisfy the ultraspinning criterion once $J$ exceeds a critical value $J_{\rm ultra}$ (for given $M$). Hence, there {\it might} be an ultraspinning instability for $J > J_{\rm crit}$ where $J_{\rm ultra} < J_{\rm crit} < J_{\rm extreme}$. Our results show that there is no instability for $D=7$ but an instability does occur for $D=9$ and, we believe, for (odd) $D>9$.

This paper is organized as follows. Section \ref{sec:background} describes the cohomogeneity-1 black holes that we shall study. In Section~\ref{sec:strategyresults}, we give a brief discussion of our approach and present our results. In Section~\ref{sec:string}, we show that thermodynamics can be used to explain many of our results.  We also explain the relevance of the Hessian $H_{ij}$ and present plots exhibiting the region of parameter space for which MP black holes are ultraspinning according to the definition above. The technical details of our work are presented in the later Sections~\ref{sec:decomposition} and \ref{sec:eigenvalue}, and in the Appendices.

{\bf Note added.} As the present paper was nearing completion, Ref.~\cite{Shibata:2009ad} appeared. This paper performs a fully nonlinear numerical evolution of a perturbed $D=5$ singly spinning MP solution. An instability was found for sufficiently large spin. It is not clear how this very interesting work relates to the instability under discussion here since $D=5$ MP solutions are never ultraspinning (according to the definition above). Furthermore, the perturbation considered in Ref.~\cite{Shibata:2009ad} breaks the symmetry in the direction of rotation of the black hole whereas we (and Ref.~\cite{Dias:2009iu}) consider only perturbations that preserve this symmetry. We suspect that Ref.~\cite{Shibata:2009ad} has found an instability of $D=5$ MP solutions of a qualitatively different form from the one discussed in this paper. It would be interesting to know whether a similar instability occurs for $D>5$.

\section{Cohomogeneity-1 Myers-Perry black holes}

\label{sec:background}

The Kerr solution was extended to higher dimensions by Myers and Perry \cite{Myers:1986un}. The Myers-Perry family can be parameterized by a mass radius parameter $r_M$ and $\lfloor (D-1)/2 \rfloor$ angular momentum parameters $a_i$. In the particular case of equal angular momenta, $a_i = a$, the solution in odd dimensions $D=2 N+3$ is cohomogeneity-1. The metric can be written as:\footnote{The radial coordinate used here can be related to the standard Boyer-Lindquist radial coordinate of \cite{Myers:1986un} through $r^2 \to r^2+a^2$.}

\be
\label{background}
ds^2 = -f(r)^2dt^2 +g(r)^2dr^2 + h(r)^2[d\psi +A_a dx^a - \Omega(r)dt]^2 + r^2 \hat{g}_{ab} dx^a dx^b\,,
\ee
\bea
g(r)^2 = \left(1- \frac{r_M^{2N}}{ r^{2N}} + \frac{r_M^{2N}a^2}{r^{2N+2}}\right)^{-1} &,& \quad h(r)^2 = r^2\left( 1+ \frac{r_M^{2N}a^2}{ r^{2N+2}} \right)\,, \nonumber \\
f(r) = \frac{r}{g(r)h(r)} &,& \quad \Omega(r) = \frac{r_M^{2N}a}{ r^{2N} h^2} \,, \nonumber
\eea
where $\hat{g}_{a b}$ is the Fubini-Study metric on $CP^{N}$ with Ricci tensor $\hat{R}_{ab} =2(N+1) \hat{g}_{ab}\,$, and $A = A_a dx^a\,$ is related to the K\"ahler form $J$ by $dA=2J$. Surfaces of constant $t$ and $r$ have the geometry of a homogeneously squashed $S^{2N+1}$, written as an $S^1$ fibred over $CP^{N}$. The fibre is parameterized by the coordinate $\psi$, which has period $2\pi$. Explicit expressions for the metric $\hat{g}_{a b}$ and K\"ahler potential $A$ of $CP^N$ can be obtained through the iterative Fubini-Study construction summarized in Appendix~\ref{cpnappendix}.

The spacetime metric satisfies $R_{\mu\nu} =0$ and the solution is asymptotically flat. The event horizon is located at $r=r_+$ (the largest real root of $g^{-2}$) and it is a Killing horizon of $\xi = \partial_t+\Omega_H \partial_\psi\,$, where the angular velocity of the horizon is given by:
\be
\label{angvel}
\Omega_H = \frac{r_M^{2N} a}{r_+^{2N+2}+r_M^{2N}a^2}.
\ee
The mass $M$ and angular momentum $J$, defined with respect to $\partial_{\psi}$, are~\cite{Gibbons:2004ai}
\be 
\label{EJmpeq}
M = \frac{A_{2N+1}}{8\pi G}r_M^{2N}\left (N+\frac{1}{2} \right)\,, \qquad J  = \frac{A_{2N+1}}{8\pi G} (N+1) r_M^{2N}a\,,
\ee
where $A_{2N+1}$ is the area of a unit ($2N+1$)-sphere.

There is an extremality bound on the angular momentum which can be expressed as
\be
\left( \frac{a}{r_+} \right)^2 \leq \left( \frac{a_{\rm ext}}{r_+} \right)^2 = N\,, \qquad \mathrm{or} \qquad \lp \frac{a}{r_M} \rp^2 \le \left( \frac{a_{\rm ext}}{r_M} \right)^2 = \frac{N}{(N+1)^{(N+1)/N}}\,.
\ee
The solution saturating this bound has a regular, but degenerate, horizon. For fixed $r_+$, or $r_M$, ultraspinning behaviour occurs for 
\be
\label{eqn:a1}
\lp \frac{a}{r_+}\rp^2  > \lp \frac{a_1}{r_+}\rp^2 \equiv \frac{1}{2}\,, \qquad \mathrm{or} \qquad \lp \frac{a}{r_M}\rp^2 >  \lp \frac{a_1}{r_M}\rp^2  \equiv \frac{1}{2^{(N+1)/N}}\,.
\ee
Note that the range $a_1 < a \le a_{\rm ext}$ (for fixed $r_M$) for which the black hole is ultraspinning becomes larger as $N$ increases, and that $a_1=a_{\rm ext}$ if $N=1$, so there is no ultraspinning behaviour for $D=5$.

\section{Strategy and Results}

\label{sec:strategyresults}

\subsection{Strategy}

Ref.~\cite{Monteiro:2009ke} introduced new numerical techniques for determining negative modes of rotating black holes. In Ref.~\cite{Dias:2009iu}, these techniques were exploited to construct the stationary zero-mode expected to indicate the onset of an ultraspinning instability of a singly rotating MP black hole. In the present paper, we shall determine the stationary zero-mode indicating the onset of instability for cohomogeneity-1 black holes. However, our main achievement is to generalize these methods to demonstrate the existence of perturbations that grow exponentially in time. 

In order to explain our approach (and that of Ref.~\cite{Dias:2009iu}), it is useful to discuss black {\it strings}. We shall be interested in a rotating uniform black string solution
\be
\label{eqn:string}
ds^2_\mathrm{string}= g_{\mu\nu}dx^\mu dx^\nu + dz^2,
\ee
where $g_{\mu\nu}$ is a cohomogeneity-1 MP metric. Consider a transverse and traceless (TT) perturbation of the string of the form 
\be
\label{glpert}
ds^2_\mathrm{string} \; \to \; ds^2_\mathrm{string}+e^{\ii kz} h_{\mu\nu} (x) dx^\mu dx^\nu\;,
\ee
The TT condition is equivalent to $h_{\mu\nu}$ being TT with respect to the MP metric:
\be
\label{eqn:TTgauge}
h^\mu_{\phantom{\mu}\mu}=0\;, \qquad \nabla^\mu h_{\mu\nu} =0\;.
\ee
The linearized Einstein equation reduces to 
\be
\label{lichn}
 (\Delta_L h)_{\mu\nu} \equiv -\nabla^\rho \nabla_\rho h_{\mu \nu} -2R_{\mu \rho \nu \sigma} h^{\rho  \sigma} = -k^2 h_{\mu\nu},
\ee
where $\Delta_L$ is the Lichnerowicz operator for the MP background. Hence perturbations with non-zero $k$ correspond to {\it negative modes} of $\Delta_L$. The boundary conditions are that $h_{\mu\nu}$ should be regular on the future horizon and vanishing at infinity.

Our strategy (following Refs.~\cite{Monteiro:2009ke,Dias:2009iu}) for studying perturbations of the black hole will be to seek a solution of (\ref{lichn}), i.e. a negative mode of the black hole, and then vary the spin of the black hole until $k$ vanishes, i.e. the negative mode becomes a zero-mode. This strategy is motivated by the availability of numerical techniques for solving eigenvalue equations of the form (\ref{lichn}). Solutions with non-zero $k$ correspond to perturbations of the black string. Therefore our method will yield results for the Gregory-Laflamme instability of rotating black strings as well as enabling us to search for black hole instabilities.

We can Fourier analyze our perturbation in the time direction, and also decompose into Fourier modes in the $\psi$-direction, i.e. we assume that the dependence on $t$ and $\psi$ is given by
\be
\label{eqn:fourier}
 h_{\mu\nu} \propto e^{-\ii\omega t + \ii m \psi},
\ee
where $m$ is an integer. As we shall explain in detail below, we can also decompose the perturbation into harmonics on $CP^N$. These can be of scalar, vector or tensor type. The tensors were considered in Ref.~\cite{Kunduri:2006qa}. We shall restrict our attention to perturbations of scalar-type, which can be expanded in terms of scalar harmonics on $CP^N$. As usual, harmonics with different eigenvalue decouple from each other. The equations satisfied by $h_{\mu\nu}$ depend only on the eigenvalue of the harmonic in question. Eigenvalues of the scalar Laplacian on $CP^N$ are labelled by a non-negative integer $\kappa$ (see Section~\ref{sec:decomposition}). Hence our perturbation is labelled by $(\omega,m,\kappa)$. 

Consider the (Lorentzian) negative mode equation (\ref{lichn}). As we have explained above, this arises from classical perturbations of a rotating black string. The usual approach to this problem is to fix $(k,m,\kappa)$ and to determine $\omega$. However, our approach, generalizing that of Refs.~\cite{Monteiro:2009ke,Dias:2009iu} will be to fix $(\omega,m,\kappa)$ and determine the possible eigenvalue(s) $-k^2$. In other words, we are determining the wavenumber $k$ for which black string perturbations with given $m$ and $\kappa$ have time-dependence associated with the given $\omega$. We shall fix the overall scale $r_M=1$ and determine the eigenvalue(s) $-k^2$ for fixed $(\omega,m,\kappa)$ as $a$ increases from $0$ to extremality. If $k$ vanishes for some value of $a$ then the associated black hole admits a zero-mode with the given values of $(\omega,m,\kappa)$ (of course it must be checked that this is not pure gauge).

In searching for an instability, we are looking for solutions of (\ref{lichn}) with ${\rm Im}(\omega)>0$. A problem with our approach is that we expect unstable modes to have complex $\omega$ in general, with the real and imaginary parts of $\omega$ related in some way. In other words, for given $m,\kappa$ and $a$, the complex quantity $\omega$ will be a function of the real quantity $k$ and hence the real and imaginary parts of $\omega$ cannot be independent. If we try to follow the above strategy for a randomly chosen complex value of $\omega$, then this will not satisfy the required relation between its real and imaginary parts, and therefore our numerical method will not output a real value of $k$. In order to locate where $k$ vanishes we would have to scan over both $a$ and, say, the real part of $\omega$. It would be difficult to do this with high accuracy.

We shall circumvent this problem by restricting attention to modes with $m=0$, i.e. modes preserving the rotational symmetry of the black hole that follows from the rigidity theorem. As we shall explain below, there are reasons to expect that unstable modes with $m=0$ will have purely imaginary $\omega$: $\omega = i \Gamma$, $\Gamma>0$, and this is confirmed by our results. From the black string perspective, for given $a$, if the string is unstable then we expect it to  be unstable for perturbations with a range of $k$, hence there will be a range of $k$ for which there exist solutions with $\Gamma>0$. Conversely, in our approach, there will be a range of values of $\Gamma>0$ for which there exists a solution for $k$.

In summary, we shall set $m=0$, $r_M=1$ and, for given $(\Gamma,\kappa,a)$ we shall determine the possible eigenvalues $-k^2$. Then we vary $a$ until the eigenvalue vanishes. We have then found a black hole that admits an unstable zero-mode with the given values of $\Gamma$ and $\kappa$.

\subsection{Results for $D=5$}

\begin{figure}[t]
\centerline{\includegraphics[width=.45\textwidth]{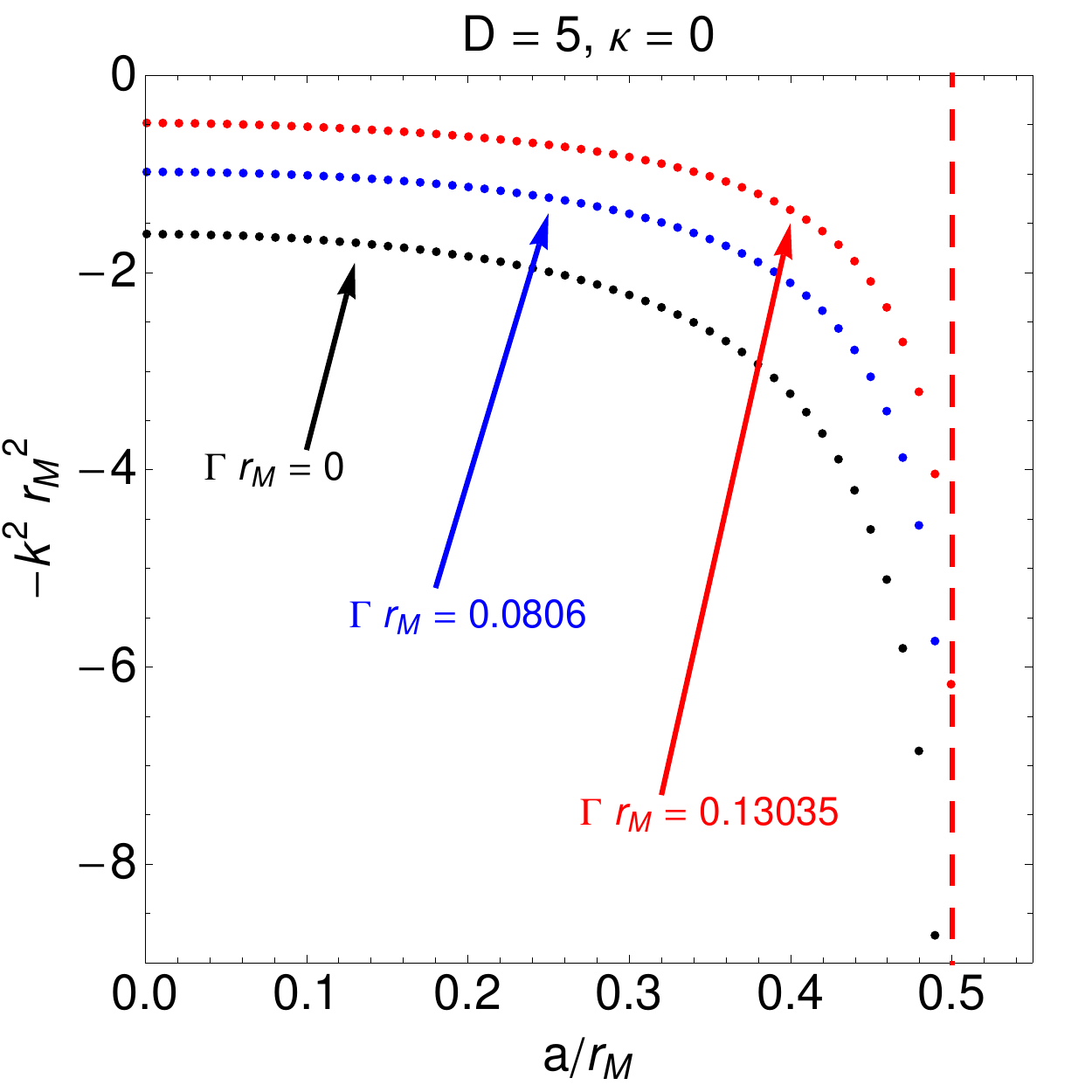}
\hspace{1cm}\includegraphics[width=.45\textwidth]{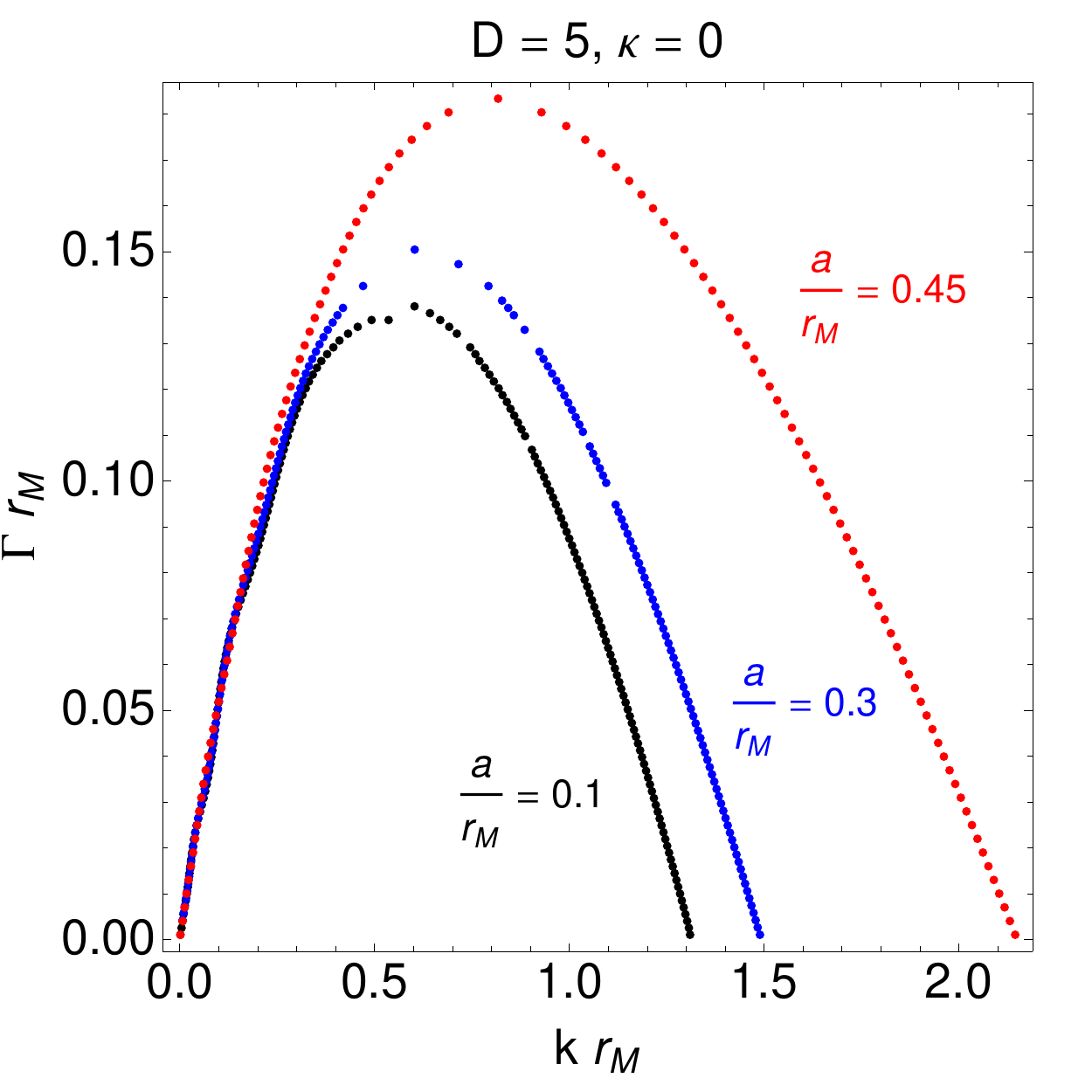}}
\caption{Results in $D=5$, $\kappa=0$. We represent this mode for fixed values of $\Gamma \,r_M$ (first graph) and $a/ r_M$ (second graph). This unstable mode of the string corresponds to well-known Gregory-Laflamme mode.}
\label{fig:5dkzero}
\end{figure}

Our expectation is that, for small $a$, the black hole will be classically stable but the associated black string will suffer a Gregory-Laflamme instability~\cite{Gregory:1993vy}. Therefore, for a range of $\Gamma$, there should exist real solutions for $k$ (corresponding to unstable perturbations of the string) but $k$ will never vanish for $\Gamma>0$, so the black hole is stable. For a static string, the Gregory-Laflamme instability is an $s$-wave perturbation, which for us translates into a $\kappa=0$ perturbation. Hence, for a rotating string, it is natural to expect this instability also to have $\kappa=0$. 

This is indeed what we find. The left plot in Fig.~\ref{fig:5dkzero} shows our result for $-k^2$ for given $a$ and $\Gamma$, with $\kappa=0$. The plot with $\Gamma=0$ corresponds to a stationary perturbation of the black string. This is the ``threshold unstable mode" at the critical wavelength beyond which the black string is unstable.
Note that the curves exist for all values of $a$, i.e. the Gregory-Laflamme instability is always present, it does not ``switch off" as $a$ increases. None of the curves extends to $k=0$ so there is no indication of any black hole instability.

In the right plot of Fig.~\ref{fig:5dkzero}, we give a more familiar plot of $\Gamma$ against $k$ for different values of $a$. For each value of $a$, we have a curve that takes the usual Gregory-Laflamme form. The maximum value of $\Gamma$ is $10-20 \%$ of $r_M$ and increases with increasing $a$. Furthermore, the range of $k$ for which there exists an instability increases, i.e. the instability persists down to shorter wavelengths as $a$ increases. Hence rotation makes the string more unstable. As usual, $\Gamma \rightarrow 0$ as $k \rightarrow 0$ and as $k \rightarrow k_c>0$. The mode with $\Gamma=k=0$ has the usual interpretation of a gauge mode \cite{Gregory:1994bj}. The mode with $\Gamma=0$ and $k=k_c$ is the threshold unstable mode associated to the onset of instability. This marks the bifurcation of a new family of non-uniform rotating black string solutions. These are the solutions constructed in Ref.~\cite{Kleihaus:2007dg}. They preserve the symmetries of the cohomogeneity-1 MP black hole but break the translational symmetry along the string.

Note that the slope $\Gamma/k$ approaches a common limiting value as $k \rightarrow 0$, independently of the value of $a$. This long-wavelength limiting behaviour is captured by the blackfold approach: it follows from Ref. \cite{Emparan:2009at} that $\Gamma/k \rightarrow 1/\sqrt{D-2}$ as $k \rightarrow 0$. This is consistent with our numerical results.

We find no solution of (\ref{lichn}) with $\kappa=1$. Therefore our results are consistent with stability of these black holes, in agreement with the results of Ref.~\cite{Murata:2008yx}.

\subsection{Results for $D=7$}

\begin{figure}[t]
\centerline{\includegraphics[width=.45\textwidth]{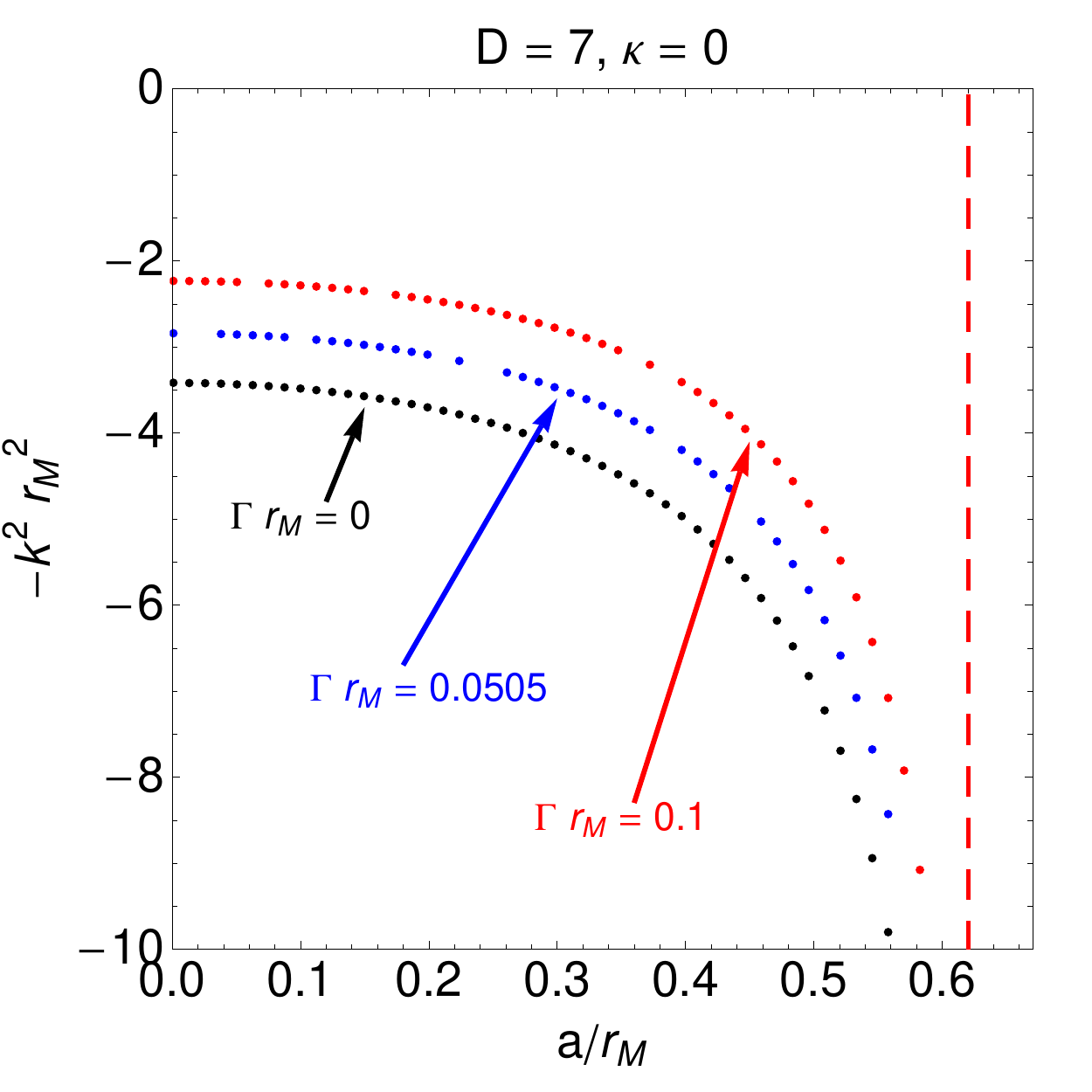}
\hspace{1cm}\includegraphics[width=.45\textwidth]{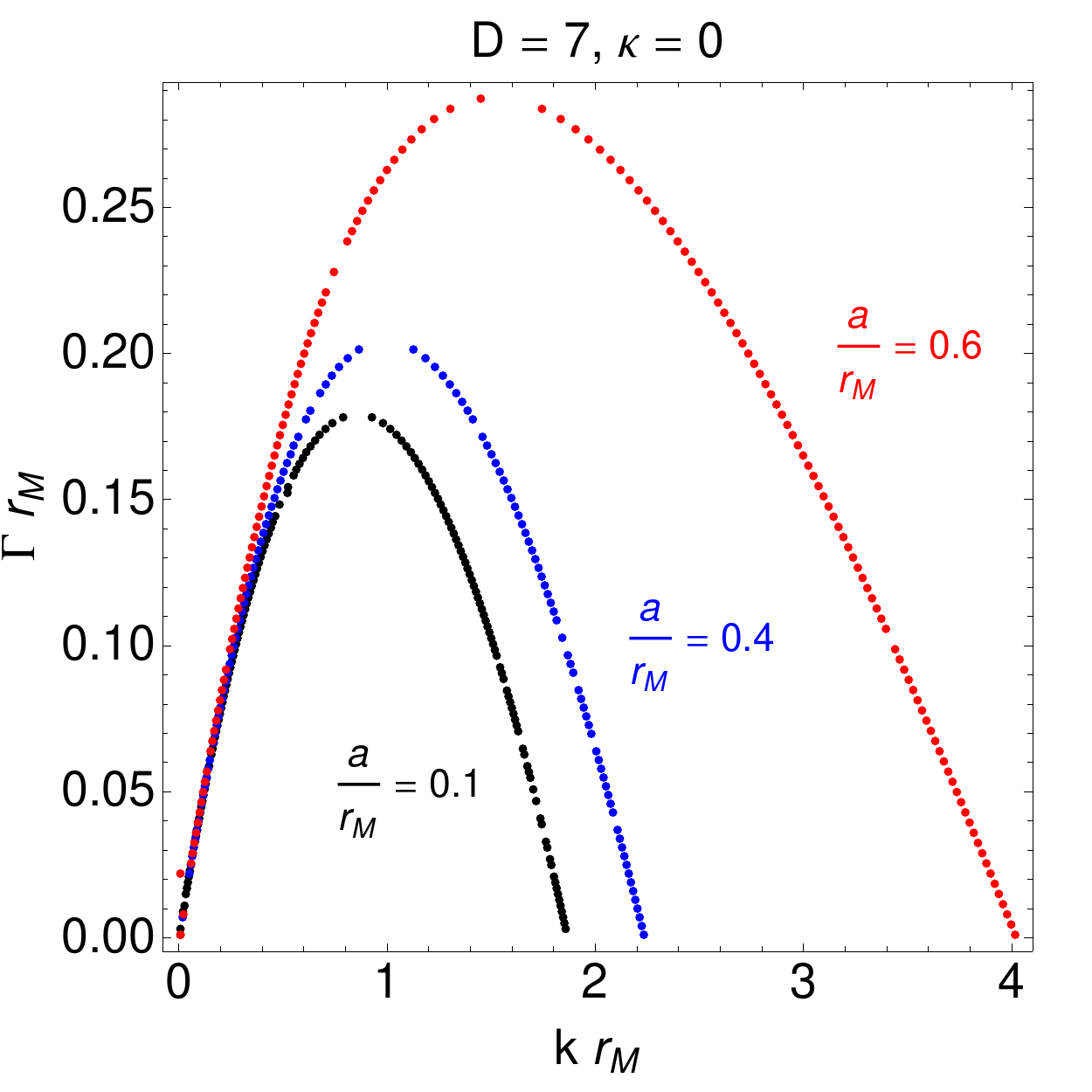}}
\caption{Results in $D=7$, $\kappa=0$. The graphs are entirely analogous to the ones in Fig.~\ref{fig:5dkzero}.}
\label{fig:7dkzero}
\end{figure}

For $\kappa=0$, we have the Gregory-Laflamme instability shown in Fig.~\ref{fig:7dkzero}. This is qualitatively the same as for $D=5$. Once again, rotation makes the string more unstable, the behaviour as $k \rightarrow 0$ is consistent with $\Gamma/k \rightarrow 1/\sqrt{D-2}$ independently of $a$, and there is a threshold mode at a critical value of $k$ at which we expect a bifurcation of a new family of non-uniform strings analogous to those constructed in Ref.~\cite{Kleihaus:2007dg}. 

\begin{figure}[t]
\centerline{\includegraphics[width=.45\textwidth]{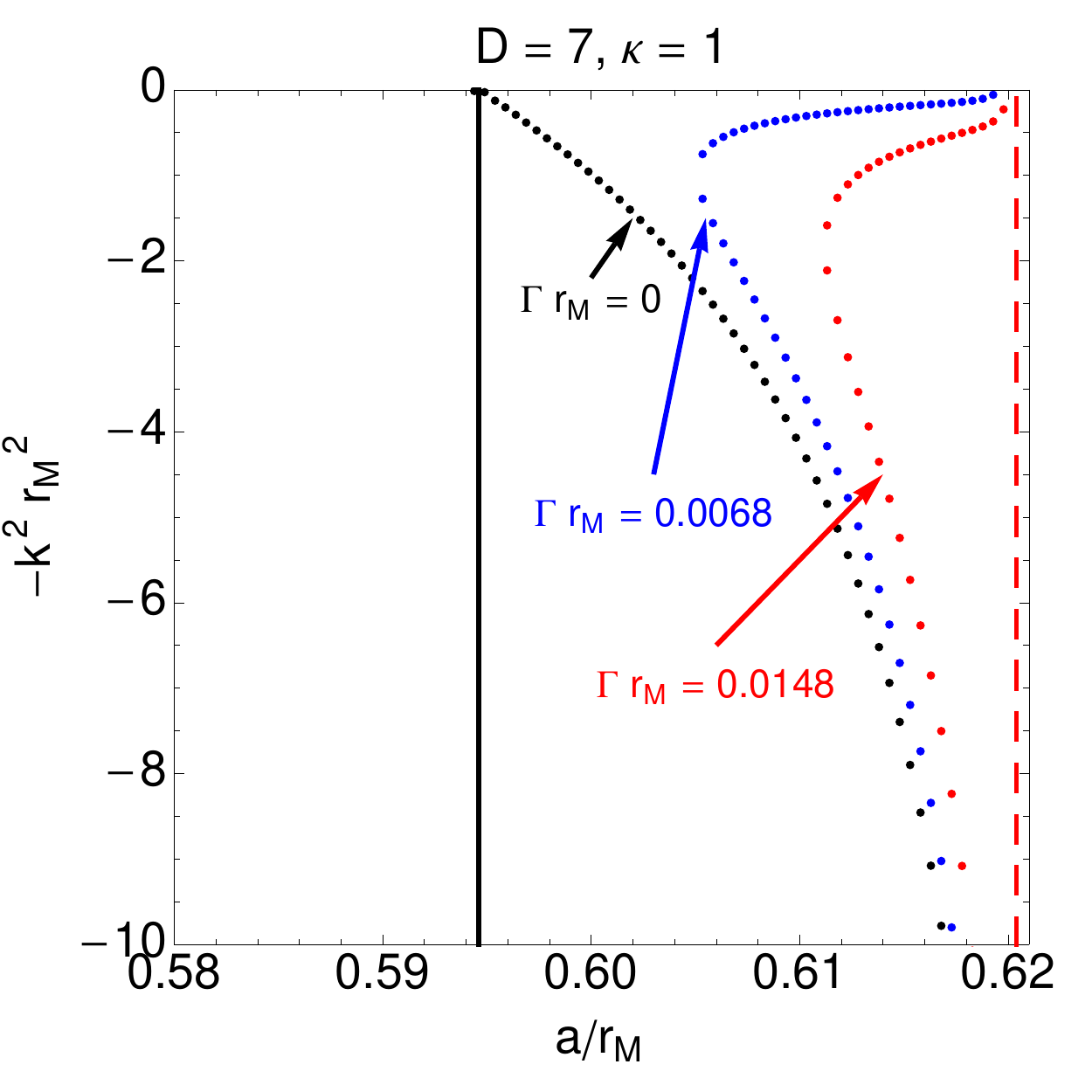}
\hspace{1cm}\includegraphics[width=.45\textwidth]{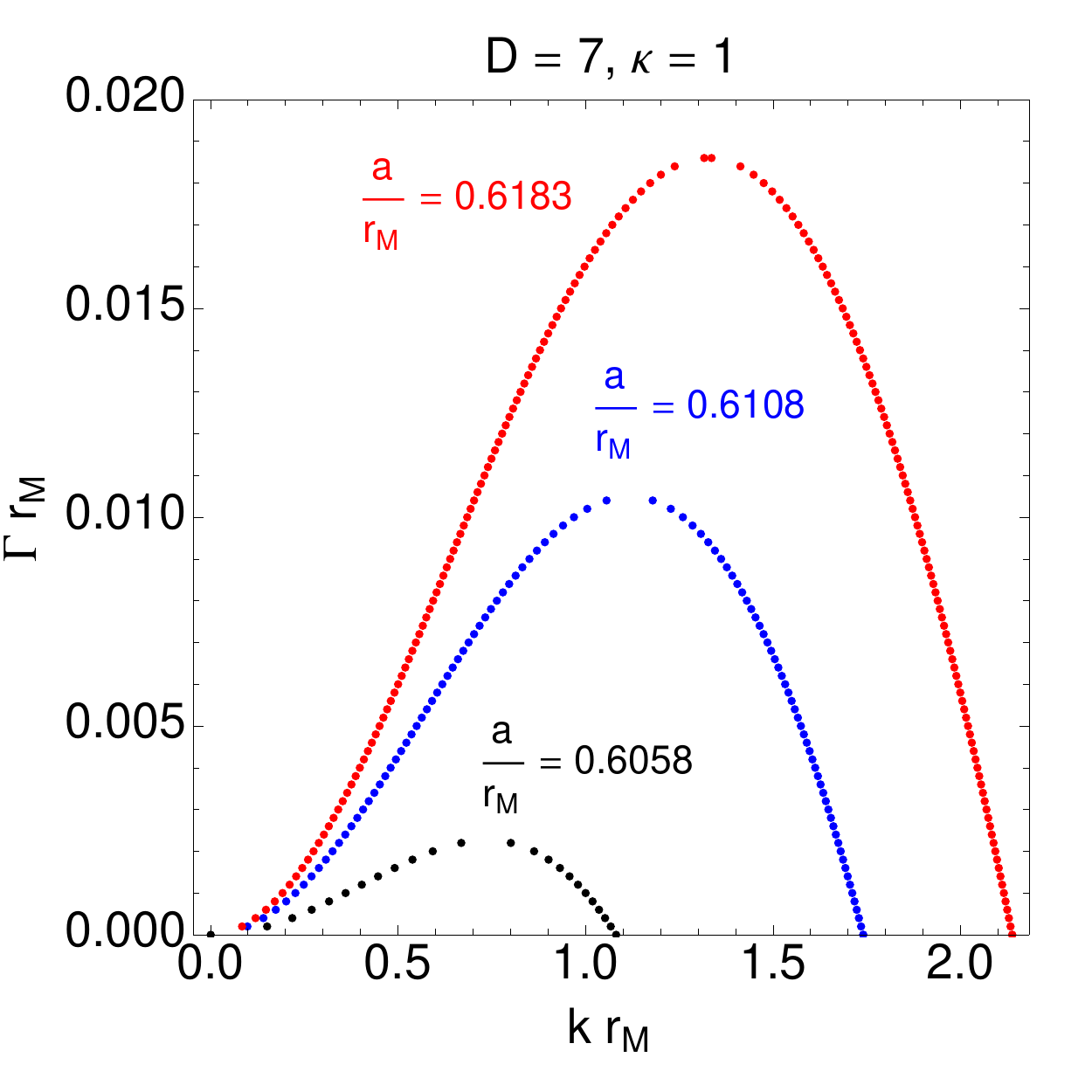}}
\caption{Results in $D=7$, $\kappa=1$. We represent this mode for fixed values of $\Gamma \,r_M$ (first graph) and $a/ r_M$ (second graph). It corresponds to a new Gregory-Laflamme instability of the rotating black string, appearing for the numerical value $(a_1/r_M)^\mathrm{num}=0.5949$. In the first graph, the vertical line to the left corresponds to the analytical prediction of \eqref{eqn:a1}, $(a_1/r_M)=0.5946\,$, and the interrupted vertical line to the right corresponds to extremality. The second graph indicates clearly that the instability of the black brane does not extend to an instability of the black hole ($k = 0$).}
\label{fig:7d}
\end{figure}

A new feature of $D=7$ is the existence of an ultraspinning regime. This occurs for $a>a_1$, where $a_1$ was defined in equation (\ref{eqn:a1}). Just as in Ref.~\cite{Dias:2009iu}, we find that a new stationary ($\Gamma=0$) negative mode of the black hole appears at this point. This negative mode has $\kappa=1$. Our numerical results are shown in  Fig.~\ref{fig:7d}. These results demonstrate that the black string has an instability in the $\kappa=1$ sector when $a>a_1$.
This is a {\it new} Gregory-Laflamme instability of the black string, distinct from the instability in the $\kappa=0$ sector. The plots of $\Gamma$ against $k$ have the same qualitative shape as for the $\kappa=0$ instability except that the slopes of the curves appear to vanish (for all $a$) as $k \rightarrow 0$.\footnote{It would be interesting to investigate whether this behaviour can be explained using blackfold methods.} Once again there is a threshold unstable mode at a critical value of $k$. Presumably this corresponds to a bifurcation to a new family of non-uniform black string solutions. In addition to breaking the symmetry along the string, this mode also breaks some of the symmetry of the black hole (typically down to that of a generic MP black hole\footnote{
This is because $\kappa=1$ harmonics are in one to one correspondence with Killing vector fields of $CP^N$: see below.}) so this new family has less symmetry than the non-uniform strings associated to the threshold unstable mode with $\kappa=0$.

Note that the $\kappa=1$ black string instability coexists with the $\kappa=0$ instability. The latter is clearly dominant since it has much larger $\Gamma$ and the instability exists for a larger range of $k$, i.e. down to shorter wavelengths.

It is important to note that there is no evidence of any instability of the black {\it hole}: none of the curves with non-zero $\Gamma$ extends to $k=0$. In the limit  $k \rightarrow 0$, solutions with non-zero $\Gamma$ approach a pure gauge mode, just as for $\kappa=0$. On the other hand, the solution with $\Gamma=0$, does not approach a gauge mode as $k \rightarrow 0$. Instead, as anticipated in Ref.~\cite{Dias:2009iu}, it corresponds simply to a variation of parameters within the MP family of solutions. We shall explain below why this must be the case, and why we find such a mode only at $a=a_1$.

For $D=7$, since the black hole is ultraspinning for $a>a_1$, there is the possibility of an ultraspinning instability appearing at $a=a_2>a_1$. However, we find no solution of equation (\ref{lichn}) for $\kappa=2$ so our results are consistent with stability of $D=7$ cohomogeneity-1 MP black holes.

\subsection{Results for $D=9$: black hole instability}

\begin{figure}[t]
\centerline{\includegraphics[width=.45\textwidth]{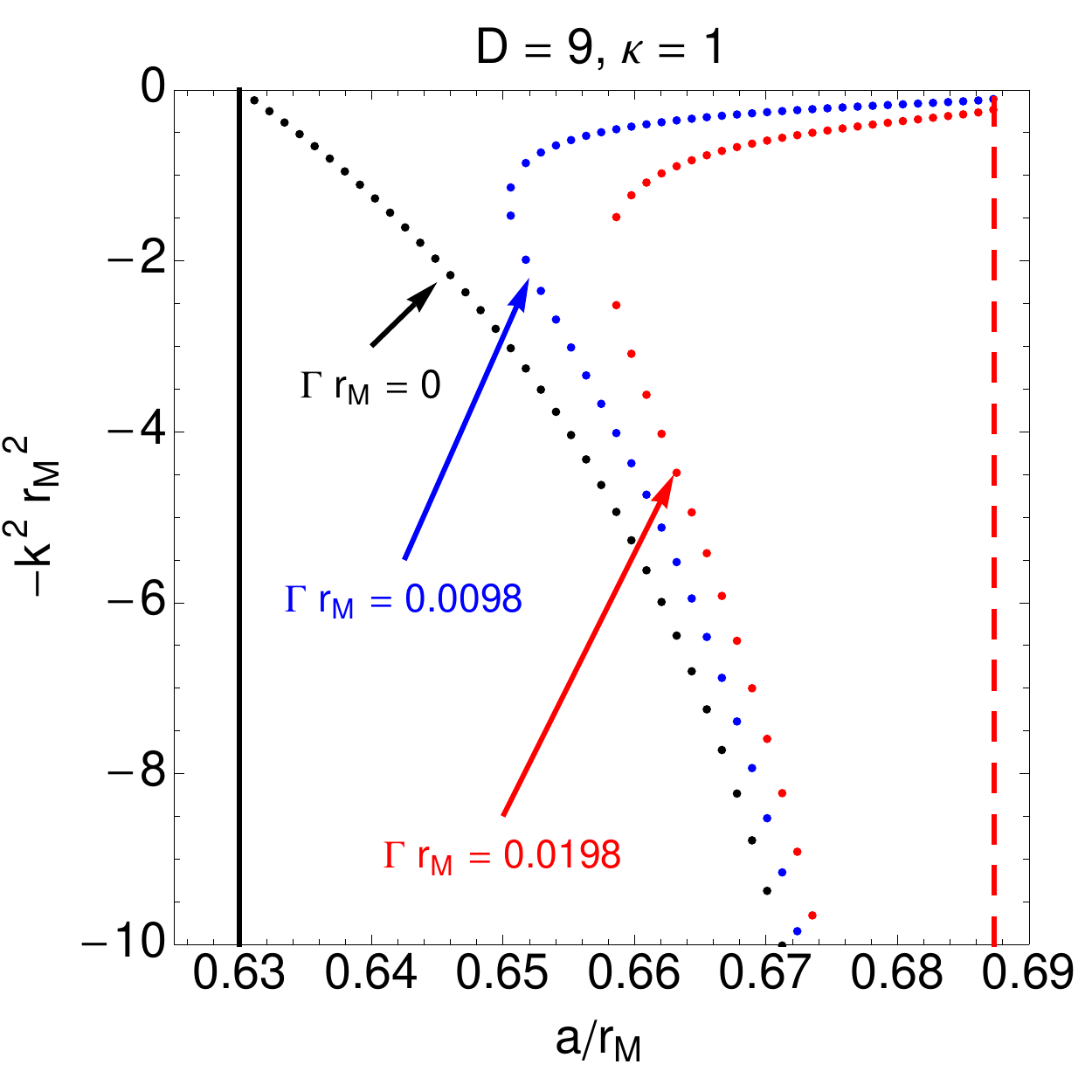}
\hspace{1cm}\includegraphics[width=.45\textwidth]{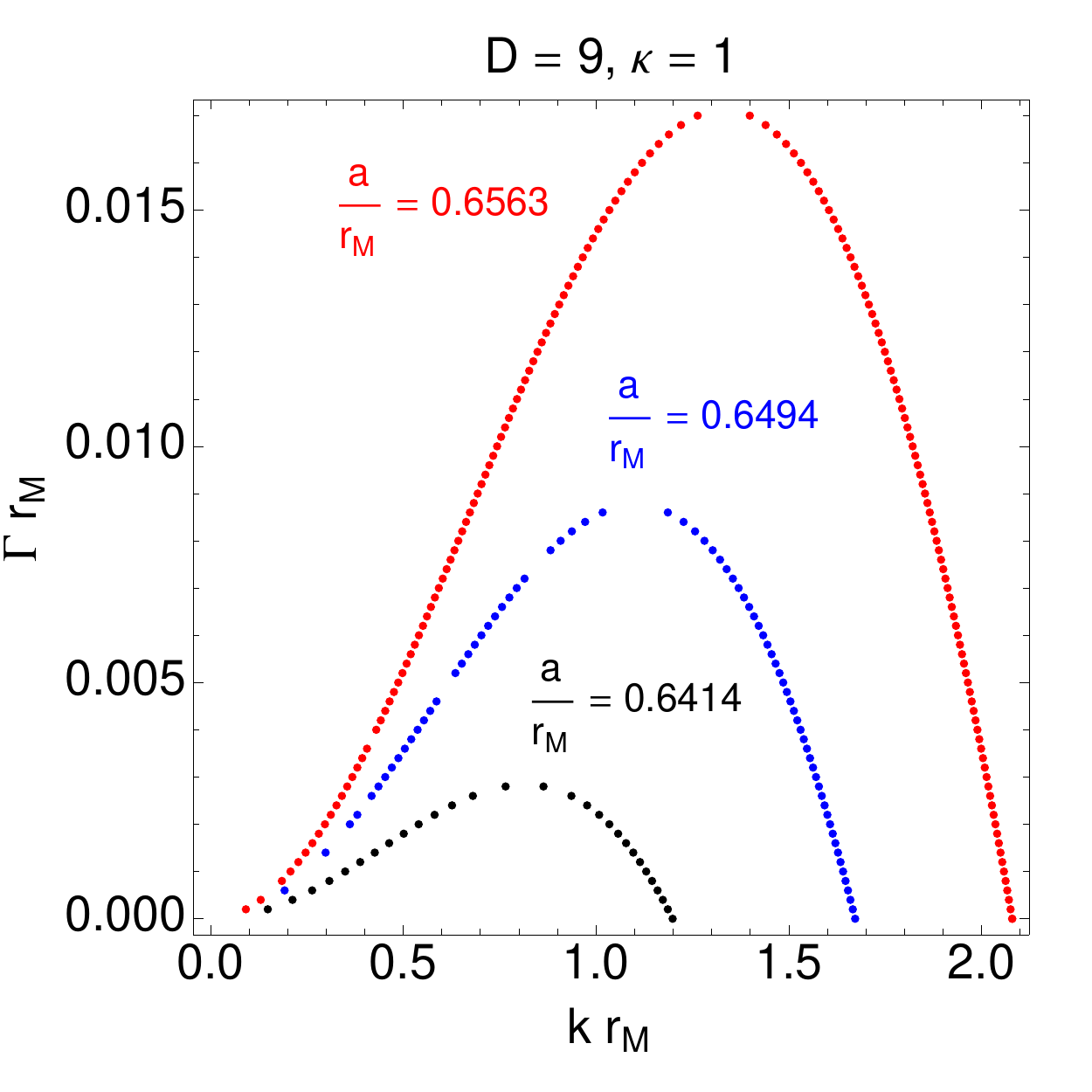}}
\caption{Results in $D=9$, $\kappa=1$. The graphs are entirely analogous to the ones in Fig.~\ref{fig:7d}.}
\label{fig:9dk1}
\end{figure}

For $\kappa=0$, we have the expected Gregory-Laflamme instability of the black string. For $\kappa=1$, we find, as for $D=7$, a new Gregory-Laflamme instability of the black string that appears at $a=a_1$. This is shown in figure Fig.~\ref{fig:9dk1}. 

\begin{figure}[t]
\centerline{\includegraphics[width=.45\textwidth]{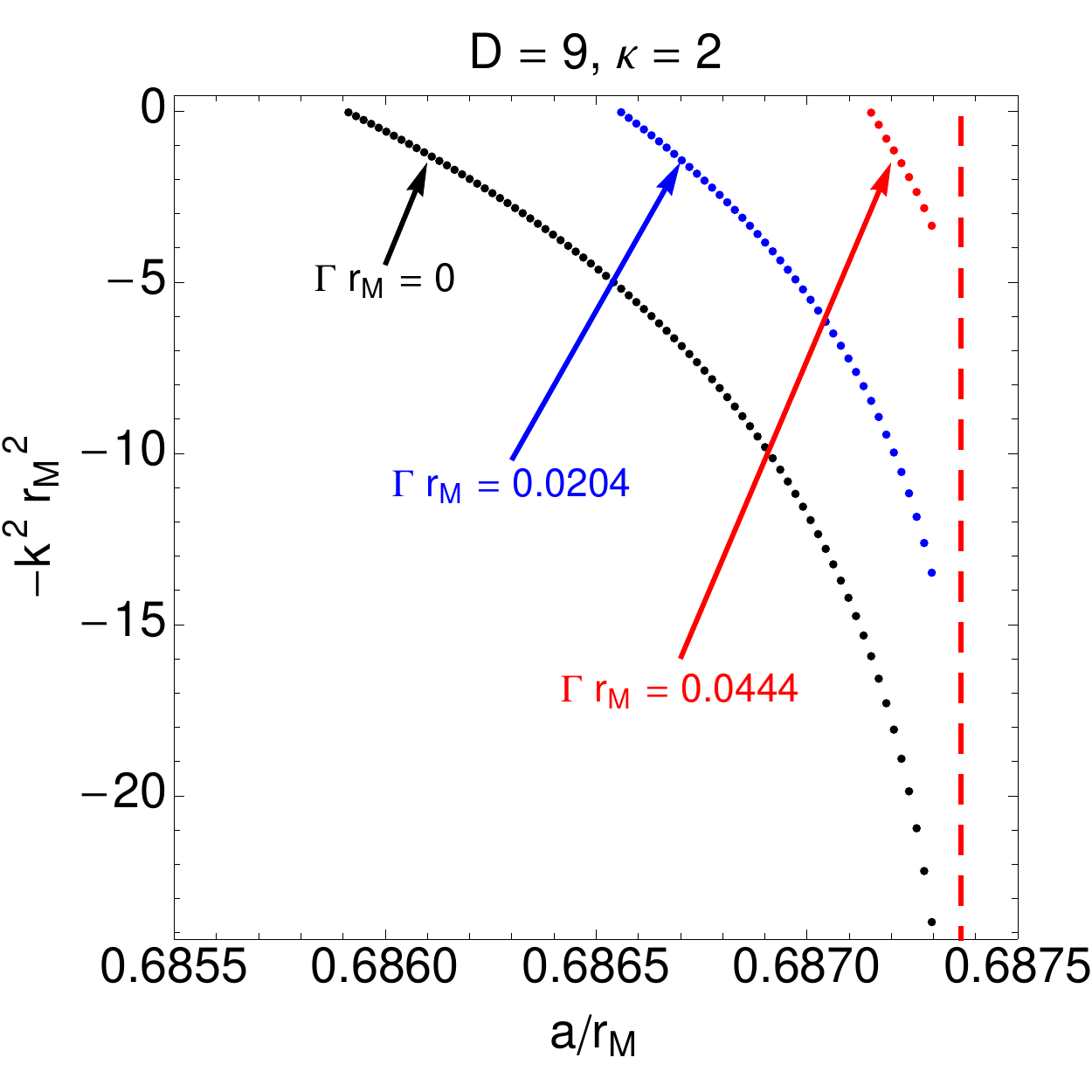}
\hspace{2cm}\includegraphics[width=.45\textwidth]{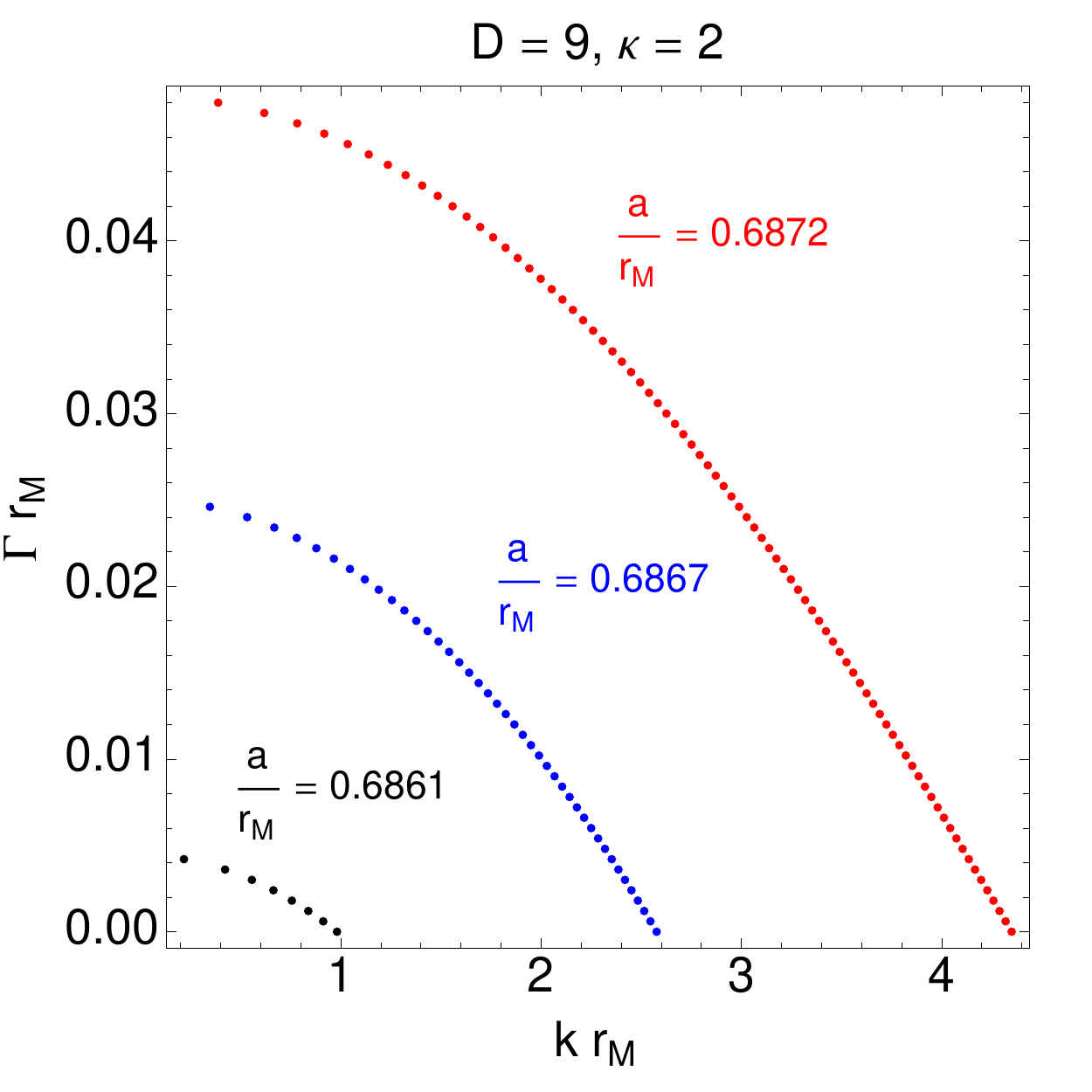}}
\caption{Results in $D=9$, $\kappa=2$. We represent this mode for fixed values of $\Gamma \,r_M$ (first graph) and $a/ r_M$ (second graph). As opposed to the $\kappa=1$ case, the time-dependent mode extends all the way to $k=0$. There is not only a new Gregory-Laflamme instability of the black string, but also an instability of the black hole, appearing at $a=a_2$, where  $a_2/r_M=0.6858 > a_1/r_M =0.6300\,$. In the first graph, the interrupted vertical line to the right corresponds to extremality. In the second graph, the curve shrinks to the origin as $a \rightarrow a_2$.}
\label{fig:9dk2}
\end{figure}

The new feature that appears for $D=9$ is an instability in the $\kappa=2$ sector, which appears at 
$a=a_2>a_1$. This is shown in Fig.~\ref{fig:9dk2}. The left plot shows a new stationary ($\Gamma=0$) negative mode which emerges from a zero-mode at $a=a_2$. We shall prove below that this zero mode cannot correspond to a variation of parameters within the MP family of solutions. Furthermore, in Appendix~\ref{sec:NoPureGauge}, we show that it cannot be a gauge mode.

For $a>a_2$, there is a new instability of the black string, corresponding to the curves with $\Gamma>0$ in the plot. But there is a qualitative difference between the left plot of Fig.~\ref{fig:9dk2} and our previous plots: the curves with $\Gamma>0$ now intersect $k=0$, i.e. we have found perturbations of the black {\it hole} that grow exponentially in time, that is, a classical instability of black holes with $a>a_2$. This is our main result. 

The onset of instability is indicated by the stationary zero-mode ($\Gamma=0$, $k=0$) at $a=a_2$. This is analogous to the mode constructed for singly spinning black holes in Ref.~\cite{Dias:2009iu}. Our main achievement is to demonstrate, for the first time, the existence of modes which grow exponentially with time when $a>a_2$.

The right plot of Fig.~\ref{fig:9dk2} shows a clear difference from our previous plots. Unlike the GL instability, we find that $\Gamma$ is maximized at $k=0$ rather than vanishing there. Hence, for the black string, the most unstable $\kappa=2$ modes are those with $k=0$, i.e. those corresponding to the black hole instability. For larger $k$, the black string instability ``switches off" in the same way as the GL instability, with a threshold mode at $k=k_c$ indicating a new family of non-uniform black string solutions. 

\begin{figure}[t]
\centerline{\includegraphics[width=.5\textwidth]{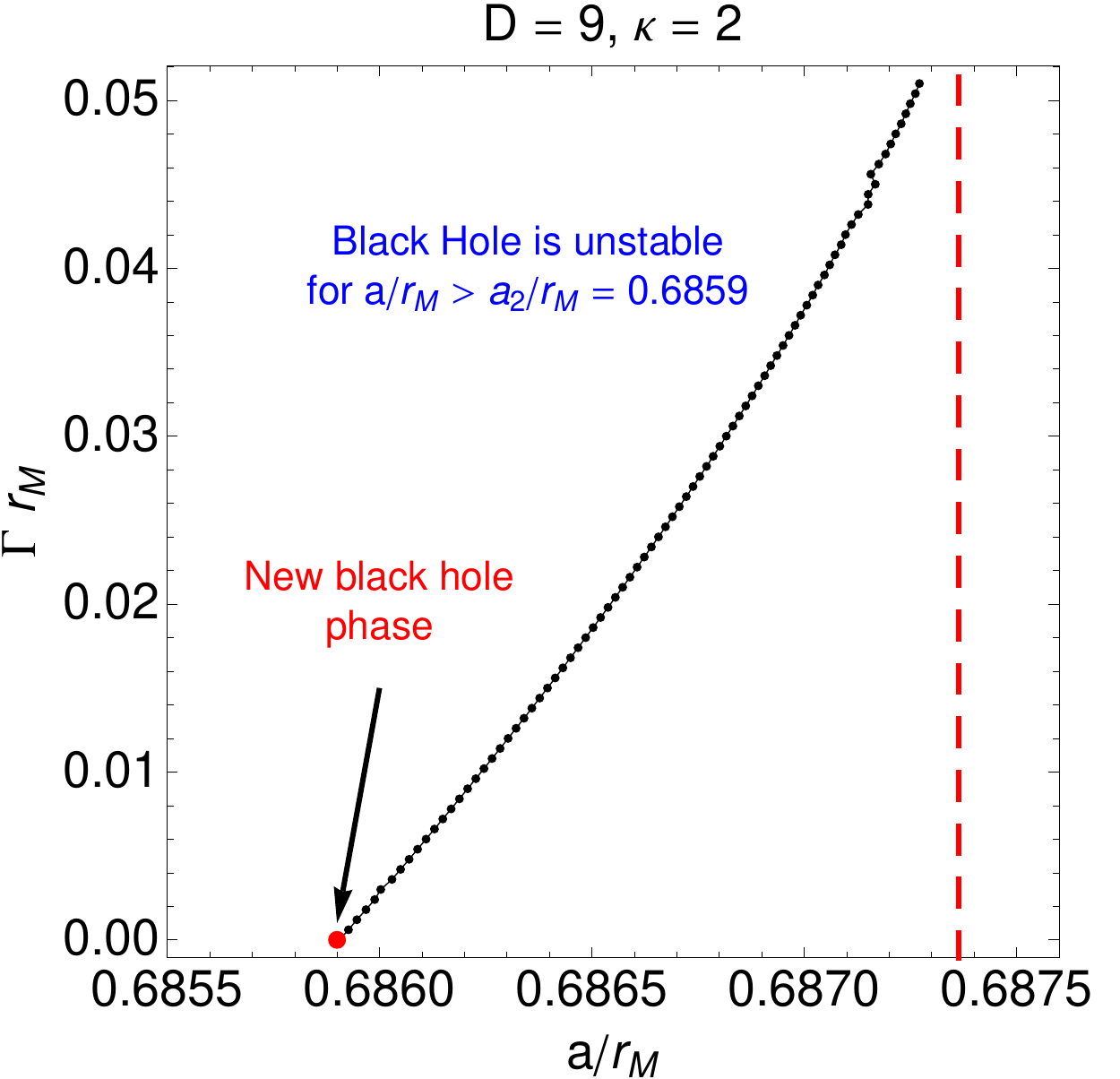}}
\caption{Results in $D=9$ for $\kappa=2$, in the limit $k=0$. We represent the black hole instability in a plot of $\Gamma \,r_M$ versus $a/ r_M$. The interrupted line corresponds to extremality. Numerical error prevents us from extending our results all the way to extremality.} 
\label{fig:9dGamma}
\end{figure}

Fig.~\ref{fig:9dGamma} presents our result for the instability time scale of the black hole as a function of its spin. For $a>a_2$, we find that $\Gamma$ increases monotonically with $a$, so extreme black holes are the most unstable. The maximum value of $\Gamma$ is just a few per cent of $r_M$, so the instability is slow compared with the GL instability of the black string.

Finally, we find no solutions of (\ref{lichn}) with $\kappa=3$.

\subsection{Rotational symmetries of higher-dimensional black holes}

As explained in the Introduction, the study of perturbations of higher-dimensional black holes can be used to investigate the possible existence of new families of black hole solutions with less symmetry than the known solutions. The idea proposed in Ref.~\cite{Reall:2002bh} is to look for a stationary zero-mode of the black hole, which is interpreted as indicating the existence of a family of solutions branching off from the known solutions.

In our case, the stationary zero-modes with $\kappa=1$, are uninteresting since (as explained below) these correspond to variations within the MP family. However, for $D=9$, we found a stationary zero-mode with $\kappa=2$ that appears at $a=a_2$, the critical value of $a$ beyond which the black hole is unstable. Therefore, we have found evidence for a new family of black hole solutions that bifurcates from the MP family at this point.\footnote{
We expect that this new family will have unequal angular momenta in general. However, this cannot be seen from our results since $\kappa=2$ modes do not change the mass or angular momenta at the linearized level (see Appendix \ref{sec:ChangeMJ}). A second order calculation would be required to determine these changes.}

How much symmetry do these new solutions have? This can be inferred from the symmetry of the $\kappa=2$ harmonics on $CP^N$. There will, of course, be a family of degenerate scalar harmonics with $\kappa=2$. Some of these will preserve some of the symmetry of $CP^N$ whereas others break it completely (this is proved in Appendix~\ref{subsec:symmetries}).\footnote{
It is helpful to think about the case of $CP^1=S^2$, for which $\kappa=\ell$, the total angular momentum quantum number. Scalar harmonics are labelled by $\ell$ and $m$. Certain modes with $\ell=2$ (i.e. quadrupole modes) may preserve some symmetry (e.g. if $m=0$ then they are axisymmetric) but generically they break all of the continuous symmetries of $S^2$.}
For a mode of the latter type, the associated metric perturbation will break completely the $SU(N+1)$ subgroup of the $R \times U(N+1)$ isometry group of the background metric. It preserves only a $R \times U(1)$ subgroup corresponding to time-translation invariance and invariance under the rotations generated by $\partial/\partial \psi$. Hence the corresponding family of new black hole solutions will possess just a single rotational symmetry.

Very recently, Ref.~\cite{Emparan:2009vd} has constructed approximate black {\it ring} solutions with just a single rotational symmetry. Our results are the first evidence for the existence of new black hole solutions with a single rotational symmetry and horizons of spherical topology.

Note that if one used a $\kappa=2$ harmonic that does preserve some of the symmetry of $CP^N$ then presumably this would give rise to a {\it different} family of new black hole solutions, with more than one rotational symmetry. Therefore, assuming that each stationary zero-mode corresponds to a new nonlinear stationary black hole solution, there must exist several new black hole solutions that bifurcate from the MP family at the same point, and these different solutions have different numbers of rotational symmetries. So how many new solutions are there?

One way of addressing this question is to determine the number of parameters in the most general $\kappa=2$ harmonic. If we take $D=9$ then $\kappa=2$ harmonics correspond to the $[2,0,2]$ representation of $SU(4)$, which is $84$ dimensional. Hence the most general $\kappa=2$ harmonic is labelled by $84$ parameters. Some such harmonics are related by acting with $SU(4)$, i.e. by rotations of the background spacetime. However, since $SU(4)$ has dimension 15, this can eliminate only 15 parameters, leaving $84-15=69$ parameters that cannot be eliminated by rotations of the background. So, up to rotations of the background, we have a family of stationary zero-modes with 69 parameters, and presumably a family of new black holes with 70 parameters, the extra parameter being the mass (or $r_M$). This is considerably more parameters than the 5 that are required to specify the $D=9$ MP solution!

\section{Thermodynamics and black string instabilities}

\label{sec:string}

\subsection{Introduction}

In this Section, we shall explain how many of our results, particularly for $\kappa=0,1$, can be explained on the basis of thermodynamic arguments. In particular, we shall explain why the stationary zero-mode that appears at $a=a_1$ must correspond to a variation of parameters within the MP family. In order to do this, we shall develop in some detail the connection between negative modes, thermodynamics, and classical stability of black strings. 

We shall consider a rotating black string with metric given by (\ref{eqn:string}) but now with $g_{\mu\nu}$ a general MP black hole, not necessarily cohomogeneity-1. 

\subsection{Rotating black strings}

The black string is expected to suffer from a classical Gregory-Laflamme instability \cite{Gregory:1993vy}, that is, there will exist perturbations which grow exponentially in time as $e^{-\ii\omega t}$ where ${\rm Im}(\omega)>0$. This is expected to be a long-wavelength instability, i.e. unstable modes have $|k|<k_c$ for some critical wavenumber $k_c$. For a {\it static} black string, unstable modes have ${\rm Re}(\omega)=0$. In the limit $k \rightarrow k_c$, unstable modes reduce to a static perturbation of the string, which we shall call the threshold unstable mode. The presence of this mode indicates the existence of a new branch of non-uniform black string solutions that bifurcates from the branch of uniform solutions \cite{Horowitz:2001cz}.

For a rotating black string, we would expect unstable modes to have ${\rm Re}(\omega) \ne 0$ in general, and so one might not expect to find stationary perturbations with $k=k_c$. Nevertheless, Ref.~\cite{Kleihaus:2007dg} has obtained numerically non-uniform rotating black string solutions that do indeed bifurcate from the uniform branch (based on cohomogeneity-1 MP solutions) at a point corresponding to a stationary perturbation. Hence stationary perturbations do indeed exist. We believe that the reason for this is that unstable modes will have ${\rm Re}(\omega)=0$ if they are invariant under the rotational symmetry of the black hole predicted by the theorems of Refs.~\cite{Hollands:2006rj,Moncrief:2008mr}, i.e. the symmetry generated by the Killing field $\Omega_i m_i$ where $m_i$ are the rotational Killing vector fields and $\Omega_i$ the associated angular velocities of the horizon. We do not have a proof of this, but our results, and the results of Refs.~\cite{Kleihaus:2007dg,Dias:2009iu}, indicate that it is true. In the limit $k \rightarrow k_c$, this gives a stationary threshold mode that preserves this symmetry. This symmetry is a necessary condition for the threshold mode to correspond to a bifurcation into a new family of non-uniform rotating black string solutions, since presumably this new family should respect the theorems of Refs.~\cite{Hollands:2006rj,Moncrief:2008mr} (although, strictly speaking, these theorems apply only to black holes, not black strings).

In summary, we expect an instability of the black string to appear for wavenumbers $|k|<k_c$. If we restrict attention to modes invariant under the symmetry generated by $\Omega_i m_i$ then unstable modes will have ${\rm Re}(\omega)=0$, and the threshold unstable mode, with $k=k_c$, will be stationary and invariant under the same symmetry. 

For cohomogeneity-1 black holes, $\Omega_i m_i$ is proportional to $\partial/\partial \psi$, so modes invariant under the symmetry generated by $\Omega_i m_i$ must have $m=0$, which is why we set $m=0$ above.

\subsection{Thermodynamic negative modes}

Equation (\ref{lichn}) shows that a threshold unstable mode of the black string corresponds to a stationary negative mode of the Lorentzian Lichnerowicz operator of the black hole. However, in order to make contact with thermodynamics, we need to relate it to a negative mode of the Lichnerowicz operator for the {\it Euclideanized} black hole geometry as was done in Ref.~\cite{Reall:2001ag} for static black strings. The Euclideanized geometry is defined by the analytic continuation $t =-\ii\,\tau$. This produces a complex metric since the MP solution is non-static. In coordinates with $m_i =\partial/\partial \phi_i$, regularity requires that we identify $(\tau,\phi_i) \sim (\tau,\phi_i + 2\pi) \sim (\tau + \beta, \phi_i + \ii\, \Omega_i \beta)$, where $\beta=1/T$ is the inverse temperature \cite{Gibbons:1976ue}.\footnote{
Note that, although the metric is complex, the manifold is real. Real coordinates can be defined by setting $\tilde{\phi_i} = \phi_i - \ii\,\Omega_i \tau$, so the identifications are $(\tau,\tilde{\phi}_i) \sim (\tau,\tilde{\phi}_i + 2\pi) \sim (\tau+\beta,\tilde{\phi}_i)$.}

If the Euclidean Lichnerowicz operator in this geometry admits a stationary negative mode, i.e. one preserved by the symmetries generated by $\partial/\partial \tau$, then, after the analytic continuation $\tau=\ii\,t$ it becomes a stationary negative mode of the Lorentzian black hole, i.e. a threshold unstable mode of the black string.


For static black strings, this correspondence between the threshold unstable mode and a Euclidean negative mode of the black hole was used in Ref.~\cite{Reall:2001ag} to justify the Gubser-Mitra conjecture regarding black string stability \cite{Gubser:2000ec,Gubser:2000mm}. In the present context, this conjecture asserts that a black string with a non-compact $z$-direction suffers a classical Gregory-Laflamme type instability if, and only if, it is locally thermodynamically unstable. Local thermodynamic stability is the statement that the  Hessian matrix
\be
 \label{hessian}
 -S_{\alpha\beta} \equiv \frac{\partial^2 (-S)}{\partial x^\alpha \partial x^\beta}
\ee
must be positive definite, where $S$ is the entropy of the black hole (entropy per unit length of the black string), which is regarded as a function of the charges $x^\alpha=(M,J_i)$ of the black hole.\footnote{This condition can be shown to be equivalent to the standard stability condition in the grand-canonical ensemble: the Hessian matrix $\partial^2 (-G)/ \partial \tilde{x}^\alpha \partial \tilde{x}^\beta$ must be positive definite, where $G=M-TS-\Omega_i J_i$ is the Gibbs free energy regarded as a function of $\tilde{x}^\alpha=(T,\Omega_i)$.} The idea is that if the Hessian fails to be positive definite then it becomes thermodynamically preferred for the mass and/or angular momentum to become non-uniformly distributed along the string.

The idea of Ref.~\cite{Reall:2001ag} is to relate local thermodynamic stability of the black string to the existence of a stationary negative mode of the Euclidean black hole. This is done by constructing a family of off-shell geometries for the Euclidean path integral for which one can show that the Euclidean action decreases in a certain direction if the black string is locally thermodynamically unstable. It then follows that there must exist a Euclidean stationary negative mode and hence a threshold unstable mode of the black string. In the present case, this argument can be phrased as follows.

Consider a Euclideanized black hole solution $\mathcal{B}(x)$ uniquely specified by parameters $x^\alpha$. Let $T(x)$, $\Omega_i(x)$, $M(x)$, etc. denote the temperature, angular velocities, mass, etc. of this solution. We can construct an off-shell generalization $\mathcal{B}(x,y)$, specified by parameters $y^\alpha$ as follows \cite{Brown:1990di} (see also Ref.~\cite{Monteiro:2009tc}). Assume the same isometries as $\mathcal{B}(x)$, i.e. Killing fields $\partial/\partial \tau$ and $\partial/\partial \phi_i$. Perform an ADM decomposition of the metric, with time coordinate $\tau$. Take the spatial geometry of $\mathcal{B}(x,y)$ to be the same as that of $\mathcal{B}(y)$. Now choose the lapse function and shift vector so that (i) $\mathcal{B}(x,y)$ has the same asymptotics as $\mathcal{B}(x)$, (ii) $\mathcal{B}(x,y)$ is regular everywhere, in particular at the bolt (Euclidean horizon), subject to the identifications $(\tau,\phi_i) \sim (\tau,\phi_i + 2\pi) \sim (\tau + \beta(x), \phi_i + \ii\, \Omega_i (x) \beta(x))$; (iii) $\mathcal{B}(x,x)=\mathcal{B}(x)$. Note that (ii) implies that $\mathcal{B}(x,y)$ is a configuration in the Euclidean path integral defined for temperature $T(x)$ and angular velocities $\Omega_i(x)$, for which the saddle point is $\mathcal{B}(x)$. Calculating the Euclidean action of $\mathcal{B}(x,y)$ using the Hamiltonian formalism gives
\be
I(x,y)=\beta(x) M(y) - S(y)-\beta(x) \Omega_i(x) J_i(y).
\ee
Condition (iii) implies that the geometry with $y=x$ satisfies the equations of motion and hence the first derivative of the action with respect to $y^\alpha$ must vanish for $y^\alpha=x^\alpha$. This is a consequence of the fact that the black hole satisfies the first law of thermodynamics,
\be
\label{firstlaw}
dM=TdS + \Omega_i J_i\,.
\ee
The second derivative of the action, i.e. the Hessian of the action, now reduces to
\be
\left( \frac{ \partial^2 I}{\partial y^\alpha \partial y^\beta} \right)_{y=x} =  \left(\beta  \frac{ \partial^2 M}{\partial x^\alpha \partial x^\beta}  - \frac{ \partial^2 S}{\partial x^\alpha \partial x^\beta}  - \beta \Omega_i \frac{ \partial^2 J_i}{\partial x^\alpha \partial x^\beta} \right),
\ee
where the RHS is evaluated at $x$. For the MP black hole, the charges $M$ and $J_i$ uniquely parameterize the solution (whereas $T$ and $\Omega_i$ do not). Hence we can choose $x^\alpha=(M,J_i)$. We then have
\be
 \left( \frac{ \partial^2 I}{\partial y^\alpha \partial y^\beta} \right)_{y=x} =  -S_{\alpha\beta}(M,J)\,.
 \ee
Therefore, if $-S_{\alpha\beta}$ fails to be positive definite for some MP black hole then the Euclidean action decreases in some direction and hence the black hole must admit a negative mode.\footnote{
Note that we have not constructed the negative mode explicitly by this argument: the linearization of $\mathcal{B}(x,y)$ around $y=x$ will give a superposition of eigenfunctions of the Lichnerowicz operator. The point is that, since the action decreases in some direction, this must involve a negative mode.}  Given that our off-shell geometries are stationary, this negative mode must also be stationary.\footnote{
The operator $\Delta_L$ commutes with $\partial/\partial \tau$ so one can work with simultaneous eigenfunctions of these operators. Eigenfunctions with different eigenvalues of the latter will be orthogonal.}
 In general, there must be at least as many negative modes as there are negative eigenvalues of this Hessian. 

We shall refer to a negative mode whose existence is predicted by this thermodynamic argument as a {\it thermodynamic negative mode}. We shall see below that there are some negative modes whose existence cannot be predicted by this thermodynamic argument, so we shall refer to these as non-thermodynamic negative modes. Although the former type are relevant for black string stability, it is the latter type that are relevant for black hole stability.


\subsection{Rotating black strings are unstable}

\label{subsec:stringinstab}

For the MP solution, indeed for any vacuum black hole solution specified by $(M,J_i)$, we can now prove that $-S_{\alpha\beta}$ always admits a negative eigenvalue. The proof goes as follows. Consider the Legendre transform of the entropy,
\be
W = S- \beta M+ \beta \Omega_i J_i\,.
\ee
The first law (\ref{firstlaw}) implies that $\partial S/\partial x^\alpha = z^\alpha\,$, and also $\partial W/\partial z^\alpha = - x^\alpha\,$, where $x^\alpha=(M,J_i)$ and $z^\alpha=(\beta,-\beta \Omega_i)$. The Hessians $-S_{\alpha\beta}$ and $W_{\alpha\beta} \equiv \partial^2 W / \partial z^\alpha \partial z^\beta$
are thus inverse matrices, and $-S_{\alpha\beta}$ has a negative engenvalue if and only if $W_{\alpha\beta}$ does. A sufficient condition for $W_{\alpha\beta}$ not being positive definite is
\be
\label{w00i}
W_{00} = \frac{\partial^2 W(z^\alpha)}{\partial \beta^2} = - \left( \frac{\partial M}{\partial \beta} \right)_{\beta \Omega} <0\,.
\ee
The Smarr relation, valid for asymptotically flat vacuum black holes, reads
\be
\label{smarr}
\frac{D-3}{D-2}\, M = T S - \Omega_i J_i  \qquad \Leftrightarrow \qquad M = -(D-2) WT\,.
\ee
Hence we have
\be
\label{w00f}
W_{00} = -(D-3) MT <0\,,
\ee
which implies that $-S_{\alpha\beta}$ always admits a negative eigenvalue. Therefore any vacuum black hole solution must admit a stationary negative mode, and
a black string based on such a black hole solution must always be classically unstable. This explains our result (for cohomogeneity-1 black holes) that the black string has an instability in the $\kappa=0$ sector for all values of $a$. We have argued that a stationary negative mode must be present for all values of $a$, and thus there is no value for $a$ for which it reduces to a {\it zero} mode. Hence there cannot be an instability of the black hole in this sector.

\subsection{Zero-modes}


We have seen how local thermodynamic instability implies the existence of a negative mode. However, we have not yet shown that this negative mode appears precisely when an eigenvalue of $-S_{\alpha\beta}$ changes sign. We shall now address this point by considering {\it zero} modes arising from variation of parameters with the MP family of solutions.

Consider, in Euclidean signature, a variation of the parameters $x^\alpha$ of the MP solution: $x^\alpha \rightarrow x^\alpha + \delta x^\alpha$. In general, this will induce a variation in $T$ and $\Omega_i$. Let the perturbed geometry be $g_{\mu\nu} + h_{\mu\nu}$. Of course, this is regular provided we make coordinate identifications associated to the new values of $T$ and $\Omega_i$. However, we are demanding that $g_{\mu\nu}$ and $h_{\mu\nu}$ must be regular {\it separately}. Regularity of $g_{\mu\nu}$ requires that $T$ and $\Omega_i$ must be those associated to the background geometry. But then $g_{\mu\nu}+h_{\mu\nu}$ cannot be regular, i.e. $h_{\mu\nu}$ cannot be regular.
Hence any variation $\delta x^\alpha$ that leads to a change (to first order) in $T$ and $\Omega_i$ must necessarily result in a $h_{\mu\nu}$ that is not regular. Therefore a variation in parameters in the MP solution gives a regular Euclidean zero-mode if, and only if, it preserves $T$ and $\Omega_i$ to first order.

If the variation preserves $T$ and $\Omega_i$ then (using the first law) it preserves $(\partial S/\partial M)_J=1/T$ and $(\partial S/\partial J_i)_M=-\Omega_i/T$, i.e. it preserves $\partial S/\partial x^\alpha$. Hence it must be a eigenvector of $S_{\alpha \beta}$ with eigenvalue zero:
\be
0= \delta(\partial_\alpha S) = \delta x^\beta\partial_\beta \partial_\alpha S = S_{\alpha \beta} \delta x^\beta.
\ee
Clearly this argument can be run backwards, so $S_{\alpha\beta}$ admits an eigenvector with eigenvalue zero if, and only if, there exists a change in the black hole parameters that preserves (to first order) $T$ and $\Omega_i$.

Consider a 1-parameter subfamily of MP black hole solutions labelled by a parameter $\lambda$ such that some eigenvalue of $-S_{\alpha\beta}$ changes from positive to negative as $\lambda$ increases through $\lambda_0$, with the signs of the other eigenvalues fixed. The above arguments shows that the solution with $\lambda=\lambda_0$ admits a stationary zero-mode arising from some variation of the parameters of the MP solution that preserves $T$ and $\Omega_i$, and solutions with $\lambda>\lambda_0$ must admit a stationary negative mode. The obvious conclusion is that the negative mode is continuously connected to the zero-mode, i.e. as $\lambda$ increases, a stationary zero-mode appears at $\lambda=\lambda_0$ and this becomes a stationary negative mode for $\lambda>\lambda_0$.

In summary, new negative modes emerge from MP solutions for which an eigenvalue of $-S_{\alpha\beta}$ changes from positive to negative. Hence a new Gregory-Laflamme instability of the associated black string emerges whenever such a change in the sign of an eigenvalue occurs. This is a refinement of the Gubser-Mitra conjecture: we are asserting not just that local thermodynamic instability implies the existence of a classical instability of the string, but that there is a distinct classical instability present for each negative eigenvalue of $-S_{\alpha\beta}$.

\subsection{Ultraspinning black holes}

Next we investigate whether $-S_{\alpha\beta}$ can admit more than one negative eigenvalue. Note that we can relate $-S_{\alpha\beta}$ to the Hessian $H_{ij}\equiv (\partial^2 (-S)/\partial J_i \partial J_j)_M = -S_{ij}$ defined in Ref.~\cite{Dias:2009iu} using the following identity:
\be
\mathrm{det}(-S_{\alpha\beta}) = -\frac{1}{(D-3) MT}\; \mathrm{det}(H_{ij})\,,
\ee
valid for asymptotically flat vacuum black holes. This is proved in Appendix~\ref{app:determinants}. It follows that, for a black hole parameterized by $(M,J_i)$, additional negative eigenvalues of $-S_{\alpha\beta}$ correspond precisely to negative eigenvalues of $H_{ij}$. 

For a MP black hole, for fixed $M$, the eigenvalues of $H_{ij}$ are all positive for small enough angular momenta. However, as some or all of the angular momenta are increased, an eigenvalue of $H_{ij}$ may become negative. If we consider the space parameterized by $J_i$ (at fixed $M$), there is some region containing the origin in which $H_{ij}$ is positive definite. We define the boundary of this region to be the {\it ultraspinning surface}. Following Ref.~\cite{Dias:2009iu}, we shall say that a given black hole is ultraspinning if it lies outside the ultraspinning surface, i.e. if $H_{ij}$ is not positive definite. From the above arguments, we know that as one crosses the ultraspinning surface, the black hole will develop a new negative mode. The associated black string will develop a {\it new} classical instability as this surface is crossed. This is in addition to the instability already present at low angular momenta. Furthermore, on the ultraspinning surface, the new negative mode must reduce to a stationary zero-mode that corresponds to a variation of parameters within the MP family of solutions.

\begin{figure}[t]
\centerline{\includegraphics[width=.45\textwidth]{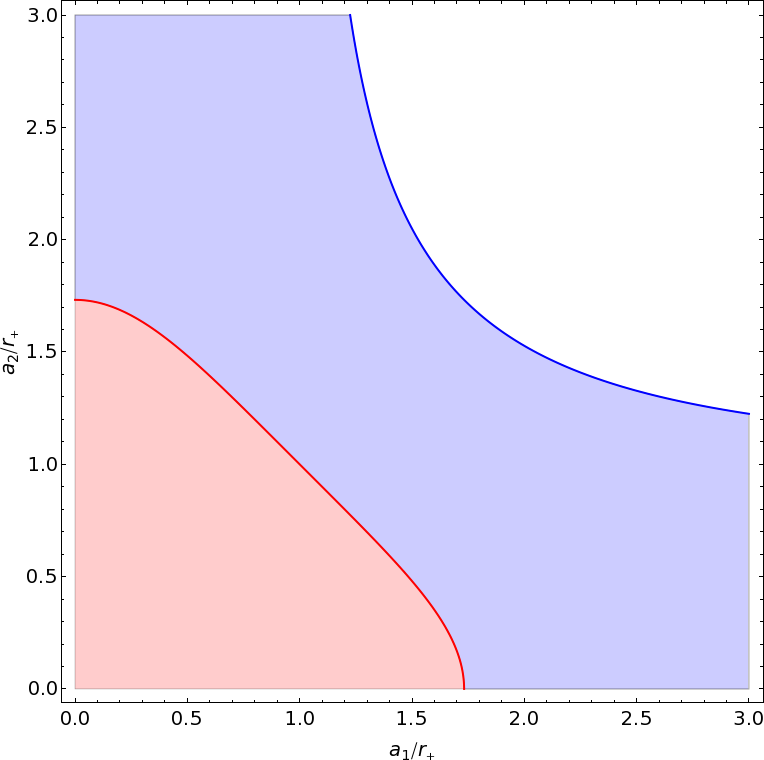}
\hspace{2cm}\includegraphics[width=.45\textwidth]{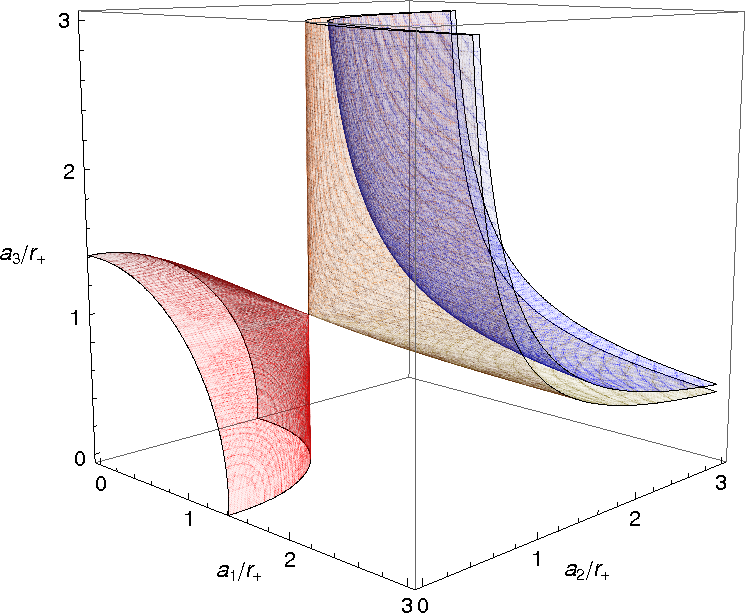}}
\caption{Parameter space for MP black holes with $D=6$ (left) and $D=7$ (right). The black hole is labelled by the horizon radius $r_+$ and the spin parameters $a_i$ which we assume to be positive. For $D=6$, the blue curve corresponds to extreme black holes. In the red region, both eigenvalues of $H_{ij}$ are positive. In the blue region, corresponding to ultraspinning black holes, one eigenvalue is positive and the other is negative. For $D=7$, the blue surface corresponds to extreme black holes. The ultraspinning surface is the red surface near the origin. Inside this surface, $H_{ij}$ is positive definite. The orange surface is where another eigenvalue of $H_{ij}$ vanishes. Between the red and orange surfaces, two eigenvalues of $H_{ij}$ are positive and one is negative. Between the orange and blue surfaces, one eigenvalue of $H_{ij}$ is positive and two are negative. Ultraspinning black holes correspond to points between the red and blue surfaces. Note that the ``cusp" where the red and orange surfaces meet has equal $a_i$, i.e. it corresponds to a cohomogeneity-1 black hole.}
\label{fig:ultraspinning}
\end{figure}

Let us now examine the form of $H_{ij}$ for MP solutions with different $D$. First $D=5$. In this case, there is no ultraspinning region \cite{Dias:2009iu}, i.e. $H_{ij}$ is always positive definite. For $D=6,7$, we find more interesting behaviour shown in Figure \ref{fig:ultraspinning}. We parameterize the black hole by the horizon radius $r_+$ and the spin parameters $a_i \sim J_i/M$ of Ref.~\cite{Myers:1986un} (rather than $M$ and $J_i$) in order to produce clearer figures.
For $D=6$, $H_{ij}$ is a $2 \times 2$ matrix. In this case, the ultraspinning ``surface" is a closed curve that encloses the origin. Inside this curve, both eigenvalues of $H_{ij}$ are positive but outside it one is positive and one is negative. Note that all extreme black holes are ultraspinning.

For $D=7$, $H_{ij}$ is a $3 \times 3$ matrix. The ultraspinning surface encloses the origin. However, now there is another surface on which a second eigenvalue of $H_{ij}$ change from positive to negative. This surface lies between the ultraspinning surface and the surface corresponding to extreme black holes, and touches the ultraspinning surface at a point with equal $a_i$. For generic $a_i$, if one gradually scales up the $a_i$ then one eigenvalue of $H_{ij}$ becomes negative as the ultraspinning surface is crossed, and another eigenvalue becomes negative as the other surface is crossed. As each surface is crossed, the black hole should develop a new stationary negative mode, and the associated black string should develop a new instability.

Some $D=7$ black holes are special, e.g. the second surface does not intersect the region of parameter space corresponding to black holes for which one of the angular momenta (i.e. one of the $a_i$'s) vanishes. For example, consider the singly spinning case. All such black holes must possess the thermodynamic negative mode predicted by Section~\ref{subsec:stringinstab}, a second negative mode should appear at the ultraspinning surface, and there should be no further thermodynamic negative modes. This is consistent with the results of Ref.~\cite{Dias:2009iu}, who found that a stationary negative mode does indeed appear at the ultraspinning surface. They also found evidence that there are infinitely many stationary {\it non-thermodynamic} negative modes that appear at larger angular momenta, the first of which corresponds to the threshold of instability of the black hole. Note that we have justified the assertion of Ref.~\cite{Dias:2009iu} that the zero-mode appearing at the ultraspinning surface does indeed correspond to a variation of parameters within the MP family.

\subsection{Cohomogeneity-1 black holes}

Another special case are the cohomogeneity-1 MP black holes studied in this paper, i.e. equal angular momenta with $D=2N+3$. In this case, $H_{ij}$ has a single eigenvalue that is always positive, associated to the eigenvector $(1,1,\ldots, 1)$, and $N$ degenerate eigenvalues (with eigenvectors related by symmetries of the background) which change from positive to negative as the ultraspinning surface is crossed, with no further changes of sign at larger angular momenta. Note that only the positive eigenvalue corresponds to a variation which preserves the equality of the angular momenta. From the above arguments, we expect that new thermodynamic negative modes will emerge only at the unique value of the angular momentum corresponding to the ultraspinning surface, there will be precisely $N$ of these, and they will break some of the symmetries of the background. 

The ultraspinning surface corresponds to $a=a_1$ (with $r_M=1$) and we found in Section~\ref{sec:strategyresults} that new stationary ($\Gamma=0$) negative modes do indeed emerge at this point. Note that our numerical calculations used Lorentzian signature, whereas the above discussion concerned Euclidean negative modes. However, in Appendix~\ref{sec:PreserveTAngVel}, we show that a regular Lorentzian negative mode (i.e. a solution of equations (\ref{eqn:TTgauge}) and (\ref{lichn})) that is stationary and has $m=0$ does indeed analytically continue to a regular stationary Euclidean negative mode.

These negative modes correspond to $\kappa=1$ harmonics. Since these are thermodynamic negative modes, we know that the zero-mode at $a=a_1$ must be simply a variation of parameters within the MP family, as was asserted in Section~\ref{sec:strategyresults}.

To see that there are precisely $N$ negative modes emerging at $a=a_1$, we use the fact that $\kappa=1$ harmonics are in 1-1 correspondence with Killing vector fields on $CP^N$, so there are $(N+1)^2-1$ such harmonics (see Appendix~\ref{cpnscalarharmonicsappendix}). However, some of these are related by rotations of $CP^N$, so we need to determine how many parameters can be eliminated by rotations. The counting is the same as for $SU(N+1)$ gauge theory with an adjoint Higgs field. Generically this breaks $SU(N+1)$ to $U(1)^{N}$ so we are left with $N$ parameters.\footnote{Thanks to David Tong for this argument.} Hence there are $N$ independent negative modes that emerge at $a=a_1$, in agreement with the prediction from thermodynamics.

Our numerical results confirm that the stationary negative mode that emerges at $a=a_1$ does indeed correspond to the onset of a new instability of the black string in the $\kappa=1$ sector, in agreement with the refinement of the Gubser-Mitra conjecture discussed above.

For $D>5$, comogeneity-1 black holes are ultraspinning for $a>a_1$ and hence might exhibit an instability of the form anticipated in Ref.~\cite{Emparan:2003sy}. Moreover, as explained at the end of section \ref{sec:background}, as we increase $D$, there is more ``space" between the ultraspinning surface and the surface of extremality for such black holes. Therefore the likelihood of an instability might be expected to increase with $D$. This is in agreement with our numerical results, which show no sign of any instability for $D=7$ but confirm that an instability is present for $D=9$. Note that this instability does indeed occur inside the ultraspinning region.

The onset of instability is associated to the appearance of a new stationary zero-mode (at $a=a_2$). This cannot correspond to a variation of parameters within the MP family (we saw above that this is possible only at $a=a_1$). Furthermore, we prove in Appendix~\ref{sec:NoPureGauge} that it cannot be a pure gauge mode. This zero-mode is continuously connected to a stationary ($\Gamma=0$) negative mode that exists for $a>a_2$. This is an example of a non-thermodynamic negative mode, i.e. one which is not associated to an eigenvalue of $-S_{\alpha\beta}$ being negative. The same behaviour was observed in Ref.~\cite{Dias:2009iu}, i.e. the onset of a classical instability of the black hole is associated to the appearance of a new stationary negative mode.

Ref.~\cite{Dias:2009iu} found that further non-thermodynamic negative modes appear as the spin of the black hole is increased still further. For us, extremality imposes an upper bound on the spin of the black hole and we do not find any further negative modes beyond the ones associated to the instability in the $\kappa=2$ sector. However, we believe that, for larger $D$, as well as an instability in the $\kappa=2$ sector there will be further negative modes in sectors with larger $\kappa$. These new negative modes will be associated to new instabilities of the black hole in sectors with larger $\kappa$. The stationary zero-modes associated to the onset of these instabilities will provide evidence for the bifurcation of new families of black hole solutions, involving a large number of parameters, and generically with just one rotational symmetry.


\section{Scalar perturbations and $CP^N$ harmonics\label{sec:decomposition}}

\subsection{Introduction}

The rest of this paper is devoted to explaining the technical details of our work. We shall start by explaining the decomposition of metric perturbations into harmonics on $CP^N$.

Metric perturbations can be decomposed into scalar, vector and tensor types according to how they transform under isometries of $CP^N$. Pertubations of different type must decouple from each other. The decomposition is explained (for a different problem) in Ref.~\cite{Martin:2008pf}. Tensor perturbations are the simplest, these were discussed in Ref.~\cite{Kunduri:2006qa}. We are interested in scalar perturbations, for which the perturbation can be expanded in scalar harmonics on $CP^N$.

We shall assume that our perturbation has been Fourier decomposed as in equation (\ref{eqn:fourier}). All of our numerical results assume $m=0$. However, for the sake of completeness, we shall derive equations that are valid for non-zero $m$. In order to do this, we must address a subtlety (already encountered in Ref.~\cite{Kunduri:2006qa}), that such a perturbation couples with charge $m$ to the 1-form $A_a$ on $CP^N$ defined in Section~\ref{sec:background}. Hence we must consider {\it charged} scalar harmonics on $CP^N$. First we shall describe these harmonics and then explain how to construct gravitational perturbations from them.

\subsection{Charged scalar-derived harmonics in $CP^N$ \label{secscalars}}

We define the gauge-covariant derivative acting on a charge-$m$ tensor field on $CP^N$ as
\begin{equation}
 \CD_a=\hat{\nabla}_a-\ii\, m\,A_a\,,
\end{equation}
where $\hat \nabla$ is the metric covariant derivative on $CP^N$.

\subsubsection{Scalars}

Charged scalar fields on $CP^N$ can be expanded in terms of charged scalar harmonics defined by
\begin{equation}
 (\CD^2+\lambda)\Y=0\,.\label{eqn:chargedS}
\end{equation}
These were studied in \cite{Hoxha:2000jf} (see summary in Appendix~\ref{cpnappendix}), where it is found that
\be
\label{eqn:eigenvalues}
\lambda=\ell(\ell+2N) -m^2\,, \qquad \ell=2\kappa +|m|\,,
\ee
with $\kappa=0,1,2,\ldots$. The modulus sign guarantees that the eigenvalue is the same for positive and negative charges, the corresponding eigenfunctions being related by complex conjugation. 

Notice that the presence of the `gauge field' $A$ leads to
\begin{equation}
\label{commutator}
 [\CD_a,\,\CD_b]\Y=-\ii\, 2\,m\,J_{ab}\,\Y\,.
\end{equation}

\subsubsection{Scalar-derived 1-forms}

Given a scalar harmonic $\Y$, we can define\footnote{Note that $\lambda=0$ if, and only if, $\kappa=m=0$, for which $\Y$ is uncharged and constant. In this case there are no scalar-derived vectors nor scalar-derived tensors.}
\begin{equation}
 \Y_a=-\frac{1}{\sqrt{\lambda}}\,\CD_a \Y\,, \label{eqn:vectorH}
\end{equation}
which transforms as a charged 1-form on $CP^N$. This can be decomposed into its $(1,0)$ and $(0,1)$ parts using the complex structure on $CP^N$. Denote these as $\Y^+_a$ and $\Y^-_a$ respectively, where 
\be
 J_a{}^b \Y^\pm_b = \mp \ii \Y^\pm_a.
\ee
We shall refer to $\Y^\pm_a$ as scalar-derived 1-form harmonics.
We find that they satisfy
\begin{equation}
\CD^2\Y_a^{\pm}=-\left[\lambda-2(N+1)\mp 4\,m\right]\Y_a^{\pm}\,\end{equation}
and
\begin{equation}
\CD^a\Y_a^\pm=
	\frac{1}{2\sqrt{\lambda}}\left(\lambda \mp 2\,m\,N\right)\,\Y\,.
\end{equation}
We shall make use of the result that Killing vectors of $CP^N$ are in one-to-one correspondence with uncharged ($m=0$) scalar harmonics with $\kappa=1$ (see e.g. \cite{Hoxha:2000jf} and our Appendix~\ref{cpnappendix}).
Given such a harmonic $\Y$, the corresponding Killing vector field is $-\ii(\Y^+_a-\Y^-_a)$.

\subsubsection{Scalar-derived tensors}

Following Ref.~\cite{Martin:2008pf}, we  decompose a symmetric tensor $\Y_{ab}$ into its Hermitian (or $(1,1)$) and anti-Hermitian components according to the eigenvalue of the map
\begin{equation}
 ({\cal J}\Y)_{ab}=J_a^{\phantom a c}J_{b}^{\phantom b d}\Y_{cd}\,.
\end{equation}
If the eigenvalue is $+1$ the corresponding eigentensor is called Hermitian, and if it is $-1$ the eigentensor is called anti-Hermitian. In the anti-Hermitian case, we can further distinguish between the $(2,0)$ and $(0,2)$ components of $\Y_{ab}\,$, which are defined by $J_a^{\phantom a c}\Y_{cb}=\mp \ii\, \Y_{ab}$ with the upper and lower signs for the $(2,0)$ and $(0,2)$ components respectively.

The following quantities form a basis for anti-hermitian scalar-derived tensors:\footnote{This follows from the scalar part of equation (39) of Ref.~\cite{Martin:2008pf}.}
\begin{equation}
 \Y_{ab}^{++}=\CD_{(a}^+\Y_{b)}^+\,,\qquad \Y_{ab}^{--}=\CD_{(a}^-\Y_{b)}^-\,.
\end{equation}
$\Y_a^{\pm}$ denotes the scalar-derived 1-form harmonics of the previous Section, and $\CD_a^{\pm}$ denotes the projection of $\CD_a$ onto its $(1,0)$ and $(0,1)$ components. Notice that the correspondence between $m=0$, $\kappa=1$ scalar harmonics and Killing vector fields on $CP^N$ implies that $\Y^{\pm\pm}_{ab}$ vanish for such harmonics.

Hermitian scalar-derived tensors can be written in terms of a trace, and a traceless part, for which the following quantities give a basis:\footnote{
In Ref.~\cite{Martin:2008pf}, hermitian tensors were converted into $(1,1)$-forms by contracting with $J^a{}_b$. The two quantities written here correspond to terms of the form $J \Y$ and (the primitive part of) $dd^c \Y$ in equation (47) of Ref.~\cite{Martin:2008pf}.}
\begin{equation}
\hat{g}_{ab} \Y, \qquad \Y_{ab}^{+-}=\CD_{(a}^+\Y_{b)}^-+\CD_{(a}^-\Y_{b)}^+-\frac{1}{2\,N}\,\hat{g}_{ab}\,(\CD\cdot \Y)\,.
\end{equation}

These tensor harmonics satisfy 
\begin{equation}
\begin{aligned}
\CD^2\Y_{ab}^{\pm\pm}&=-\left[\lambda-4(N+3) \mp 8\,m\right]\Y_{ab}^{\pm\pm}\,,\\
\CD^2\Y_{ab}^{+-}&=-\left(\lambda-4\,N\right)\Y_{ab}^{+-}\,,
\end{aligned}
\end{equation}
with
\begin{equation}
\begin{aligned}
\CD^c\Y_{ca}^{\pm\pm}&=-\frac{\lambda-4(N+1) \mp 2\,m(N+2)}{2}\,\Y_a^\pm \,,\\
\CD^c\Y_{ca}^{+-}&=-\frac{N-1}{2\,N}\left[(\lambda+2\,m\,N)\Y_a^+
	+(\lambda-2\,m\,N)\Y_a^-\right]\,.
\end{aligned}
\end{equation}
 

\subsection{Decomposition of perturbations in scalar-derived harmonics}

Let us consider now the perturbations of the full spacetime metric. We introduce the orthonormal basis
\begin{equation}
\label{vielbein}
 e^{(0)}=f\,dt\,,\quad e^{(1)}=g\,dr\,,\quad e^{(2)}=h\,\left(d\psi+A-\Omega\,dt\right)\,,\quad
e^{(i)}=r\,\hat e^{(i)}\,,
\end{equation}
where $\hat e^{(i)}$ is the tetrad of the $CP^N$ manifold. The dual basis is then
\begin{equation}
 e_{(0)}=\frac{1}{f}\left(\partial_t+\Omega\,\partial_\psi\right)\,,\quad 
e_{(1)}=\frac{1}{g}\,\partial_r\,,\quad e_{(2)}=\frac{1}{h}\,\partial_\psi\,,\quad
e_{(i)}=\frac{1}{r}\left[\hat e_{(i)}-\langle A,\hat e_{(i)}\rangle\,\partial_\psi\right]\,.
\end{equation}
Take $e^{(A)}=\{e^{(0)},e^{(1)},e^{(2)}\}$ and a coordinate basis $dx^a$ on $CP^N$. The components $h_{AB}$ of the metric perturbation transform as scalars under isometries of $CP^N$ and can therefore be decomposed using scalar harmonics on $CP^N$. Similarly, since we are restricting attention to scalar-type perturbations, components of the form $h_{Aa}$ and $h_{ab}$ can be decomposed using scalar-derived 1-forms and scalar-derived tensors on $CP^N$:
\begin{equation}
\begin{aligned}
h_{AB}&=f_{AB}\,\Y\,,\\
h_{Aa}&=r\left(f^+_A\,\Y^+_a+f^-_A\,\Y^-_a\right)\,,\\
h_{ab}&=-\frac{r^2}{\sqrt\lambda}\left(
H^{++}\,\Y^{++}_{ab}+H^{--}\, \Y^{--}_{ab}+H^{+-}\,\Y^{+-}_{ab}\right)
+r^2\,H_L\,\hat{g}_{ab}\Y\,,
\end{aligned}
\label{eqn:hmetric}
\end{equation}
where $f^\pm_A=\{W^\pm,X^\pm,Z^\pm\}$, and the functions multiplying the harmonics depend only on ($t,r,\psi$) and not on the coordinates of $CP^N$. The real spacetime metric perturbation is given by $\textrm{Re}\left(h_{\mu\nu}\right)$. Since $\partial_t$ and $\partial_\psi$ are Killing vectors of the background solution, we will Fourier expand all of these functions in $t$ and $\psi$, i.e. we assume a dependence $e^{-\ii\omega t+\ii m\psi}$. It remains to determine the dependence of these functions on $r$. The stability problem will thus be reduced to a system of linear ordinary differential equations. We give these differential equations in Appendices~\ref{gaugeconditions} and \ref{Lichnerowicz}.

\subsection{Boundary conditions \label{secBC}}

The metric perturbations must be regular on the future event horizon ${\cal H}^+$. This boundary condition can be imposed by considering a basis which is regular on ${\cal H}^+$, since the components of the perturbation in that basis must be regular. Let us change to the ingoing Eddington-Finkelstein coordinates that are regular at ${\cal H}^+$:
\begin{equation}
 dt\to dv-\frac{g}{f}\,dr\,,\qquad d\psi\to d\varphi -\frac{\Omega\,g}{f}\,dr\,,
\end{equation}
and consider the basis $\{ dv, dr, d\varphi +A -\Omega dv, dx^a \}$. Denote the components of the metric perturbation with respect to this new basis with a bar (e.g. $f_{\bar{0}\bar{0}}$, $\bar{W}^+$). Our boundary condition is that these components should be smooth functions of $(v,r,\phi,x^a)$ at the horizon.\footnote{In other words, we demand that the tensor field $h_{\mu\nu}$ should be smooth at ${\cal H}^+$. This is stronger than the statement that the metric perturbation should be regular at the horizon e.g. it excludes the possibility that $h_{\mu\nu}$ is singular at ${\cal H}^+$ in a certain gauge but can be made regular by a gauge transformation. For example, we show in Appendix \ref{sec:PreserveTAngVel} that, in the transverse traceless gauge, a perturbation with $\omega=m=0$ that satisfies our boundary condition cannot change the temperature or angular velocity of the black hole. It follows that a variation in the parameters of the MP solution that {\it does} change $T$ or $\Omega_H$ will not give a perturbation $h_{\mu\nu}$ that is smooth at the horizon in this gauge.}

For this class of black holes, the horizon is located at the largest real root $r=r_+$ of $\Delta=g(r)^{-2}$. For a non-extreme black hole, near the horizon,
$\Delta(r)=\Delta'(r_+)(r-r_+)+O[(r-r_+)^2]$, with $\quad \Delta'(r_+)>0$.
Using the relation $f(r)=r/(g(r)\,h(r))$, we find that, near the horizon, the metric components in the original basis are related to the components in the new basis by
\begin{equation}
\label{bchor}
\begin{aligned}
& f_{00}\approx \frac{h(r_+)^2}{r_+^2\Delta'(r_+)} \,\frac{f_{\bar{0}\bar{0}}}{r-r_+}\,,\qquad f_{01}-f_{00} = \frac{h(r_+)}{r_+} \, f_{\bar{0}\bar{1}}\,, \\ 
&  f_{00}-2 f_{01}+ f_{11}\approx \Delta'(r_+) \, f_{\bar{1}\bar{1}}\; (r-r_+)\,, \\
& f_{02} \approx \frac{1}{r_+\sqrt{\Delta'(r_+)}} \, \frac{f_{\bar{0}\bar{2}}}{\sqrt{r-r_+}}\,,\qquad 
 f_{12}-f_{02}\approx \frac{\sqrt{\Delta'(r_+)}}{h(r_+)}  \, f_{\bar{1}\bar{2}} \sqrt{r-r_+}\,, \\
& f_{22}= \frac{1}{h(r_+)^2} \, f_{\bar{2}\bar{2}}\,, \qquad Z^{\pm}= \frac{1}{r_+ h(r_+)} \, \bar{Z}^{\pm} \,,  \\
&W^\pm \approx \frac{h(r_+)}{r_+\sqrt{\Delta'(r_+)}} \,\frac{\bar{W}^{\pm}}{\sqrt{r-r_+}}\,,\qquad
X^\pm - W^\pm \approx \frac{\sqrt{\Delta'(r_+)}}{r_+} \, \bar{X}^{\pm} \sqrt{r-r_+}\,.
\end{aligned}
\end{equation}
The functions $H^{++}$, $H^{--}$, $H^{+-}$, $H_L$ associated to the components of the metric perturbation on $CP^N$ are the same in the two bases.

Since the components in the new basis should be smooth at the horizon, the above expressions give us boundary conditions on the behaviour of the components in the old basis. In imposing these boundary conditions, it is important to remember that, near the horizon,
\be
\label{expohor}
e^{-\ii \omega v + \ii m \varphi} \approx e^{-\ii \omega t + \ii m \psi} \left( \frac{r-r_+}{r_+} \right)^{-\ii \alpha (\omega-m\Omega_H)},
\ee
where $\alpha \equiv h(r_+)/(r_+ \Delta'(r_+))$ is positive (for non-extreme black holes). Hence, for example, the radial dependence of $f_{00}$ near the horizon must be
\be
 f_{00} \propto  \left( \frac{r-r_+}{r_+} \right)^{-1-\ii \alpha (\omega-m\Omega_H)}F(r),
\ee
where $F(r)$ is smooth at $r=r_+$.

When solving numerically the stability equations, it will be necessary to work with the combinations that maximize the information on the boundary conditions. For instance, one should work with $f_{02}$ and $f_{12}-f_{02}$, instead of considering only the leading behaviour of $f_{02}$ and $f_{12}$, otherwise the information that $f_{12}(r_+)-f_{02}(r_+)=0$ is lost.

As for the behaviour of the perturbations at spatial infinity $r \to \infty$, we are interested in boundary conditions that preserve the asymptotic flatness of the spacetime. For perturbations of the black string, the equations of motion \eqref{lichn} then imply that all the functions vanish exponentially for large $r$.

\section{The eigenvalue problem \label{sec:eigenvalue}}

The ansatz for the metric perturbation $h_{\mu\nu}$ is given by Eq.~\eqref{eqn:hmetric}. We list in Appendix~\ref{Lichnerowicz} the components of the Lichnerowicz eigenvalue equation \eqref{lichn} in the tetrad basis \eqref{vielbein}. These consist of sixteen coupled second order ordinary differential equations, each one being second order only in one of the perturbation functions, e.g. \eqref{lichne:f00} is second order in $f_{00}$. However, six of these functions can be solved for in terms of the ten remaining functions and their first derivatives when we impose the transverse traceless (TT) gauge conditions, listed in Appendix~\ref{gaugeconditions}.

Notice that the TT conditions completely fix the gauge in Eq.~(\ref{lichn}) when $k>0$, since the action of the Lichnerowicz operator on a gauge mode is trivial, $\Delta_L \nabla_{(\mu} \xi_{\nu)}=0$. As for $k=0$, there are two distinct cases. The first is the limit $k \to 0$ for which $\Gamma \to 0$, as can be seen on the right plots of Figs.~\ref{fig:5dkzero}, \ref{fig:7dkzero}, \ref{fig:7d} and \ref{fig:9dk1}. The limiting perturbation $k=0$ is an unphysical pure gauge mode, as happens in the original Gregory-Laflamme case \cite{Gregory:1994bj}. The second is the much more interesting stationary perturbation $k_c=0$ marking the onset of a new Gregory-Laflamme instability when the rotation increases. In the left plots of Figs.~\ref{fig:7d}, \ref{fig:9dk1} and \ref{fig:9dk2}, this is the threshold mode of the curve $\Gamma=0$. In the right plots of the same Figures, it would correspond to the graph squeezing into the origin for a critical value of the rotation. These stationary perturbations are physical since there is no gauge ambiguity. The TT conditions require that any gauge vector $\xi_\mu$ is a harmonic 1-form, satisfying $\nabla^\mu \xi_\mu =0$ and $\nabla^\rho \nabla_\rho \xi_{\mu} =0$. In Appendix~\ref{sec:NoPureGauge}, we show that no regular harmonic 1-forms exist in this case: they lead to pure gauge metric perturbations that diverge either at the boundary $r=r_+$ or at infinity $r \to \infty$. A proof along the same lines, but much more cumbersome, can be given for the modes which represent the actual instability of the black holes, i.e. the exponential growth of the perturbations with time ($k=0$ and $\Gamma >0$).

The tetrad basis \eqref{vielbein} is very convenient in the explicit derivation of the TT gauge conditions and the Lichnerowicz equations. However, for the actual implementation of the numerical problem, it is convenient to choose perturbation functions that make the final equations more amenable to numerics, e.g. it is useful to avoid using expressions involving square roots. It is also helpful to define combinations of the original perturbation functions which can be solved for algebraically through the TT gauge conditions. Both features are respected if we consider the perturbations in a related basis such that:
\be
\label{eq:newfunc}
\begin{array}{l}
\mathfrak{f}_{00} = f_{00}\,f^2-2 f_{02}\,f\, h\, \Omega +f_{22} \,h^2\, \Omega^2\,, \qquad
\mathfrak{f}_{01} = f_{01}\, f \,g-f_{12}\, g\, h\, \Omega\,, \qquad
\mathfrak{f}_{02} = f_{02}\, f\, h-f_{22}\, f^2\, \Omega \,, \\ \\
\mathfrak{f}_{11} = f_{11}\, g^2 \,, \qquad \qquad
\mathfrak{f}_{12} = f_{12}\, g\,h\,, \qquad \qquad
\mathfrak{f}_{22} = f_{22}\, h^2\,, \\ \\
\mathfrak{f}_{0} = -\frac{1}{2} \,r\, \big( (W^+ +W^-)\,f -(Z^+ +Z^-)\,h\, \Omega \big) \,, \qquad
\tilde{\mathfrak{f}}_{0} =- \ii \, \frac{1}{2} \,r\, \big( (W^+ -W^-)\,f -(Z^+ -Z^-)\,h\, \Omega \big) \,, \\ \\
\mathfrak{f}_{1} = -\frac{1}{2} \,r \, g\, (X^+ +X^-)\,, \qquad
\tilde{\mathfrak{f}}_{1} =- \ii \, \frac{1}{2} \,r \, g\, (X^+ -X^-)\,, \\ \\
\mathfrak{f}_{2} = -\frac{1}{2} \,r \, h \,(Z^+ +Z^-)\,, \qquad
\tilde{\mathfrak{f}}_{2} = -\ii \, \frac{1}{2} \,r \, h\, (Z^+ -Z^-)\,, \\ \\
P = \frac{1}{4}\, (2 H^{+-}-H^{++}-H^{--})\,, \qquad
Q = \ii\,\frac{1}{2}\, (H^{++}-H^{--})\,, \\ \\
U = \frac{1}{4}\, (2 H^{+-}+H^{++}+H^{--})\,, \qquad
V = H_L + \frac{1}{4 N}\, \left(H^{+-}-\frac{1}{2}\, (H^{++}+H^{--}) \right)\,.
\end{array}
\ee
Now, we solve for six of these functions ($\mathfrak{f}_{00}$, $\mathfrak{f}_{0}$, $\mathfrak{f}_{2}$, $Q$, $U$, $V$) in terms of the ten remaining functions and their first derivatives by imposing the TT gauge conditions in Appendix~\ref{gaugeconditions}. Upon this substitution, the corresponding second order Lichnerowicz equations will become third order. The ten equations which are second order in $\mathfrak{f}_{01}$, $\mathfrak{f}_{02}$, $\mathfrak{f}_{11}$, $\mathfrak{f}_{12}$, $\mathfrak{f}_{22}$, $\tilde{\mathfrak{f}}_{0}$, $\mathfrak{f}_{1}$, $\tilde{\mathfrak{f}}_{1}$, $\tilde{\mathfrak{f}}_{2}$, $P$, will remain second order. They will constitute the system of equations to be solved numerically\footnote{The explicit representation of the system would be far too cumbersome. The interested reader can obtain it from the TT conditions in Appendix~\ref{gaugeconditions}, the equations of Appendix~\ref{Lichnerowicz} and the relations \eqref{eq:newfunc}.}. A non-trivial consistency check on the gauge choice procedure is that the ten final second order equations must solve the six third order equations (e.g. a third order equation is a derivative of a second order one). We verified explicitly that this is the case.

The final system will be solved using a spectral numerical method. The application of the method is simpler for Dirichlet boundary conditions. We consider then the following perturbation functions:
\be
\label{eq:qs}
\begin{array}{l}
\displaystyle{q_1 =  \left( 1-\frac{r_+}{r} \right)^{\ii \alpha(\omega-m\Omega_H)+3}}\, \mathfrak{f}_{11} \,, \qquad \qquad
\displaystyle{q_2 =  \left( 1-\frac{r_+}{r} \right)^{\ii \alpha(\omega-m\Omega_H)+1}}\, \mathfrak{f}_{22} \,, \\
\displaystyle{q_3 =  \left( 1-\frac{r_+}{r} \right)^{\ii \alpha(\omega-m\Omega_H)+2}}\, \mathfrak{f}_{01} \,, \qquad \qquad
\displaystyle{q_4 =  \left( 1-\frac{r_+}{r} \right)^{\ii \alpha(\omega-m\Omega_H)+2}}\, \mathfrak{f}_{1} \,, \\
\displaystyle{q_5 =  \left( 1-\frac{r_+}{r} \right)^{\ii \alpha(\omega-m\Omega_H)+2}}\, \tilde{\mathfrak{f}}_{1} \,, \qquad \qquad \;\,
\displaystyle{q_6 =  \left( 1-\frac{r_+}{r} \right)^{\ii \alpha(\omega-m\Omega_H)+1}}\, \tilde{\mathfrak{f}}_{2} \,, \\ 
\displaystyle{q_7 =  \left( 1-\frac{r_+}{r} \right)^{\ii \alpha(\omega-m\Omega_H)+1}}\, P \,, \\
\displaystyle{q_8 =  \left( 1-\frac{r_+}{r} \right)^{\ii \alpha(\omega-m\Omega_H)}}\, \left\{\mathfrak{f}_{02}+\Omega_H\,\mathfrak{f}_{22} +
\frac{r_+}{\alpha\,\Omega_H} \left( 1-\frac{r_+}{r} \right) \left[ \mathfrak{f}_{01}-\frac{r_+}{\alpha} \left( 1-\frac{r_+}{r} \right)\mathfrak{f}_{11} \right]\right\} \,, \\
\displaystyle{q_9 =  \left( 1-\frac{r_+}{r} \right)^{\ii \alpha(\omega-m\Omega_H)+1}}\, \left\{\mathfrak{f}_{12}+ \frac{1}{\Omega_H}\,\left[ \mathfrak{f}_{01} - \frac{r_+}{\alpha} \left( 1-\frac{r_+}{r} \right) \mathfrak{f}_{11} \right]\right\} \,, \\
\displaystyle{q_{10} =  \left( 1-\frac{r_+}{r} \right)^{\ii \alpha(\omega-m\Omega_H)}}\, \left\{ \tilde{\mathfrak{f}}_{0} + \Omega_H\, \tilde{\mathfrak{f}}_{2} -\frac{r_+}{\alpha} \left( 1-\frac{r_+}{r} \right) \tilde{\mathfrak{f}}_{1} \right\}\,, \\
\end{array}
\ee
which vanish linearly at the horizon location $r=r_+$. This behaviour can be verified in the expressions \eqref{bchor} and \eqref{expohor}. The particular combinations chosen for $q_8$, $q_9$ and $q_{10}$ encode the total information about the boundary conditions imposed by regularity, as argued in the end of Section~\ref{secBC}. For the numerical implementation, it is convenient to use the variable
\be
y=1-\frac{r_+}{r}
\ee
instead of the radial coordinate $r$, since $y$ is dimensionless and bounded, $0\leq y \leq 1\,$. The functions represented above vanish at infinity $r=\infty$ ($y=1$) since the large $r$ behaviour of the solutions of \eqref{lichn} is exponential, $e^{\pm k r}$, and regularity at infinity imposes the minus sign. The $k=0$ case will be obtained as the limit $k \to 0$.

The system of ten second order ordinary differential equations is ready to be solved numerically. We now describe the numerical method that we employed to solve it. Amongst the most popular choices, finite difference methods, finite element methods and spectral methods, we decided to use spectral methods. These methods are particularly useful to solve a system of coupled ordinary differential equations (ODEs), or coupled partial differential equations (PDEs), to high accuracy on a simple finite domain, as long as the data defining the problem are smooth. To our knowledge, the first application of these methods to general relativity was given in \cite{Monteiro:2009ke,Dias:2009iu}.

Typically, in spectral methods one tries to approximate a given function, defined on a finite domain, as a finite sum of algebraic polynomials $p(y) = \sum_{i=0}^{\mathcal N} a_i y^i$, i.e. one constructs an interpolation polynomial. At first sight, it seems that the best idea is to use polynomial interpolation in an equidistant grid with ${\mathcal N}+1$ points. However, it turns out that this is catastrophic, because of the Runge phenomenon. Not only this approximation does not converge in general, as ${\mathcal N}\to\infty$, but also the approximation gets worse at a rate that can be as large as $2^{\mathcal N}$. The correct approach is to perform the interpolation in a non-uniform grid. There are many sets of points that are effective, we used the so called Chebyshev points given by
\begin{equation}
y_j = \frac{a+b}{2}+\frac{a-b}{2}\cos \left(\frac{j \pi}{{\mathcal N}}\right),\qquad
j\in\{0,1,\ldots,{\mathcal N}\},
\label{eq:ap1}
\end{equation}
for $y_j\in[a,b]$. Not only does this set of points avoids the Runge phenomenon, by clustering points near the boundary, but it has another well-known advantage over uniform grids: it typically leads to methods with exponential accuracy. However, because we are effectively approximating a given function as a sum of polynomials, we must require our functions to be analytic.

The procedure to solve differential equations is in the same spirit as standard quantum mechanics. Consider a complicated system of $n$ coupled linear ODEs with variable coefficients,
\begin{equation}
\sum_{\beta=1}^n H_{\alpha\beta}\; q_\beta^{(k)} = -k^2 \sum_{\beta=1}^n
T_{\alpha\beta}\;q_\beta^{(k)},\qquad \alpha\in\{0,1,\ldots,n\},
\label{eq:ap2}
\end{equation}
where each $H_{\alpha\beta}$ is a second order operator in $y$, each $T_{\alpha\beta}$ is a scalar function and $\{k,q_\beta^{(k)}\}$ are the eigenvalues and eigenfunctions that we want to determine. We now perform our approximation and discretise the $[a,b]$ interval in a set of discrete points given by Eq.~(\ref{eq:ap1}). Each $q_\beta^{(k)}$ is then approximated by a vector, $\uline{q}_\beta^{(k)}$, whose entries are the values of the eigenfunctions we want to determine at $y_j$. Following this procedure, one represents derivatives with respect to $y$ by matrices, $D_{\mathcal N}$, that act on the vectors $\uline{q}_\beta^{(k)}$, mixing adjacent points (see p.53 of \cite{Trefethen} for an explicit construction of such matrices). After this approach is complete, each $H_{\alpha\beta}$ and $T_{\alpha\beta}$ are transformed into square matrices $\uuline{H}_{\alpha\beta}$  and $\uuline{T}_{\alpha\beta}$, respectively, of dimension $({\mathcal N}+1)\times ({\mathcal N}+1)$.

All we are left with are the boundary conditions. However we conveniently choose to work with the functions defined in Eq.~(\ref{eq:qs}), in terms of which the boundary conditions are just of the Dirichlet type. One can easily impose this type of boundary conditions by setting the first and last element of each $\uline{q}_\beta^{(k)}$ to zero, which amounts to delete the first and last elements of each eigenvector, and to eliminate the first and last column and row of each matrix $\uuline{H}_{\alpha\beta}$ and $\uuline{T}_{\alpha\beta}$ \cite{Trefethen}. In the end we are left with the following system of linear algebraic equations
\begin{equation}
\sum_{\beta=1}^n\uuline{\hat{H}}_{\alpha\beta}\; \uline{\hat{q}}_\beta^{(k)} =
-k^2
\sum_{\beta=1}^n\uuline{\hat{T}}_{\alpha\beta}\;\uline{\hat{q}}_\beta^{(k)},
\label{eq:ap3}
\end{equation}
where $\uuline{\hat{H}}_{\alpha\beta}$ and $\uuline{\hat{T}}_{\alpha\beta}$ are obtained from $\uuline{H}_{\alpha\beta}$  and $\uuline{T}_{\alpha\beta}$ by deleting their first and last column and row, and thus are square $({\mathcal N}-1)\times ({\mathcal N}-1)$ matrices. Finally, the system of equations Eq.~(\ref{eq:ap3}) can be recast in a more convenient form,
\begin{equation}
\left[
\begin{array}{ccc}
\uuline{\hat{H}}_{11} & \ldots & \uuline{\hat{H}}_{1n}
\\
\vdots & \ddots & \vdots
\\
\uuline{\hat{H}}_{n1} & \ldots & \uuline{\hat{H}}_{nn}
\end{array}\right]
\left[\begin{array}{c}
\uline{\hat{q}}_1^{(k)}
\\
\vdots
\\
\uline{\hat{q}}_{n}^{(k)}
\end{array}\right]=-k^2\left[
\begin{array}{ccc}
\uuline{\hat{T}}_{11} & \ldots & \uuline{\hat{T}}_{1n}
\\
\vdots & \ddots & \vdots
\\
\uuline{\hat{T}}_{n1} & \ldots & \uuline{\hat{T}}_{nn}
\end{array}\right]
\left[\begin{array}{c}
\uline{\hat{q}}_1^{(k)}
\\
\vdots
\\
\uline{\hat{q}}_{n}^{(k)}
\end{array}\right],
\end{equation}
which is just a standard generalized eigenvalue problem in $-k^2$ of dimension $n ({\mathcal N}-1)$.

The success of the application of spectral allocation methods to our system of ODEs hinges on two key properties that our system of ODEs possess, namely ellipticity of the second order operator, meaning that we are dealing with a boundary value problem, and the fact that our system reduces to a generalized eigenvalue problem in $-k^2$.

\section{Acknowledgements}

We are grateful to Roberto Emparan for valuable discussions and comments on a draft. PF would like to thank Veronika Hubeny, Mukund Rangamani, Simon Ross, Paul Sutcliffe and especially James Lucietti for useful discussions.  OJCD acknowledges financial support provided by the European Community through the Intra-European Marie Curie contract PIEF-GA-2008-220197. PF is supported by an STFC rolling grant. RM and JES acknowledge support from the Funda\c c\~ao para a Ci\^encia e Tecnologia (FCT-Portugal) through the grants SFRH/BD/22211/2005 (RM) and SFRH/BD/22058/2005 (JES). HSR is a Royal Society University Research Fellow. This work was partially funded by FCT-Portugal through projects PTDC/FIS/64175/2006, PTDC/FIS/098962/2008, PTDC/FIS/099293/2008, CERN/FP/83508/2008 and CERN/FP/109306/2009. This is preprint DCPT-10/03.

\appendix

\section{Determinants of the thermodynamic Hessians \label{app:determinants}}

We wish to show that
\be
\mathrm{det}(-S_{\alpha\beta}) = -\frac{1}{(D-3) MT} \,\mathrm{det}(H_{ij})\,,
\ee
where $S_{\alpha\beta} \equiv \partial^2 S / \partial x^\alpha \partial x^\beta$, with $x^\alpha=(M,J_i)$, and $H_{ij} \equiv \partial^2 (-S) / \partial J_j \partial J_j = -S_{ij}$. The first step is to use the determinant identity
\be
\mathrm{det}\left(  \begin{array}{cc}
A & B_i \\
B_j & C_{ij}
\end{array}
\right) = (A - C_{kl}^{-1} B_k B_l) \;\mathrm{det}(C_{ij})\,.
\ee
In our case, the left-hand side is $\mathrm{det}(-S_{\alpha\beta})$, and $C_{ij}=H_{ij}$. We have then $\mathrm{det}(-S_{\alpha\beta}) = \sigma\, \mathrm{det}(H_{ij})$, with
\be
\sigma= -S_{00} - H^{-1}_{ij} S_{0i} S_{0j} = - \left(\frac{\partial \beta}{\partial M}\right)_J + \left(\frac{\partial (\beta \Omega_j)}{\partial J_i}\right)_M^{-1} \left(\frac{\partial (\beta \Omega_i)}{\partial M}\right)_J \left(\frac{\partial (\beta \Omega_j)}{\partial M}\right)_J\,.
\ee
where we have used the first law of thermodynamics $dS= \beta dM -\beta \Omega_i dJ_i$. Since 
\be
\left(\frac{\partial (\beta \Omega_j)}{\partial J_i}\right)_M^{-1} = \left(\frac{\partial J_j}{\partial (\beta \Omega_i)}\right)_M \qquad \mathrm{and} \qquad
\left(\frac{\partial J_j}{\partial M}\right)_{\beta \Omega} = \left(\frac{\partial J_j}{\partial (\beta \Omega_i)}\right)_M \left(\frac{\partial (\beta \Omega_j)}{\partial M}\right)_J\,,
\ee
we get
\be
\sigma= - \left(\frac{\partial \beta}{\partial M}\right)_J + \left(\frac{\partial J_j}{\partial M}\right)_{\beta \Omega} \left(\frac{\partial (\beta \Omega_j)}{\partial M}\right)_J\,.
\ee
The identity
\be
\left(\frac{\partial \beta}{\partial M}\right)_J = \left(\frac{\partial \beta}{\partial M}\right)_{\beta \Omega} - \left(\frac{\partial \beta}{\partial J_j}\right)_M \left(\frac{\partial J_j}{\partial M}\right)_{\beta \Omega}
\ee
and the $S_{0i}=S_{i0}$ ``Maxwell relation'', $\left(\partial (\beta \Omega_j) / \partial M \right)_J = \left(\partial \beta / \partial J_j \right)_M $, further imply that
\be
\sigma= - \left(\frac{\partial \beta}{\partial M}\right)_{\beta \Omega} = - \left(\frac{\partial M}{\partial \beta}\right)_{\beta \Omega}^{-1} \,.
\ee
Following the steps taken between Eqs.~\eqref{w00i} and \eqref{w00f}, we finally obtain
\be
\sigma = -\frac{1}{(D-3) MT} <0\,.
\ee
Those steps require that the Smarr relation \eqref{smarr} is valid, i.e. they apply only to asymptotically flat vacuum black holes.


\section{The geometry of $CP^N$ \label{cpnappendix}}

\subsection{The Fubini-Study construction}

We review here the Fubini-Study construction of the Einstein-K\"ahler metric and K\"ahler potential on $CP^N$ \cite{Hoxha:2000jf}\footnote{We use the coordinates $\{R_N,\Psi_N\}$ that are related with the coordinates $\{\xi,\widetilde{\tau} \}$ of \cite{Hoxha:2000jf} through the coordinate transformation $\sin^2\xi=R_N^2/(1+R_N^2)$ and $\widetilde{\tau}=\Psi_N/2$.}. This construction allows us to generate iteratively the $CP^N$ metric and potential from the knowledge of the metric and potential of $CP^{N-1}$.

Take the $\mathbb{C}^{N+1}$ manifold with complex coordinates $Z^A$ and flat metric
\beq
ds_{2N+2}^2 = dZ^A\, d\overline{Z}_A\,,\label{FlatMetric}
\eeq
where the index $A$ runs as $A=(0,\alpha)$, with $1\leq\alpha\leq N$. Introduce $N$ inhomogeneous coordinates $\zeta^\alpha = Z^\alpha/Z^0$ in the patch where $Z^0\neq 0\,$, such that
\beqa
Z^0=e^{\im \tau}\, |Z^0|\,,\qquad Z^\alpha=Z^0 \,\zeta^\alpha =R_N
\, u^\alpha \qquad Z^A\, \overline{Z}_A=r^2\,,\qquad
f=1+\zeta^\alpha\, \overline{\zeta}^{\bar\alpha}=1+R_N^2
\,.\label{Def:zeta}
\eeqa
Furthermore, introduce a new set of $(N-1)$ inhomogeneous coordinates $v^i$ ($0\leq i\leq N-1$) such that
\beq
u^N=e^{\im \Psi_N/2}\, |u^N|\,,\qquad  u^i=u^N v^i \qquad \hbox{with}\qquad u^\alpha \, \bar u^{\bar\alpha}= 1 \,.\label{Def:v}
\eeq
The flat metric on $\mathbb{C}^{N+1}$ can then be written as
\beqa
ds_{2N+2}^2 = dr^2 + r^2\, d\Omega_{2N+1}^2\,, \qquad
\hbox{where}\qquad d\Omega_{2N+1}^2 = (d\tau + A_{(N)})^2 +
d\Sigma_N^2\, \label{Hopf}
\eeqa
is the metric on the unit sphere $S^{2N+1}$, and $d\Sigma_N^2$ is the unit $CP^N$ metric. Written in this way we see that $S^{2N+1}$ is a Hopf fibration of $S^1$ over $CP^N$. In \eqref{Hopf}, $A_{(N)}$ is the $CP^N$ K\"ahler potential. Explicitly, the  $CP^N$  metric and K\"ahler potential are given by ($R_N\geq 0$ and $0\leq \Psi_N \leq  4\pi$)
\begin{eqnarray}
&& d\Sigma_{N}^2 = \hat{g}_{ab} dx^a dx^b = \frac{dR_N^2}{\lp 1+R_N^2\rp^2}
  +\frac{1}{4}\frac{R_N^2}{\lp 1+R_N^2\rp^2}\,\lp d\Psi_N+2A_{(N-1)} \rp^2 +
\frac{R_N^2}{ 1+R_N^2}\, d\Sigma_{N-1}^2\,, \nonumber \\
&&
 A_{(N)}=\frac{1}{2}\frac{R_N^2}{1+R_N^2}\lp d\Psi_N + 2
 A_{(N-1)} \rp\,,
\label{iterativeCPn}
\end{eqnarray}
in terms of the Fubini-Study metric, $d\Sigma_{N-1}^2$, and K\"ahler potential, $A_{(N-1)}$, on the unit $CP^{N-1}$,
\begin{eqnarray}
&& d\Sigma_{N-1}^2 = f_{N-1}^{-1}\, dv^i\, d\bar v^{\bar\imath}\, -
f_{N-1}^{-2}\, |\bar v^{\bar \imath}\, dv^i|^2\,, \nonumber \\
&& A_{(N-1)}= \frac{1}{2} \, \im\, f_{N-1}^{-1}\, \lp v^i\, d\bar
v^{\bar \imath} - \bar v^{\bar \imath}\, dv^i \rp\,, \qquad f_{N-1} = 1 +
v^i\, \bar v^{\bar \imath} \,. \label{iterativeCPn:2}
\end{eqnarray}

By definition, the K\"ahler form on $CP^{N}$, $J_{N}=\frac{1}{2}dA_{N}\,$, is covariantly conserved, $\hat{\nabla}_a J_{(N)}^{bc}=0\,$, and satisfies $J_a^{\:\:b}J_{bc}=-\hat{g}_{ac}\,$.

The lesson from this analysis is that starting from the $CP^1$ fields we can construct iteratively the $CP^N$ geometry as well as the complex coordinates $Z^A$ that define the embedding of $CP^N$ in $\mathbb{C}^{N+1}$. $CP^1$ is isomorphic to the 2-sphere $S^2$, its metric and K\"ahler potential being given by
\beq
d\Sigma_{1}^2=\frac{1}{4} \lp d\theta^2 +\sin^2\theta d\phi^2\rp \qquad \mathrm{and} \qquad
A_{(1)}=\frac{1}{2} \cos\theta \,d\phi\,.
\eeq
Examples of the embedding in $\mathbb{C}^{N+1}$ may be elucidative. For $CP^2$, parameterized by $(\theta,\phi,R_2,\Psi_2)$, the map is given by
\beq
\lp Z^0,Z^1,Z^2\rp= \frac{r  e^{i\tau}}{\sqrt{1+R_2^2}}\lp 1,R_2 \cos\frac{\theta}{2} \,e^{i\,\frac{1}{2}\,(\Psi_2+\phi)},
R_2 \sin\frac{\theta}{2} \,e^{i\,\frac{1}{2}\,(\Psi_2-\phi)} \rp,
\eeq
while for $CP^3$, parameterized by $(\theta,\phi,R_2,\Psi_2,R_3,\Psi_3)$, the map is
\beq \lp Z^0,Z^1,Z^2,Z^3\rp= \frac{r  e^{i\tau}}{\sqrt{1+R_3^2}} \lp 1,
\frac{R_3 e^{i\,\frac{1}{2}\,\Psi_3}}{\sqrt{1+R_2^2}} R_2\,\cos\frac{\theta}{2} \,e^{i\,\frac{1}{2}\,(\Psi_2+\phi)},
\frac{R_3 e^{i\,\frac{1}{2}\,\Psi_3}}{\sqrt{1+R_2^2}} R_2 \sin\frac{\theta}{2} \,e^{i\,\frac{1}{2}\,(\Psi_2-\phi)} \rp.
\eeq

In order to reproduce the results in Section~\ref{secscalars}, it is useful to recall that, for $CP^N$,
\be
\hat{R}_{abcd} = \hat{g}_{ac} \hat{g}_{bd} - \hat{g}_{ad} \hat{g}_{bc} + J_{ac} J_{bd} - J_{ad} J_{bc} + 2 J_{ab} J_{cd}\,.
\ee

\subsection{Scalar harmonics and Killing vectors \label{cpnscalarharmonicsappendix}}

We review here the systematic way to construct all scalar harmonics and Killing vector fields on $CP^N$ \cite{Hoxha:2000jf}. The isometry group of $CP^N$ is $SU(N+1)$. Let $T_{A_1\cdots A_p}{}^{B_1\cdots B_q}$ be a constant Hermitean $SU(N+1)$ tensor, which is symmetric in the index set $\{A_1,\ldots,A_p\}$ and the index set $\{B_1,\ldots,B_q\}$, and traceless in any contraction between an $A_i$ and a $B_i$ index. This defines the $(p,q)$ representation of $SU(N+1)$. The charged scalar harmonics are then given by
\beq
\mathbb{Y}=T_{A_1\cdots A_p}{}^{B_1\cdots B_q}\, Z^{A_1}\cdots Z^{A_p}\, \overline{Z}_{B_1}\cdots \overline{Z}_{B_q}\,,
\label{Harmonics}
\eeq
and satisfy the Laplacian \eqref{eqn:chargedS} for $\lambda=2[2pq+N(p+q)]$. We have $\kappa=\mathrm{max}\{p,q\}$ and $m=p-q$. Uncharged scalar harmonics have $\kappa=p=q$ and $\lambda=4\kappa(\kappa+N)$.

The Killing vectors on an Einstein-K\"ahler space can be constructed from the uncharged scalar harmonics with $\kappa=1\,$, which have eigenvalue $\lambda=4(1+N)$. Indeed all the Killing vectors $\xi_{(i)}$ of $CP^N$ are generated by the relation
\beq 
\xi^a_{(i)} = J_{(N)}^{ab}\, \partial_b \mathbb{Y}_{\kappa=1,(i)}^{m=0}\,\,,\label{Killing}
\eeq
where $i=1,\ldots,N(N+2)\,$. That is, setting to zero all but one of the constant components of the arbitrary Hermitean traceless tensor $T_A^{\:\:B}$ we get a $\kappa=1$, $m=0$ scalar harmonic on $CP^N$ through \eqref{Harmonics}. Repeating the exercise for all possible combinations, we generate the $N(N+2)$ uncharged scalar
harmonics $\mathbb{Y}_{\kappa=1,(i)}^{m=0}\,$, and the associated $(N+1)^2-1=N(N+2)$ Killing vectors through \eqref{Killing}. There are $N$ linearly independent Killing vectors which commute with all the others, thus generating the Cartan subgroup $U(1)^N$ of $SU(N+1)$.

\subsection{Symmetries of $\kappa=2$ harmonics}

\label{subsec:symmetries}

The symmetry group $SU(N+1)$ is broken, at least partially, by any linear perturbation $h_{ab}$ satisfying $\mathcal{L}_\xi h_{ab}\neq 0\,$, where $\xi$ is one of the $N(N+2)$ Killing vectors of $CP^N$. For scalar type perturbations, if $\mathcal{L}_\xi \Y \neq 0$ for some $\xi$ then the symmetry associated to $\xi$ is broken by the perturbation.

We explained above how to construct the Killing vectors from the uncharged $\kappa=1$ scalar harmonics. Consider now the most general linear combination of Killing vectors $K=\sum_i c_i\, \xi_{(i)}$, with $i=1,\ldots,N(N+2)\,$. The entire symmetry group $SU(N+1)$ is broken by $h_{ab}$ if the only solution to $\mathcal{L}_K \mathbb{Y} =0$ is  $c_i=0$ for all $i$, i.e. $K=0$. 
There are (uncharged) $\kappa=2$ harmonics on $CP^3$ ($D=9$) for which this is true. For reference, we present here a particular example in the coordinate system used above: a family of $m=0$, $\kappa=2$ harmonics with three non-zero continuous parameters, $\beta_1$, $\beta_2$ and $\beta_3$,
\beqa
& \displaystyle{ \Y_{\kappa=2}^{m=0} \;=\; \beta_1 \; \frac{R_3 R_2}{\left(1+R_3^2\right)^2} \sqrt{\frac{1+\cos\theta}{1+R_2^2}} \left[1-\frac{R_3^2 R_2^2 (1+\cos\theta)}{2 \left(1+R_2^2\right)}\right] e^{\frac{1}{2} \ii (\Psi_3 +\Psi_2+\phi)} } \qquad \qquad \nonumber \\
& \displaystyle{ + \beta_2 \; \frac{R_3^2 R_2 \sqrt{1-\cos\theta}}{\left(1+R_3^2\right)^2 \left(1+R_2^2\right)} \left[1-\frac{R_3^2}{2 \left(1+R_2^2\right)}\right] 
e^{\frac{1}{2} \ii (\phi-\Psi_2)} + \beta_3 \; \frac{ R_3^2  \;e^{i \Psi_3 }}{\left(1+R_3^2\right)^2 \left(1+R_2^2\right)} }\,.
\eeqa
If one of $\beta_1,\beta_2,\beta_3$ vanishes then this is still a $\kappa=2$ harmonic but it preserves some symmetry.

\section{Traceless-transverse gauge conditions \label{gaugeconditions}}

The tetrad components of the TT gauge conditions \eqref{eqn:TTgauge} are
\begin{equation}
\label{TTexplicit1}
 -f_{00}+f_{11}+f_{22}+2\,N\,H_L=0\,,
\end{equation}
and
\begin{subequations}
\label{TTexplicit2}
\begin{align}
&\, \frac{\ii}{f}(\omega-m\,\Omega)\,f_{00}+\frac{1}{g}\left(\frac{2\,f'}{f}+\frac{h'}{h}+\frac{2\,N}{r}+\frac{\partial}{\partial r}\right)f_{01}+\frac{\ii\,m}{h}\,f_{02}-\frac{h\,\Omega'}{f\,g}\,f_{12}\nonumber\\
&~~~+\frac{1}{2\,r\,\sqrt\lambda}\left[W^+(\lambda-2\,m\,N)+W^-(\lambda+2\,m\,N)\right]=0\,,\\
&\,\frac{\ii}{f}(\omega-m\,\Omega)\,f_{01}+\frac{f'}{f\,g}\,f_{00}-\frac{h\,\Omega'}{f\,g}\,f_{02}+\frac{1}{g}\left(\frac{f'}{f}+\frac{h'}{h}+\frac{2\,N}{r}+\frac{\partial}{\partial r}\right)f_{11}+\frac{\ii\,m}{h}\,f_{12}-\frac{h'}{g\,h}\,f_{22}\nonumber\\
&~~~-\frac{2\,N}{r\,g}\,H_L
+\frac{1}{2\,r\,\sqrt\lambda}\left[X^+(\lambda-2\,m\,N)+X^-(\lambda+2\,m\,N)\right]=0\,,\\
&\,\frac{\ii}{f}(\omega-m\,\Omega)\,f_{02}+\frac{1}{g}\left(\frac{f'}{f}+\frac{2\,h'}{h}+\frac{2\,N}{r}+\frac{\partial}{\partial r}\right)f_{12}+\frac{\ii\,m}{h}\,f_{22}\nonumber\\
&~~~+\frac{1}{2\,r\,\sqrt\lambda}\left[Z^+(\lambda-2\,m\,N)+Z^-(\lambda+2\,m\,N)\right]=0\,,\\
&\,\frac{\ii}{f}(\omega-m\,\Omega)\,W^+
+\frac{1}{g}\left(\frac{f'}{f}+\frac{h'}{h}+\frac{2\,N+1}{r}+\frac{\partial}{\partial r}\right)\,X^++\ii\left(\frac{m}{h}+\frac{2\,h}{r^2}\right)Z^+-\frac{\sqrt\lambda}{r}\,H_L\nonumber\\
&~~~+\frac{1}{2\,r\,\sqrt\lambda}
\left[\Big(\lambda-4(N+1)-2m(N+2)\Big)H^{++}
+\frac{(N-1)(\lambda+2\,m\,N)}{N}\,H^{+-}\right]=0\,,\\
&\,\frac{\ii}{f}(\omega-m\,\Omega)\,W^-
+\frac{1}{g}\left(\frac{f'}{f}+\frac{h'}{h}+\frac{2\,N+1}{r}+\frac{\partial}{\partial r}\right)\,X^-
+\ii\left(\frac{m}{h}-\frac{2\,h}{r^2}\right)Z^--\frac{\sqrt\lambda}{r}\,H_L\nonumber\\
&~~~+\frac{1}{2\,r\,\sqrt\lambda}\left[\Big(\lambda-4(N+1)+2m(N+2)\Big)H^{--}
+\frac{(N-1)(\lambda-2\,m\,N)}{N}\,H^{+-}\right]=0\,.
\end{align}
\end{subequations}

\section{Lichnerowicz eigenvalue equations \label{Lichnerowicz}}

In the following, we list the sixteen non-trivial components of the linearized equations \eqref{lichn},
\begin{equation*}
 (\Delta_L h)_{\mu\nu} \equiv -\nabla^\rho \nabla_\rho h_{\mu \nu} -2R_{\mu \rho \nu \sigma} h^{\rho  \sigma} = -k^2 h_{\mu\nu},
\end{equation*}
in the tetrad basis \eqref{vielbein}, given the ansatz \eqref{eqn:hmetric} for the metric perturbation. Notice that the TT gauge conditions have not been explicitly considered yet. The list of components is:

\begin{subequations}
\begin{align}
%
%
\label{lichne:f00}
&-\bigg[\frac{1}{g}\dr\frac{1}{g}\dr+\frac{1}{g^2}\left(\frac{f'}{f}+\frac{h'}{h}+\frac{2\,N}{r}\right)\dr-\frac{\lambda}{r^2}-\frac{m^2}{h^2}+\frac{(\omega-m\,\Omega)^2}{f^2}\nonumber\\
&\hspace{2cm}-\frac{2}{g^2}\bigg(\bigg(\frac{f'}{f}\bigg)^2-\frac{1}{2}\bigg(\frac{h\,\Omega'}{f}\bigg)^2\bigg)\bigg]f_{00}\nonumber\\
&\hspace{1cm}+2\,\ii\bigg[\frac{2f'(\omega-m\Omega)}{f^2g}+\frac{m\,\Omega'}{fg}\bigg]f_{01}+2\bigg[\frac{h\,\Omega'}{f\,g^2}\,\dr+\frac{1}{2\,g}\,\dr\left(\frac{h\,\Omega'}{f\,g}\right)+\frac{N\,h\,\Omega'}{fg^2r}\bigg]f_{02}\nonumber\\
&\hspace{1cm}+\bigg[\left(\frac{h\,\Omega'}{fg}\right)^2-\frac{2}{g}\dr\left(\frac{f'}{fg}\right)\bigg]f_{11}-\frac{2}{g^2}\bigg[\frac{f'h'}{f\,h}+\frac{1}{2}\left(\frac{h\,\Omega'}{f}\right)^2\bigg]f_{22}-\frac{4\,N\,f'}{fg^2r}\,H_L = - k^2\, f_{00}\,, \\ 
%
%
&-\bigg[\frac{1}{g}\frac{\partial}{\partial r}\frac{1}{g}\frac{\partial}{\partial r}+\frac{1}{g^2}\left(\frac{f'}{f}+\frac{h'}{h}+\frac{2\,N}{r}\right)\frac{\partial}{\partial r}-\frac{\lambda}{r^2}-\frac{m^2}{h^2}+\frac{(\omega-m\,\Omega)^2}{f^2}\nonumber\\
&\hspace{2cm}-\frac{1}{g^2}\bigg(2\bigg(\frac{f'}{f}\bigg)^2+\bigg(\frac{h'}{h}\bigg)^2+\frac{2\,N}{r^2}\bigg)+\frac{2}{g}\frac{\partial}{\partial r}\left(\frac{f'}{f\,g}\right)\bigg]f_{01}\nonumber\\
&\hspace{1cm}+\bigg[\frac{h\,\Omega'}{f\,g^2}\,\dr+\frac{3}{2\,g}\,\dr\left(\frac{h\,\Omega'}{f\,g}\right)+\frac{h\,\Omega'}{f\,g^2}\left(-\frac{f'}{f}+\frac{h'}{h}+\frac{N}{r}\right)\bigg]f_{12}\nonumber\\
&\hspace{1cm}+\ii\left[\frac{2f'(\omega-m\Omega)}{f^2g}+\frac{m\,\Omega'}{f\,g}\right](f_{00}+f_{11})
-\ii\left[\frac{h\,\Omega'(\omega-m\Omega)}{f^2g}-\frac{2\,m\,h'}{h^2\,g}\right]f_{02}\nonumber\\
&\hspace{1cm}+\frac{1}{r^2\,g\sqrt\lambda}\left[W^+(\lambda-2\,m\,N)+W^-(\lambda+2\,m\,N)\right] =  - k^2\, f_{01}\,, \\
%
%
&-\bigg[\frac{1}{g}\frac{\partial}{\partial r}\frac{1}{g}\frac{\partial}{\partial r}+\frac{1}{g^2}\left(\frac{f'}{f}+\frac{h'}{h}+\frac{2\,N}{r}\right)\frac{\partial}{\partial r}-\frac{\lambda}{r^2}-\frac{m^2}{h^2}+\frac{(\omega-m\,\Omega)^2}{f^2}\nonumber\\
&\hspace{2cm}-\frac{1}{g^2}\bigg(\bigg(\frac{f'}{f}-\frac{h'}{h}\bigg)^2-2\bigg(\frac{h\,\Omega'}{f}\bigg)^2\bigg)-\frac{2\,N\,h^2}{r^4}\bigg]f_{02}\nonumber\\
&\hspace{1cm}+\bigg[\frac{h\,\Omega'}{f\,g^2}\,\dr+\frac{1}{2\,g}\,\dr\left(\frac{h\,\Omega'}{f\,g}\right)+\frac{h\,\Omega'}{f\,g^2}\bigg(\frac{f'}{f}+\frac{h'}{h}+\frac{N}{r}\bigg)\bigg]f_{00}+\ii\bigg[\frac{h\,\Omega'(\omega-m\Omega)}{f^2g}-\frac{2\,m\,h'}{h^2g}\bigg]f_{01}\nonumber\\
&\hspace{1cm}-\bigg[\frac{1}{g}\dr\bigg(\frac{h\,\Omega'}{fg}\bigg)-\frac{h\,\Omega'}{f\,g^2}\bigg(\frac{f'}{f}-\frac{h'}{h}\bigg)\bigg]f_{11}+\ii\bigg[\frac{2f'(\omega-m\Omega)}{f^2g}+\frac{m\,\Omega'}{f\,g}\bigg]f_{12}\nonumber\\
&\hspace{1cm}+\bigg[\frac{h\,\Omega'}{f\,g^2}\,\dr+\frac{1}{2\,g}\,\dr\left(\frac{h\,\Omega'}{f\,g}\right)+\frac{N\,h\,\Omega'}{fg^2r}\bigg]f_{22}\nonumber\\
&\hspace{1cm}-\frac{\ii\,h}{r^3\sqrt\lambda}\left[W^+(\lambda-2\,m\,N)-W^-(\lambda+2\,m\,N)\right]-\frac{2\,N\,h\,\Omega'}{fg^2r}\,H_L = - k^2\, f_{02}\,, \\
%
%
&-\bigg[\frac{1}{g}\frac{\partial}{\partial r}\frac{1}{g}\frac{\partial}{\partial r}+\frac{1}{g^2}\left(\frac{f'}{f}+\frac{h'}{h}+\frac{2\,N}{r}\right)\frac{\partial}{\partial r}-\frac{\lambda-2(N+1)-2\,m}{r^2}-\frac{m^2}{h^2}+\frac{(\omega-m\,\Omega)^2}{f^2}\nonumber\\
&\hspace{2cm}-\frac{1}{g^2}\bigg(\bigg(\frac{f'}{f}-\frac{1}{r}\bigg)^2-\frac{1}{2}\bigg(\frac{h\,\Omega'}{f}\bigg)^2\bigg)-\frac{2\,h^2}{r^4}\bigg]W^+\nonumber\\
&\hspace{1cm}+\bigg[\frac{h\,\Omega'}{f\,g^2}\,\dr+\frac{1}{2\,g}\,\dr\left(\frac{h\,\Omega'}{f\,g}\right)+\frac{(N+1)\,h\,\Omega'}{fg^2r}\bigg]Z^+\nonumber\\
&\hspace{1cm}+\ii\bigg[\frac{2f'(\omega-m\Omega)}{f^2g}+\frac{m\,\Omega'}{f\,g}-\frac{2\,h^2\,\Omega'}{f\,g\,r^2}\bigg]X^++\frac{2\sqrt\lambda}{r^2\,g}\,f_{01}+\frac{\ii\,2\,h\,\sqrt{\lambda}}{r^3}\,f_{02} = - k^2\, W^+ \,, \\
%
%
&-\bigg[\frac{1}{g}\frac{\partial}{\partial r}\frac{1}{g}\frac{\partial}{\partial r}+\frac{1}{g^2}\left(\frac{f'}{f}+\frac{h'}{h}+\frac{2\,N}{r}\right)\frac{\partial}{\partial r}-\frac{\lambda-2(N+1)+2\,m}{r^2}-\frac{m^2}{h^2}+\frac{(\omega-m\,\Omega)^2}{f^2}\nonumber\\
&\hspace{2cm}-\frac{1}{g^2}\bigg(\bigg(\frac{f'}{f}-\frac{1}{r}\bigg)^2-\frac{1}{2}\bigg(\frac{h\,\Omega'}{f}\bigg)^2\bigg)-\frac{2\,h^2}{r^4}\bigg]W^-\nonumber\\
&\hspace{1cm}+\bigg[\frac{h\,\Omega'}{f\,g^2}\,\dr+\frac{1}{2\,g}\,\dr\left(\frac{h\,\Omega'}{f\,g}\right)+\frac{(N+1)\,h\,\Omega'}{fg^2r}\bigg]Z^-\nonumber\\
&\hspace{1cm}+\ii\bigg[\frac{2f'(\omega-m\Omega)}{f^2g}+\frac{m\,\Omega'}{f\,g}+\frac{2\,h^2\,\Omega'}{f\,g\,r^2}\bigg]X^-+\frac{2\sqrt\lambda}{r^2\,g}\,f_{01}-\frac{\ii\,2\,h\,\sqrt{\lambda}}{r^3}\,f_{02} = - k^2\, W^- \,, \\
%
%
&-\bigg[\frac{1}{g}\frac{\partial}{\partial r}\frac{1}{g}\frac{\partial}{\partial r}+\frac{1}{g^2}\left(\frac{f'}{f}+\frac{h'}{h}+\frac{2\,N}{r}\right)\frac{\partial}{\partial r}-\frac{\lambda}{r^2}-\frac{m^2}{h^2}+\frac{(\omega-m\,\Omega)^2}{f^2}\nonumber\\
&\hspace{2cm}-\frac{2}{g^2}\bigg(\bigg(\frac{f'}{f}\bigg)^2+\bigg(\frac{h'}{h}\bigg)^2-\frac{1}{2}\bigg(\frac{h\,\Omega'}{f}\bigg)^2+\frac{2\,N}{r^2}\bigg)\bigg]f_{11}\nonumber\\
&\hspace{1cm}-2\bigg[\frac{1}{g}\dr\bigg(\frac{f'}{f\,g}\bigg)-\frac{1}{2}\bigg(\frac{h\,\Omega'}{f\,g}\bigg)^2\bigg]f_{00}
+2\,\ii\bigg[\frac{2f'(\omega-m\Omega)}{f^2g}+\frac{m\,\Omega'}{f\,g}\bigg]f_{01}\nonumber\\
&\hspace{1cm}+2\bigg[\frac{1}{g}\,\dr\left(\frac{h\,\Omega'}{f\,g}\right)-\frac{h\,\Omega'}{f\,g^2}\bigg(\frac{f'}{f}-\frac{h'}{h}\bigg)\bigg]f_{02}-2\,\ii\bigg[\frac{h\,\Omega'(\omega-m\Omega)}{f^2g}-\frac{2\,m\,h'}{h^2g}\bigg]f_{12}\nonumber\\
&\hspace{1cm}+2\bigg[\frac{1}{g}\,\dr\left(\frac{h'}{h\,g}\right)+\frac{1}{2}\bigg(\frac{h\,\Omega'}{f\,g}\bigg)^2\bigg]f_{22}
+\frac{2}{r^2\,g\,\sqrt\lambda}\left[X^+(\lambda-2\,m\,N)+X^-(\lambda+2\,m\,N)\right]\nonumber\\
&\hspace{1cm}-\frac{4\,N}{g^2\,r}\bigg(\frac{g'}{g}+\frac{1}{r}\bigg)H_L = - k^2\, f_{11} \,, \\
%
%
&-\bigg[\frac{1}{g}\frac{\partial}{\partial r}\frac{1}{g}\frac{\partial}{\partial r}+\frac{1}{g^2}\left(\frac{f'}{f}+\frac{h'}{h}+\frac{2\,N}{r}\right)\frac{\partial}{\partial r}-\frac{\lambda}{r^2}-\frac{m^2}{h^2}+\frac{(\omega-m\,\Omega)^2}{f^2}\nonumber\\
&\hspace{2cm}-\frac{1}{g^2}\bigg(\bigg(\frac{f'}{f}\bigg)^2+2\bigg(\frac{h'}{h}\bigg)^2-2\bigg(\frac{h\,\Omega'}{f}\bigg)^2+\frac{2\,N}{r^2}\bigg)-\frac{2\,N\,h^2}{r^4}+\frac{2}{g}\dr\bigg(\frac{h'}{h\,g}\bigg)\bigg]f_{12}\nonumber\\
&\hspace{1cm}+\bigg[\frac{h\,\Omega'}{f\,g^2}\,\dr-\frac{1}{2\,g}\,\dr\left(\frac{h\,\Omega'}{f\,g}\right)+\frac{h\,\Omega'}{f\,g^2}\bigg(\frac{2\,f'}{f}+\frac{N}{r}\bigg)\bigg]f_{01}\nonumber\\
&\hspace{1cm}+\ii\bigg[\frac{2f'(\omega-m\Omega)}{f^2g}+\frac{m\,\Omega'}{f\,g}\bigg]f_{02}+\ii\bigg[\frac{h\,\Omega'(\omega-m\Omega)}{f^2g}-\frac{2\,m\,h'}{h^2g}\bigg](f_{11}-f_{22})\nonumber\\
&\hspace{1cm}-\frac{\ii\,h}{r^3\sqrt\lambda}\left[X^+(\lambda-2\,m\,N)-X^-(\lambda+2\,m\,N)\right]
+\frac{1}{r^2\,g\,\sqrt\lambda}\left[Z^+(\lambda-2\,m\,N)+Z^-(\lambda+2\,m\,N)\right] \nonumber\\
&\hspace{1cm} = - k^2\, f_{12} \,, \\
%
%
&-\bigg[\frac{1}{g}\frac{\partial}{\partial r}\frac{1}{g}\frac{\partial}{\partial r}+\frac{1}{g^2}\left(\frac{f'}{f}+\frac{h'}{h}+\frac{2\,N}{r}\right)\frac{\partial}{\partial r}-\frac{\lambda-2(N+1)-2\,m}{r^2}-\frac{m^2}{h^2}+\frac{(\omega-m\,\Omega)^2}{f^2}\nonumber\\
&\hspace{2cm}-\frac{1}{g^2}\bigg(\bigg(\frac{f'}{f}\bigg)^2+\bigg(\frac{h'}{h}\bigg)^2-\frac{1}{2}\bigg(\frac{h\,\Omega'}{f}\bigg)^2+\frac{2\,N+3}{r^2}\bigg)-\frac{2\,h^2}{r^4}-\frac{2\,g'}{g^3\,r}\bigg]X^+\nonumber\\
&\hspace{1cm}+\ii\bigg[\frac{2f'(\omega-m\Omega)}{f^2g}+\frac{h\,\Omega'}{f\,g}\bigg(\frac{m}{h}-\frac{2\,h}{r^2}\bigg)\bigg]W^+
+\ii\bigg[-\frac{h\,\Omega'(\omega-m\Omega)}{f^2g}+\frac{2\,m\,h'}{h^2\,g}-\frac{4\,h}{r^2\,g}\bigg(\frac{h'}{h}-\frac{2}{r}\bigg)\bigg]Z^+\nonumber\\
&\hspace{1cm}+\frac{1}{r^2\,g\,\sqrt\lambda}\bigg[\big(\lambda-4(N+1)-2\,m(N+2)\big)H^{++}+\frac{(N-1)(\lambda+2\,m\,N)}{N}\,H^{+-}-2\,\lambda\,H_L\bigg]\nonumber\\
&\hspace{1cm}+\frac{2\,\sqrt\lambda}{r^2\,g}\,f_{11}+\frac{\ii\,2\,h\,\sqrt\lambda}{r^3}\,f_{12} = - k^2\, X^+ \,, \\
%
%
&-\bigg[\frac{1}{g}\frac{\partial}{\partial r}\frac{1}{g}\frac{\partial}{\partial r}+\frac{1}{g^2}\left(\frac{f'}{f}+\frac{h'}{h}+\frac{2\,N}{r}\right)\frac{\partial}{\partial r}-\frac{\lambda-2(N+1)+2\,m}{r^2}-\frac{m^2}{h^2}+\frac{(\omega-m\,\Omega)^2}{f^2}\nonumber\\
&\hspace{2cm}-\frac{1}{g^2}\bigg(\bigg(\frac{f'}{f}\bigg)^2+\bigg(\frac{h'}{h}\bigg)^2-\frac{1}{2}\bigg(\frac{h\,\Omega'}{f}\bigg)^2+\frac{2\,N+3}{r^2}\bigg)-\frac{2\,h^2}{r^4}-\frac{2\,g'}{g^3\,r}\bigg]X^-\nonumber\\
&\hspace{1cm}+\ii\bigg[\frac{2f'(\omega-m\Omega)}{f^2g}+\frac{h\,\Omega'}{f\,g}\bigg(\frac{m}{h}+\frac{2\,h}{r^2}\bigg)\bigg]W^-
+\ii\bigg[-\frac{h\,\Omega'(\omega-m\Omega)}{f^2g}+\frac{2\,m\,h'}{h^2\,g}+\frac{4\,h}{r^2\,g}\bigg(\frac{h'}{h}-\frac{2}{r}\bigg)\bigg]Z^-\nonumber\\
&\hspace{1cm}+\frac{1}{r^2\,g\,\sqrt\lambda}\bigg[\big(\lambda-4(N+1)+2\,m(N+2)\big)H^{--}+\frac{(N-1)(\lambda-2\,m\,N)}{N}\,H^{+-}-2\,\lambda\,H_L\bigg]\nonumber\\
&\hspace{1cm}+\frac{2\,\sqrt\lambda}{r^2\,g}\,f_{11}-\frac{\ii\,2\,h\,\sqrt\lambda}{r^3}\,f_{12} = - k^2\, X^- \,, \\
%
%
&-\bigg[\frac{1}{g}\frac{\partial}{\partial r}\frac{1}{g}\frac{\partial}{\partial r}+\frac{1}{g^2}\left(\frac{f'}{f}+\frac{h'}{h}+\frac{2\,N}{r}\right)\frac{\partial}{\partial r}-\frac{\lambda}{r^2}-\frac{m^2}{h^2}+\frac{(\omega-m\,\Omega)^2}{f^2}\nonumber\\
&\hspace{2cm}-\frac{2}{g^2}\bigg(\bigg(\frac{h'}{h}\bigg)^2-\frac{1}{2}\bigg(\frac{h\,\Omega'}{f}\bigg)^2\bigg)-\frac{4\,N\,h^2}{r^4}\bigg]f_{22}\nonumber\\
&\hspace{1cm}-\frac{2}{g^2}\bigg[\frac{f'\,h'}{f\,h}+\frac{1}{2}\left(\frac{h\,\Omega'}{f}\right)^2\bigg]f_{00}
+2\bigg[\frac{h\,\Omega'}{f\,g^2}\,\dr+\frac{1}{2\,g}\,\dr\left(\frac{h\,\Omega'}{f\,g}\right)+\frac{h\,\Omega'}{fg^2}\bigg(\frac{f'}{f}+\frac{h'}{h}+\frac{N}{r}\bigg)\bigg]f_{02}\nonumber\\
&\hspace{1cm}+\bigg[\frac{2}{g}\,\dr\left(\frac{h'}{h\,g}\right)+\bigg(\frac{h\,\Omega'}{f\,g}\bigg)^2\bigg]f_{11}
+2\,\ii\bigg[\frac{h\,\Omega'(\omega-m\Omega)}{f^2g}-\frac{2\,m\,h'}{h^2\,g}\bigg]f_{12}\nonumber\\
&\hspace{1cm}-\frac{\ii\,2\,h}{r^3\sqrt\lambda}\left[Z^+(\lambda-2\,m\,N)-Z^-(\lambda+2\,m\,N)\right]
+4\,N\bigg(\frac{h'}{h\,g^2\,r}-\frac{2\,h^2}{r^4}\bigg)H_L = - k^2\, f_{22} \,, \\
%
%
&-\bigg[\frac{1}{g}\frac{\partial}{\partial r}\frac{1}{g}\frac{\partial}{\partial r}+\frac{1}{g^2}\left(\frac{f'}{f}+\frac{h'}{h}+\frac{2\,N}{r}\right)\frac{\partial}{\partial r}-\frac{\lambda-2(N+1)-2\,m}{r^2}-\frac{m^2}{h^2}+\frac{(\omega-m\,\Omega)^2}{f^2}\nonumber\\
&\hspace{2cm}-\frac{1}{g^2}\bigg(\bigg(\frac{h'}{h}-\frac{1}{r}\bigg)^2-\frac{1}{2}\bigg(\frac{h\,\Omega'}{f}\bigg)^2\bigg)-\frac{2(N+3)\,h^2}{r^4}\bigg]Z^+\nonumber\\
&\hspace{1cm}+\bigg[\frac{h\,\Omega'}{f\,g^2}\,\dr+\frac{1}{2\,g}\,\dr\left(\frac{h\,\Omega'}{f\,g}\right)+\frac{h\,\Omega'}{fg^2}\bigg(\frac{f'}{f}+\frac{h'}{h}+\frac{N-1}{r}\bigg)\bigg]W^+\nonumber\\
&\hspace{1cm}+\ii\bigg[\frac{h\,\Omega'(\omega-m\Omega)}{f^2g}-\frac{2\,m\,h'}{h^2\,g}+\frac{4\,h}{r^2g}\bigg(\frac{h'}{h}-\frac{2}{r}\bigg)\bigg]X^+\nonumber\\
&\hspace{1cm}-\frac{\ii\,h}{r^3\sqrt\lambda}\bigg[\big(\lambda-4(N+1)-2\,m(N+2)\big)H^{++}-\frac{(N-1)(\lambda+2\,m\,N)}{N}\,H^{+-}+2\,\lambda\,H_L\bigg]\nonumber\\
&\hspace{1cm}+\frac{2\,\sqrt\lambda}{r^2\,g}\,f_{12}+\frac{\ii\,2\,h\,\sqrt\lambda}{r^3}\,f_{22}= - k^2\, Z^+ \,, \\
%
%
&-\bigg[\frac{1}{g}\frac{\partial}{\partial r}\frac{1}{g}\frac{\partial}{\partial r}+\frac{1}{g^2}\left(\frac{f'}{f}+\frac{h'}{h}+\frac{2\,N}{r}\right)\frac{\partial}{\partial r}-\frac{\lambda-2(N+1)+2\,m}{r^2}-\frac{m^2}{h^2}+\frac{(\omega-m\,\Omega)^2}{f^2}\nonumber\\
&\hspace{2cm}-\frac{1}{g^2}\bigg(\bigg(\frac{h'}{h}-\frac{1}{r}\bigg)^2-\frac{1}{2}\bigg(\frac{h\,\Omega'}{f}\bigg)^2\bigg)-\frac{2(N+3)\,h^2}{r^4}\bigg]Z^-\nonumber\\
&\hspace{1cm}+\bigg[\frac{h\,\Omega'}{f\,g^2}\,\dr+\frac{1}{2\,g}\,\dr\left(\frac{h\,\Omega'}{f\,g}\right)+\frac{h\,\Omega'}{fg^2}\bigg(\frac{f'}{f}+\frac{h'}{h}+\frac{N-1}{r}\bigg)\bigg]W^-\nonumber\\
&\hspace{1cm}+\ii\bigg[\frac{h\,\Omega'(\omega-m\Omega)}{f^2g}-\frac{2\,m\,h'}{h^2\,g}-\frac{4\,h}{r^2g}\bigg(\frac{h'}{h}-\frac{2}{r}\bigg)\bigg]X^-\nonumber\\
&\hspace{1cm}+\frac{\ii\,h}{r^3\sqrt\lambda}\bigg[\big(\lambda-4(N+1)+2\,m(N+2)\big)H^{--}-\frac{(N-1)(\lambda-2\,m\,N)}{N}\,H^{+-}+2\,\lambda\,H_L\bigg]\nonumber\\
&\hspace{1cm}+\frac{2\,\sqrt\lambda}{r^2\,g}\,f_{12}-\frac{\ii\,2\,h\,\sqrt\lambda}{r^3}\,f_{22}= - k^2\, Z^- \,, \\
%
%
&\frac{1}{\sqrt\lambda}\bigg[\frac{1}{g}\frac{\partial}{\partial r}\frac{1}{g}\frac{\partial}{\partial r}+\frac{1}{g^2}\left(\frac{f'}{f}+\frac{h'}{h}+\frac{2\,N}{r}\right)\frac{\partial}{\partial r}-\frac{\lambda-4(N+1+m)}{r^2}-\frac{m^2}{h^2}+\frac{(\omega-m\,\Omega)^2}{f^2}\bigg]H^{++}
\nonumber\\
&\hspace{1cm}-\frac{4}{r^2\,g}\,X^+-\frac{\ii\,4\,h}{r^3}\,Z^+ = k^2\,\frac{1}{\sqrt\lambda} \, H^{++} \,, \\
%
%
&\frac{1}{\sqrt\lambda}\bigg[\frac{1}{g}\frac{\partial}{\partial r}\frac{1}{g}\frac{\partial}{\partial r}+\frac{1}{g^2}\left(\frac{f'}{f}+\frac{h'}{h}+\frac{2\,N}{r}\right)\frac{\partial}{\partial r}-\frac{\lambda-4(N+1-m)}{r^2}-\frac{m^2}{h^2}+\frac{(\omega-m\,\Omega)^2}{f^2}\bigg]H^{--}
\nonumber\\
&\hspace{1cm}-\frac{4}{r^2\,g}\,X^-+\frac{\ii\,4\,h}{r^3}\,Z^- = k^2\,\frac{1}{\sqrt\lambda} \, H^{--} \,, \\
%
%
&\frac{1}{\sqrt\lambda}\bigg[\frac{1}{g}\frac{\partial}{\partial r}\frac{1}{g}\frac{\partial}{\partial r}+\frac{1}{g^2}\left(\frac{f'}{f}+\frac{h'}{h}+\frac{2\,N}{r}\right)\frac{\partial}{\partial r}-\frac{\lambda-4\,N}{r^2}-\frac{m^2}{h^2}+\frac{(\omega-m\,\Omega)^2}{f^2}+\frac{4}{r^2}\bigg(1-\frac{2\,h^2}{r^2}\bigg)\bigg]H^{+-}\nonumber\\
&\hspace{1cm}-\frac{2}{r^2\,g}\,(X^++X^-)
+\frac{2\,\ii\,h}{r^3}\,(Z^+-Z^-)= k^2\,\frac{1}{\sqrt\lambda} \, H^{+-} \,, \\
%
%
&-\bigg[\frac{1}{g}\frac{\partial}{\partial r}\frac{1}{g}\frac{\partial}{\partial r}+\frac{1}{g^2}\left(\frac{f'}{f}+\frac{h'}{h}+\frac{2\,N}{r}\right)\frac{\partial}{\partial r}-\frac{\lambda-4(N+1)}{r^2}-\frac{m^2}{h^2}+\frac{(\omega-m\,\Omega)^2}{f^2}
-\frac{4\,N}{g^2\,r^2}-\frac{8\,h^2}{r^4}\bigg]H_L\nonumber\\
&\hspace{1cm}-2\bigg[\frac{f'}{f\,g^2\,r}\,f_{00}+\frac{1}{g^2\,r}\bigg(\frac{g'}{g}+\frac{1}{r}\bigg)f_{11}-\frac{h\,\Omega'}{f\,g^2\,r}\,f_{02}-\bigg(\frac{h'}{h\,g^2\,r}-\frac{2\,h^2}{r^4}\bigg)f_{22}\bigg]\nonumber\\
&\hspace{1cm}-\frac{1}{r^2\,g\,N\,\sqrt\lambda}\left[X^+(\lambda-2\,m\,N)+X^-(\lambda+2\,m\,N)\right]\nonumber\\
&\hspace{1cm}+\frac{\ii\,h}{r^3\,N\,\sqrt\lambda}\left[Z^+(\lambda-2\,m\,N)-Z^-(\lambda+2\,m\,N)\right]= - k^2\, H_L \,.
\end{align}
\end{subequations}

 \section{Properties of stationary axisymmetric modes   \label{sec:stationaryModes}}

In the main body of the paper, we have presented our numerical
results for general axisymmetric  time dependent scalar
perturbations. Our numerical code has a continuous limit as
$\omega=\ii\, \Gamma \rightarrow 0$. Therefore this particular case
was already included in our discussion. However, in this appendix we
want to have a closer look into stationary axisymmetric modes.
The reasons are: (i) we developed an independent code
for this particular case which confirms the results from
our general code with time dependence; (ii) we proved that the
stationary zero-modes (with $\omega=0$ and $k=k_c=0$) are not pure gauge
modes; (iii) we confirmed that our stationary perturbations
preserve the angular velocity and temperature of the background
geometry; (iv) we found which stationary zero-mode perturbations
can change the mass and angular momenta of the background solution;
finally (v) we want to give special attention to the stationary modes,
and not just to the time dependent instability, since these may
indicate bifurcation points to new branches of black hole solutions.

 \subsection{The sub-sector of stationary perturbations  \label{sec:statModes2}}

As described in Section~\ref{sec:eigenvalue}, the stability problem
of axisymmetric  perturbations with time dependence consists of a
system of 16 Lichnerowicz eigenvalue equations for 16 unknown
functions. Choosing the TT gauge reduces the problem to a system of
6 TT gauge conditions and 10 Lichnerowicz equations. The procedure
is consistent because, as explained in that Section, the latter 10
equations automatically imply, through the 6 gauge conditions, that
the other 6 Lichnerowicz eigenvalue equations are satisfied.

When we consider the axisymmetric stationary sub-sector of the perturbations, i.e. $m=0$, $\omega=\ii\, \Gamma=0$ we find
that the initial system of  16 Lichnerowicz equations decouples into
a subsystem of 10 equations for 10 functions and another subsystem
with 6 equations involving only the remaining 6 perturbation
functions. Moreover, we can use the stationarity condition, $\omega=\ii\, \Gamma=0$, to further simplify our system of equations. To see how this is accomplished, let us introduce
the harmonics associated with the time and azimuthal Killing
directions as $\mathbb{S}=e^{-i\omega t}e^{im \psi}$. We can then
decompose the perturbations according to how they transform under the $\{t,\psi\}$ Killing
isometries. For example, the scalar-derived vector perturbations satisfy
$h_{A\bar{b}}\sim f_A \,\partial_b\mathbb{S}$ for $b=t,\psi$ and the index $A$
running over the radial and $CP^N$ coordinates (the bar in $\bar{b}$
denotes that the 1-form basis is really $\{dt,d\psi+A-\Omega dt\}$).
These perturbations, $h_{At}$ and $h_{A \bar{\psi}}$, must
then vanish when we set $\omega=0$ and $m=0$. In our dictionary,
this amounts to say that we expect the following: $\mathfrak{f}_{01}=
\mathfrak{f}_{12}=\mathfrak{f}_0=\mathfrak{f}_2=0$. The original 10
time dependent Lichnerowicz equations then imply that
$\tilde{\mathfrak{f}}_1=0$ when we set $\omega=0$. Similarly, the
original 6 TT gauge conditions imply that $Q=0$. We are then led to the following conditions
\beq
\mathfrak{f}_{01}=\mathfrak{f}_{12}=\mathfrak{f}_0=\mathfrak{f}_2=\tilde{\mathfrak{f}}_1=Q=0\,,
\label{killed}
\eeq
or, using the map \eqref{eq:newfunc}: $f_{01}=f_{12}=0$, $W^-=-W^+$,
$X^-=X^+$, $Z^-=-Z^+$,  and $H^{--}=H^{++}$.

With (\ref{killed}) the original system reduces to a closed system
of 3 TT gauge conditions and 7 Lichnerowicz equations. In this
axisymmetric stationary case, the boundary conditions at the horizon
(\ref{bchor}) reduce to
\begin{equation}
\label{BCsStationary}
\begin{aligned}
& f_{00}= -\frac{h(r_+)}{r_+} \, f_{\bar{0}\bar{1}}\,,
\qquad  
f_{00}+f_{11}\approx 
\Delta'(r_+) \, f_{\bar{1}\bar{1}}\; (r-r_+)\,, \\
& f_{02} \approx -\frac{\sqrt{\Delta'(r_+)}}{h(r_+)}  \,
f_{\bar{1}\bar{2}} \sqrt{r-r_+}\,, \qquad 
f_{22}= \frac{1}{h(r_+)^2} \, f_{\bar{2}\bar{2}}\,, \qquad Z^{\pm}= \frac{1}{r_+ h(r_+)} \, \bar{Z}^{\pm} \,,  \\
&W^\pm \approx \frac{h(r_+)}{r_+\sqrt{\Delta'(r_+)}}
\,\bar{W}'^{\pm}\sqrt{r-r_+}\,,\qquad X^\pm - W^\pm \approx
\frac{\sqrt{\Delta'(r_+)}}{r_+} \, \bar{X}^{\pm} \sqrt{r-r_+}\,,
\end{aligned}
\end{equation}
where we used $\bar{W}^{\pm}=\bar{W'}^{\pm}(r-r_+)$ and one further
has $\bar{W'}^{-}=-\bar{W'}^{+}$, $\bar{X}^{-}=\bar{X}^{+}$ and
$\bar{Z}^{-}=-\bar{Z}^{+}$. It is important to emphasize that
(\ref{killed}) already encode the information that the perturbations
are in the TT gauge, and that the boundary conditions
(\ref{BCsStationary}) are compatible with the TT gauge.

We have done an explicit search of the stationary modes using only
the subsystem of 3 TT gauge conditions and 7 Lichnerowicz equations
described above subject to (\ref{BCsStationary}). We recover
independently the same results we obtain when we set $\omega=0$ in
our time dependent code.

 \subsection{Stationary zero-modes are not pure gauge  \label{sec:NoPureGauge}}

As discussed in Section~\ref{sec:eigenvalue}, perturbations with
$k>0$ in the TT gauge have  all the gauge freedom fixed. However, TT
perturbations with $k=0 \neq k_c$ and $\Gamma=0$ are pure gauge modes \cite{Gregory:1994bj}.
In this subsection, we want to confirm that our stationary
axisymmetric zero-modes with $k=k_c=0$ cannot be pure gauge modes. Given that for any residual gauge
freedom the gauge parameter would be constrained to be a harmonic 1-form, we will
prove that there is no regular harmonic 1-form that could generate our perturbations.

Consider the effect of a scalar gauge transformation on the metric perturbations. The most general scalar-type gauge parameter can be decomposed as
\begin{equation}
 \xi=e^{-\ii\omega t+\ii m\psi}\big[\xi_0(r)\Y\,e^{(0)}+\xi_{1}(r)\Y\,e^{(1)}+\xi_{2}(r)\Y\,e^{(2)}+ r \big(\xi^+(r)
 \Y^+_a +\xi^-(r)\Y^-_a \big) dx^a\big]\,.
\end{equation}
Under a gauge transformation,
\begin{equation}
 h_{\mu\nu}\to h_{\mu\nu}+2\,\nabla_{(\mu}\xi_{\nu)}\,,
\end{equation}
the tetrad components of the metric perturbations transform as
\begin{equation}
\begin{aligned}
&f_{00}\to
f_{00}-2\bigg[\frac{\ii(\omega-m\,\Omega)}{f}\,\xi_0+\frac{f'}{f\,g}\,\xi_1\bigg]\,,
\qquad
f_{01} \to f_{01}+\frac{1}{g}\bigg(\dr-\frac{f'}{f}\bigg)\xi_0-\frac{\ii(\omega-m\,\Omega)}{f}\,\xi_1\,,\\
&f_{02}\to
f_{02}+\frac{\ii\,m}{h}\,\xi_0-\frac{h\,\Omega'}{f\,g}\,\xi_1-\frac{\ii(\omega-m\,\Omega)}{f}\,\xi_2\,,
\qquad
f_{11}\to f_{11}+\frac{2}{g}\,\dr\,\xi_1\,,\\
&f_{12}\to
f_{12}-\frac{h\,\Omega'}{f\,g}\,\xi_0+\frac{\ii\,m}{h}\,\xi_1+\frac{1}{g}\bigg(\dr-\frac{h'}{h}\bigg)\xi_2\,,\qquad
f_{22}\to f_{22}+2\bigg[\frac{h'}{h\,g}\,\xi_1+\frac{\ii\,m}{h}\,\xi_2\bigg]\,,\\
&W^+\to
W^+-\bigg[\frac{\sqrt\lambda}{r}\,\xi_0+\frac{\ii(\omega-m\,\Omega)}{f}\,\xi^+\bigg]\,,\quad
W^-\to
W^--\bigg[\frac{\sqrt\lambda}{r}\,\xi_0+\frac{\ii(\omega-m\,\Omega)}{f}\,\xi^-\bigg]\,,\\
&X^+\to X^+-\bigg[\frac{\sqrt\lambda}{r}\,\xi_1-\frac{1}{g}\bigg(\dr-\frac{1}{r}\bigg)\xi^+\bigg]\,,\quad X^-\to X^--\bigg[\frac{\sqrt\lambda}{r}\,\xi_1-\frac{1}{g}\bigg(\dr-\frac{1}{r}\bigg)\xi^-\bigg]\,,\\
&Z^+\to
Z^+-\bigg[\frac{\sqrt\lambda}{r}\,\xi_2-\ii\bigg(\frac{m}{h}+\frac{2\,h}{r^2}\bigg)\xi^+\bigg]\,,\quad
Z^-\to Z^--\bigg[\frac{\sqrt\lambda}{r}\,\xi_2-\ii\bigg(\frac{m}{h}-\frac{2\,h}{r^2}\bigg)\xi^-\bigg]\,,\\
&H_L\to H_L+\frac{2}{r}\bigg[\frac{1}{g}\,\xi_1+\frac{1}{4\,N\,\sqrt\lambda}\big(\xi^+(\lambda-2\,m\,N)+\xi^-(\lambda+2\,m\,N)\big)\bigg]\,,\\
&H^{+-}\to H^{+-}-\frac{\sqrt\lambda}{r}(\xi^++\xi^-)\,,\quad
H^{++}\to H^{++}-\frac{2\,\sqrt\lambda}{r}\,\xi^+\,,\quad H^{--}\to
H^{--}-\frac{2\,\sqrt\lambda}{r}\,\xi^-\,.
\end{aligned}
\end{equation}
Our stationary axisymmetric perturbations must satisfy (\ref{killed}) which requires that a potentially dangerous gauge parameter $\xi$ must satisfy
\begin{equation}
 \xi_0(r)=\xi_{2}(r)=0\,, \quad\hbox{and} \qquad
 \xi^-(r)=\xi^+(r)\,. \label{KillComp}
\end{equation}
We now prove that a parameter $\xi$ obeying these conditions cannot generate a pure gauge metric perturbation that is regular. By regularity we mean that the gauge transformation cannot diverge at the horizon $r=r_+$ nor at the asymptotic boundary $r \to \infty$.

A TT gauge perturbation generated by $\xi$ must satisfy the conditions $\nabla_\mu\xi^\mu=0$ and $\Box\xi_\nu=0$. If we introduce the antisymmetric tensor $F_{\mu\nu}=\nabla_{[\mu}\xi_{\nu]}$, for a Ricci flat background, these conditions reduce to $\nabla_\mu\xi^\mu=0$ and $\nabla_\mu F^{\mu\nu}=0$ which read simply
  \beq
 \partial_\mu\lp \sqrt{-g}\xi^\mu \rp=0\,,\qquad \partial_\mu\lp \sqrt{-g}F^{\mu\nu}
 \rp=0\,. \label{NoPureGaugeEqs}
 \eeq
Using $\sqrt{-g}=r^{2N+1}\sqrt{\hat{g}}$ and Eq.~(\ref{eqn:chargedS}), the first of the equations above requires that
 \beq
 \xi^+(r)=\frac{1}{\lambda r^{2N-1}}\,\partial_r\left[ r^{2N+1} g(r)^{-2}\xi_1(r)
 \right]\,,
 \eeq
where the background function $g(r)$ is defined in (\ref{background}) and $\lambda$ is the $CP^N$ eigenvalue (\ref{eqn:eigenvalues}). The second family of equations in (\ref{NoPureGaugeEqs}) further demands that $\left(\xi^+\right)^{\prime}(r)=\xi_1(r)$. Introducing the new
variable
 \beq
 \xi_1(r)= r^{-(N+3/2)g^2(r)}\,\chi(r)\,,
 \eeq
the solution of (\ref{NoPureGaugeEqs}) must then solve
 \beq
 \chi^{\prime\prime}(r)=V(r)\chi(r)\,,  \qquad \hbox{with}\quad V(r)=\frac{1}{r^2}\left[\lp N^2-\frac{1}{4}\rp+\lambda
 g^2(r)\right]>0\,. \label{Qeq}
 \eeq

To study the regularity of the associated gauge transformation, we need the asymptotic behavior of $\chi(r)$ at the horizon and at infinity. The solution to Eq.~\eqref{Qeq} can be obtained in these regions by considering the dominant contributions of $V(r)$ or by doing a Frobenius analysis. Using (\ref{eqn:eigenvalues}), we find that
\begin{eqnarray}
&&\chi(r){\bigl|}_h\sim a_0(r-r_+)\,\quad \mathrm{or} \quad \chi(r){\bigl|}_h\sim a_0 \,, \label{BC:Q1} \\
&& \chi(r){\bigl|}_{\infty}\sim b_0 r^{\frac{1}{2}\pm
(2\kappa+N)}\,.
 \label{BC:Q2}
\end{eqnarray}
Recall that $\delta h_{\mu\nu}=-\mathcal{L}_\xi g_{\mu\nu}$. We have
to  discard the second possibility in \eqref{BC:Q1} because it would
generate a dependence $f_{11}{\big|}_h \sim (r-r_+)^{-2}$, not
compatible with the TT boundary conditions \eqref{BCsStationary}. On
the other hand, we have to discard the solution with the positive
sign in \eqref{BC:Q2} because it generates a perturbation that grows
faster than the unperturbed background metric as $r\to 0$. The
appropriate boundary conditions for \eqref{Qeq} are thus
 \beq
 \chi(r){\bigl|}_h\sim a_0(r-r_+)\,,\qquad \chi(r){\bigl|}_{\infty}\sim b_0 r^{\frac{1}{2}- (2\kappa+N)}
 \,. \label{BC:Qfinal}
 \eeq

We can now complete our proof. Notice that
\begin{eqnarray}
 0\leq
\int_{r_+}^{\infty}\chi^{\prime}(r)^2&=&\chi(r)\chi^{\prime}(r){\biggl|}_{r_+}^{\infty}-\int_{r_+}^{\infty}\chi(r)\chi^{\prime\prime}(r)
\nonumber\\
 &=& -\int_{r_+}^{\infty}V(r)\chi(r)^2 \leq 0 \,, \label{nullQ}
\end{eqnarray}
where we used \eqref{BC:Q2} and \eqref{Qeq}. But these relations can be satisfied only for \beq
 \chi(r)=0 \qquad \Rightarrow \qquad \xi_1(r)=0\,, \quad \hbox{and}
 \quad \xi^+(r)=0 \,, \label{BC:Qend}
 \eeq
which in addition to (\ref{KillComp}) proves that our regular zero
mode perturbations in the TT gauge cannot be pure gauge modes.

 \subsection{Stationary modes preserve the temperature and the angular velocities of the solution  \label{sec:PreserveTAngVel}}

In Section~\ref{sec:string}, we discussed the connection between the
classical instability of the black hole and its thermodynamics. This
contact was built on the claim that the stationary and axisymmetric
modes that we study preserve the temperature and the
angular velocities of the background solution. Here we will prove
that this is indeed the case.

We will first compute the angular velocity and the temperature of
the unperturbed background solution using standard Euclidean
methods. Our strategy then is to check that our TT metric
perturbation $h_{\mu\nu} dx^\mu dx^\nu$ is a regular symmetric 2-tensor, when expressed
in coordinates where the background metric is regular, which confirms
that they preserve the angular velocity and temperature.

We start with the computation of the background angular velocity and
temperature. This is done performing the standard three steps in the
background solution: i) a coordinate transformation to coordinates
$(t,\tilde{\psi})$ that corotate with the black horizon, ii) a Wick rotation
of the time coordinate so that we work with the Euclidean solution,  and iii) a change to a new radial
coordinate that zooms the geometry in the near-horizon region. That
is, we perform the coordinate transformations
 \be
 \tilde{\psi}=\psi-\Omega_H t\,, \qquad t=-i\tau \,, \qquad
 r=r_+ +\frac{\Delta'(r_+)}{4}\,\rho^2\,,
  \eeq
with $\Omega_H$ being the angular velocity of the background
solution \eqref{angvel}. A final coordinate
transformation,
 \be
 \tilde{\tau}=2\pi T_H \tau\,, \qquad \hbox{with} \qquad T_H= \frac{r_+ \Delta'(r_+)}{4\pi
 h(r_+)}\,,
  \eeq
sets the period of $\tau$ to be the inverse of the horizon
temperature and avoids a conical singularity at the horizon. The
Euclidean sector of the near horizon region of the background
solution (\ref{background}) then reads
 \be
\label{NHbackground} ds^2_E \simeq \rho^2 \,d\tilde{\tau}^2 +
d\rho^2 + h(r_+)^2[d\tilde{\psi} +A_a dx^a]^2 + r_+^2 \hat{g}_{ab}
dx^a dx^b\,.
 \ee
which is a manifestly regular geometry. Indeed the polar coordinate
singularity can be removed by a coordinate transformation into
cartesian coordinates, $\tilde{\tau}=\hbox{Arctan}(y/x)$ and
$\rho=\sqrt{x^2+y^2}$. This concludes our computation of the angular
velocity and temperature of the background black hole.

To study the regularity of the perturbations, start by introducing the
manifestly regular 1-forms,
 \be \label{NHperturbations}
E^{\tilde{\tau}}=\rho^2 d\tilde{\tau}=x\,dy-y\,dx\,,\qquad E^{\rho}=\rho\,
d\rho=x\,dx+y\,dy\,.
 \ee
Consider now our TT metric perturbation $h_{\mu\nu} dx^\mu dx^\nu$.
After using the boundary conditions (\ref{BCsStationary}), which
satisfy the TT gauge conditions, we get
\begin{equation}
\begin{aligned}
h_{\mu\nu}\,dx^\mu\,dx^\nu&\simeq
 \frac{h(r_+)}{r_+} f_{\bar{0}\bar{1}}\Y \lp \rho^2 \,d\tilde{\tau}^2
 + d\rho^2\rp +f_{\bar{2}\bar{2}}\Y  \lp d\tilde{\psi} +A_a dx^a \rp^2
 + {\rm i}\frac{\Delta'(r_+)}{h(r_+)} f_{\bar{1}\bar{2}}\Y E^{\tilde{\tau}} \lp d\tilde{\psi} +A_a dx^a \rp \\
&~~~+ \frac{\Delta'(r_+)^2}{4}  f_{\bar{1}\bar{1}}\Y (E^{\rho})^2 -4 {\rm i}\frac{r_+ h(r_+)}{\Delta'(r_+)}E^{\tilde{\tau}}\lp \bar{W}'^{+}_a + \bar{W}'^{-}_a \rp dx^a + 2 r_+E^{\rho} \lp \bar{X}_a^+ + \bar{X}_a^-\rp dx^a\\
&~~~+ 2 \lp d\tilde{\psi} +A_a dx^a \rp \lp \bar{Z}_a^+ +
\bar{Z}_a^-\rp dx^a\\
&~~~+r_+^2\left[-\frac{1}{\sqrt{\lambda}}\left(H_{ab}^{++}+H_{ab}^{--}+H_{ab}^{+-}\right)+
\tilde H_L\,\hat g_{ab}\right]dx^adx^b\,.
\end{aligned}
\label{eqn:Regularhmetric}
\end{equation}
Notice that the first term on the right-hand side ensures that there is no conical singularity if $\tilde{\tau}$ has the same periodicity as the background metric. Furthermore, the remaining dependence on $\tilde{\tau}$ and $\rho$ is given by the manifestly regular 1-forms $E^{\tilde{\tau}}$ and $E^{\rho}$. Hence, $h_{\mu\nu} dx^\mu dx^\nu$ is a regular two tensor in the
cartesian coordinates $(x,y,\tilde{\psi},x^a)$ where the background
metric is regular, which confirms that the perturbations indeed preserve the
angular velocity and temperature.

 \subsection{Modes that can change the mass and angular momenta  \label{sec:ChangeMJ}}

The stationary zero-modes can potentially change the mass and angular momenta of the geometry. In this subsection, we find which of our perturbations with $\omega=0$, $m=0$ and $k=0$ can change these conserved charges.

The change on the conserved charges associated to a Killing generator $\xi$ introduced by a perturbation $h_{\mu\nu}$ can be defined via a surface boundary integral as (for TT perturbations) \cite{Abbott:1981ff,Barnich:2001jy}
\begin{eqnarray}
   Q_\xi[h,g]=-\frac{1}{32 \pi G} \int_{\partial \Sigma} \epsilon_{\alpha\beta\mu\nu} \big[
  \xi_\sigma \nabla^\nu h^{\mu\sigma}
 - h^{\nu\sigma} \nabla_\sigma \xi^\mu  
  + \frac{1}{2} h^{\sigma\nu}(\nabla^\mu\xi_\sigma
+ \nabla_\sigma\xi^\mu)\big] dx^\alpha\wedge dx^\beta \,.
\label{chargeXi}
\end{eqnarray}
The conserved charges of interest are the energy, for $\xi=-\partial/\partial t$, and the angular momenta associated with the $\lfloor (D-1)/2 \rfloor$ $U(1)$ Killing vectors $\xi=\partial/\partial \Psi_i$. The corresponding changes are denoted by $\mathcal{E}$ and $\mathcal{J}_i$, respectively. The surface integral is over a constant time hypersurface at asymptotic infinity, $\partial \Sigma$.

To compute the charges of our perturbations, we need the asymptotic behaviour of our solutions. This can be obtained from a Frobenius analysis of the Lichnerowicz equations at $r \to \infty$. We are interested in boundary conditions that preserve the asymptotic flatness of the spacetime, i.e. that decay (strictly) faster than the background geometry. We find that the behaviour of the regular perturbations is such that they decay at infinity according to
\begin{eqnarray}
 &&  \mathfrak{f}_{11}\approx \mathcal{O}\lp r^{-4-2\kappa}\rp\,, \quad \mathfrak{f}_{12}\approx\mathcal{O}\lp r^{-2\kappa}\rp\,, \quad
\mathfrak{f}_{22}\approx\mathcal{O}\lp r^{-2\kappa}\rp\,, \quad
\mathfrak{f}_1\approx\mathcal{O}\lp r^{-3-2\kappa}\rp\,, \nonumber\\
&& \tilde{\mathfrak{f}}_0\approx\mathcal{O}\lp r^{-2\kappa}\rp\,,
\quad \tilde{\mathfrak{f}}_2\approx\mathcal{O}\lp r^{-2\kappa}\rp\,,
\quad P\approx\mathcal{O}\lp r^{-2-2\kappa}\rp\,.
\label{FroebInfinity}
\end{eqnarray}

We then find the generic behaviour for the changes in the conserved charges,
\begin{eqnarray}
 && \mathcal{E}= E_0 \, r^{-2\kappa}\lp 1+ \mathcal{O}\lp r^{-1}\rp \rp\,,\nonumber\\ \label{ChargesPerturb}
 && \mathcal{J}_{\Psi_i}= A_i\, r^{2-2\kappa} + B_i\, r^{-2\kappa}
+ \mathcal{O}\lp r^{-1-2\kappa}\rp \,,
\end{eqnarray}
where  $E_0,A_i,B_i$ are functions of $\kappa$ and $r_+$ and, in particular, $A_i=B_i=0$ for $\kappa=0$.

We thus conclude that the $\kappa=0$ modes are the only ones that decay sufficiently slowly to change the mass of the geometry. Moreover these modes cannot change the angular momenta. The only modes that change the angular momenta are those with $\kappa=1$. All other modes with $\kappa\geq 2$ do not change the mass nor the angular momenta. We have explicitly done these computations for the $D=7$ case.


\bibliographystyle{hplain.bst}
\bibliography{bibliography}

\begin{thebibliography}{10}

\bibitem{Abbott:1981ff}
L.~F. Abbott and Stanley Deser.
\newblock {Stability of Gravity with a Cosmological Constant}.
\newblock {\em Nucl. Phys.}, B195:76, 1982.

\bibitem{Barnich:2001jy}
Glenn Barnich and Friedemann Brandt.
\newblock {Covariant theory of asymptotic symmetries, conservation laws and
  central charges}.
\newblock {\em Nucl. Phys.}, B633:3--82, 2002, hep-th/0111246.

\bibitem{Brown:1990di}
J.~David Brown, Erik~A. Martinez, and Jr. York, James~W.
\newblock {Rotating black holes, complex geometry, and thermodynamics}.
\newblock {\em Annals N. Y. Acad. Sci.}, 631:225--234, 1991.

\bibitem{Dias:2009iu}
Oscar J.~C. Dias, Pau Figueras, Ricardo Monteiro, Jorge~E. Santos, and Roberto
  Emparan.
\newblock {Instability and new phases of higher-dimensional rotating black
  holes}.
\newblock {\em Phys. Rev.}, D80:111701, 2009, 0907.2248.

\bibitem{Emparan:2009cs}
Roberto Emparan, Troels Harmark, Vasilis Niarchos, and Niels~A. Obers.
\newblock {Blackfolds}.
\newblock {\em Phys. Rev. Lett.}, 102:191301, 2009, 0902.0427.

\bibitem{Emparan:2009at}
Roberto Emparan, Troels Harmark, Vasilis Niarchos, and Niels~A. Obers.
\newblock {Essentials of Blackfold Dynamics}.
\newblock 2009, 0910.1601.

\bibitem{Emparan:2009vd}
Roberto Emparan, Troels Harmark, Vasilis Niarchos, and Niels~A. Obers.
\newblock {New Horizons for Black Holes and Branes}.
\newblock 2009, 0912.2352.

\bibitem{Emparan:2003sy}
Roberto Emparan and Robert~C. Myers.
\newblock {Instability of ultra-spinning black holes}.
\newblock {\em JHEP}, 09:025, 2003, hep-th/0308056.

\bibitem{Gibbons:1976ue}
G.~W. Gibbons and S.~W. Hawking.
\newblock Action integrals and partition functions in quantum gravity.
\newblock {\em Phys. Rev.}, D15:2752--2756, 1977.

\bibitem{Gibbons:2004ai}
G.~W. Gibbons, M.~J. Perry, and C.~N. Pope.
\newblock {The first law of thermodynamics for Kerr - anti-de Sitter black
  holes}.
\newblock {\em Class. Quant. Grav.}, 22:1503--1526, 2005, hep-th/0408217.

\bibitem{Gibbons:2002pq}
Gary Gibbons and Sean~A. Hartnoll.
\newblock {A gravitational instability in higher dimensions}.
\newblock {\em Phys. Rev.}, D66:064024, 2002, hep-th/0206202.

\bibitem{Gregory:1993vy}
R.~Gregory and R.~Laflamme.
\newblock {Black strings and p-branes are unstable}.
\newblock {\em Phys. Rev. Lett.}, 70:2837--2840, 1993, hep-th/9301052.

\bibitem{Gregory:1994bj}
Ruth Gregory and Raymond Laflamme.
\newblock {The Instability of charged black strings and p-branes}.
\newblock {\em Nucl. Phys.}, B428:399--434, 1994, hep-th/9404071.

\bibitem{Gubser:2001ac}
Steven~S. Gubser.
\newblock {On non-uniform black branes}.
\newblock {\em Class. Quant. Grav.}, 19:4825--4844, 2002, hep-th/0110193.

\bibitem{Gubser:2000ec}
Steven~S. Gubser and Indrajit Mitra.
\newblock {Instability of charged black holes in anti-de Sitter space}.
\newblock 2000, hep-th/0009126.

\bibitem{Gubser:2000mm}
Steven~S. Gubser and Indrajit Mitra.
\newblock {The evolution of unstable black holes in anti-de Sitter space}.
\newblock {\em JHEP}, 08:018, 2001, hep-th/0011127.

\bibitem{Hollands:2006rj}
Stefan Hollands, Akihiro Ishibashi, and Robert~M. Wald.
\newblock {A Higher Dimensional Stationary Rotating Black Hole Must be
  Axisymmetric}.
\newblock {\em Commun. Math. Phys.}, 271:699--722, 2007, gr-qc/0605106.

\bibitem{Horowitz:2001cz}
Gary~T. Horowitz and Kengo Maeda.
\newblock {Fate of the black string instability}.
\newblock {\em Phys. Rev. Lett.}, 87:131301, 2001, hep-th/0105111.

\bibitem{Hoxha:2000jf}
P.~Hoxha, R.~R. Martinez-Acosta, and C.~N. Pope.
\newblock {Kaluza-Klein consistency, Killing vectors, and Kaehler spaces}.
\newblock {\em Class. Quant. Grav.}, 17:4207--4240, 2000, hep-th/0005172.

\bibitem{Ishibashi:2003ap}
Akihiro Ishibashi and Hideo Kodama.
\newblock {Stability of higher-dimensional Schwarzschild black holes}.
\newblock {\em Prog. Theor. Phys.}, 110:901--919, 2003, hep-th/0305185.

\bibitem{Kleihaus:2007dg}
Burkhard Kleihaus, Jutta Kunz, and Eugen Radu.
\newblock {Rotating nonuniform black string solutions}.
\newblock {\em JHEP}, 05:058, 2007, hep-th/0702053.

\bibitem{Kodama:2003jz}
Hideo Kodama and Akihiro Ishibashi.
\newblock {A master equation for gravitational perturbations of maximally
  symmetric black holes in higher dimensions}.
\newblock {\em Prog. Theor. Phys.}, 110:701--722, 2003, hep-th/0305147.

\bibitem{Kodama:2009bf}
Hideo Kodama, R.~A. Konoplya, and Alexander Zhidenko.
\newblock {Gravitational stability of simply rotating Myers-Perry black holes:
  tensorial perturbations}.
\newblock 2009, 0904.2154.

\bibitem{Kunduri:2006qa}
Hari~K. Kunduri, James Lucietti, and Harvey~S. Reall.
\newblock {Gravitational perturbations of higher dimensional rotating black
  holes: Tensor Perturbations}.
\newblock {\em Phys. Rev.}, D74:084021, 2006, hep-th/0606076.

\bibitem{Martin:2008pf}
Jonathan~E. Martin and Harvey~S. Reall.
\newblock {On the stability and spectrum of non-supersymmetric AdS(5) solutions
  of M-theory compactified on Kahler-Einstein spaces}.
\newblock {\em JHEP}, 03:002, 2009, 0810.2707.

\bibitem{Moncrief:2008mr}
Vincent Moncrief and James Isenberg.
\newblock {Symmetries of Higher Dimensional Black Holes}.
\newblock {\em Class. Quant. Grav.}, 25:195015, 2008, 0805.1451.

\bibitem{Monteiro:2009tc}
Ricardo Monteiro, Malcolm~J. Perry, and Jorge~E. Santos.
\newblock {Thermodynamic instability of rotating black holes}.
\newblock {\em Phys. Rev.}, D80:024041, 2009, 0903.3256.

\bibitem{Monteiro:2009ke}
Ricardo Monteiro, Malcolm~J. Perry, and Jorge~E. Santos.
\newblock {Semiclassical instabilities of Kerr-AdS black holes}.
\newblock {\em Phys. Rev.}, D81:024001, 2010, 0905.2334.

\bibitem{Murata:2008yx}
Keiju Murata and Jiro Soda.
\newblock {Stability of Five-dimensional Myers-Perry Black Holes with Equal
  Angular Momenta}.
\newblock {\em Prog. Theor. Phys.}, 120:561--579, 2008, 0803.1371.

\bibitem{Myers:1986un}
Robert~C. Myers and M.~J. Perry.
\newblock {Black Holes in Higher Dimensional Space-Times}.
\newblock {\em Ann. Phys.}, 172:304, 1986.

\bibitem{Reall:2001ag}
Harvey~S. Reall.
\newblock {Classical and thermodynamic stability of black branes}.
\newblock {\em Phys. Rev.}, D64:044005, 2001, hep-th/0104071.

\bibitem{Reall:2002bh}
Harvey~S. Reall.
\newblock {Higher dimensional black holes and supersymmetry}.
\newblock {\em Phys. Rev.}, D68:024024, 2003, hep-th/0211290.

\bibitem{Shibata:2009ad}
Masaru Shibata and Hirotaka Yoshino.
\newblock {Nonaxisymmetric instability of rapidly rotating black hole in five
  dimensions}.
\newblock 2009, 0912.3606.

\bibitem{Stergioulas:2009zz}
N.~Stergioulas.
\newblock {Numerical simulations of black hole formation}.
\newblock {\em Lect. Notes Phys.}, 769:177--208, 2009.

\bibitem{Teukolsky:1972my}
S.~A. Teukolsky.
\newblock {Rotating black holes - separable wave equations for gravitational
  and electromagnetic perturbations}.
\newblock {\em Phys. Rev. Lett.}, 29:1114--1118, 1972.

\bibitem{Trefethen}
L.~N. Trefethen.
\newblock {\em Spectral Methods in MATLAB}.
\newblock SIAM, Philadelphia, 2000.

\bibitem{Whiting:1988vc}
Bernard~F. Whiting.
\newblock {Mode Stability of the Kerr Black Hole}.
\newblock {\em J. Math. Phys.}, 30:1301, 1989.

\bibitem{Wiseman:2002zc}
Toby Wiseman.
\newblock {Static axisymmetric vacuum solutions and non-uniform black strings}.
\newblock {\em Class. Quant. Grav.}, 20:1137--1176, 2003, hep-th/0209051.

\end{thebibliography}

\end{document}